\documentclass{arxiv}

\usepackage{txfonts}
\usepackage{graphicx}
\usepackage{amssymb} 
\usepackage{xspace}
\usepackage{hyperref}
\usepackage{multirow}
\usepackage{colortbl}
\usepackage{xcolor}
\usepackage{graphicx}
\usepackage{booktabs}
\usepackage{caption}
\usepackage{hyperref}
\usepackage{academicons}  
\usepackage{xcolor}

\newcommand\Msixty{M60$^{*}$\,}
\newcommand\sgra{Sgr\,A$^{*}$\,}
\newcommand{\aips}{\texttt{AIPS}\xspace}
\newcommand{\orcid}[1]{\href{https://orcid.org/#1}{\textcolor{gray}{\aiOrcid}}}

\title{A Jet from a Nearly Dormant Black Hole}

\author{Xiaopeng Cheng$^{1}$\thanks{These authors contributed equally to this work}\thanks{xcheng0808@gmail.com},
Hai Yang$^{2}$\thanks{hyang@camk.edu.pl}\footnotemark[1],
Jun Yang$^{3}$\thanks{jun.yang@chalmers.se},
Xiaofeng Li$^{4}$,
Feng Yuan$^{5}$\thanks{fyuan@fudan.edu.cn},
Rusen Lu$^{6,7,8}$,
Hyunwook Ro$^{1}$,
Bong-Won Sohn$^{1,9}$,
Lulu Fan$^{10,11,12,13}$,
Yihang Zhang$^{10,11}$,
Wen Chen$^{14,15}$,
Niu Liu$^{16,17}$,
John E. Conway$^{2}$,
Taehyun Jung$^{1}$}

\begin{document}

\maketitle

\begin{affiliations}
 \item Korea Astronomy and Space Science Institute (KASI), 776 Daedeokdae-ro, Yuseong-gu, Daejeon 305-348, Korea
 \item Nicolaus Copernicus Astronomical Center, Polish Academy of Sciences, Bartycka 18, PL-00-716 Warszawa, Poland
 \item Department of Physics and Astronomy, Chalmers University of Technology, Onsala Space Observatory SE-43992 Onsala, Sweden
 \item School of Computer Science and Information Engineering, Changzhou Institute of Technology, Changzhou, 213002, China
 \item Center for Astronomy and Astrophysics and Department of Physics, Fudan University, Shanghai 200438, P.R.China
 \item Shanghai Astronomical Observatory, Chinese Academy of Sciences, 80 Nandan Road, Shanghai 200030, People's Republic of China
 \item Key Laboratory of Radio Astronomy and Technology, Chinese Academy of Sciences, A20 Datun Road, Chaoyang District, Beijing 100101, People's Republic of China
 \item Max-Planck-Institut für Radioastronomie, Auf dem Hügel 69, D-53121 Bonn, Germany
 \item University of Science and Technology, Gajeong-ro 217, Yuseong-gu, Daejeon 34113, Republic of Korea
 \item Department of Astronomy, University of Science and Technology of China, Hefei 230026, China
 \item School of Astronomy and Space Science, University of Science and Technology of China, Hefei 230026, China
 \item College of Physics, Guizhou University, Guiyang 550025, China
 \item Deep Space Exploration Laboratory, Hefei 230088, China
 \item Yunnan Observatories, Chinese Academy of Sciences, Kunming 650216, Yunnan, China
 \item State Key Laboratory of Radio Astronomy and Technology, Yunnan Observatories, Chinese Academy of Sciences, Kunming 650216, China
 \item School of Astronomy and Space Science, Nanjing University, Nanjing 210023, China
 \item Key Laboratory of Modern Astronomy and Astrophysics(Ministry of Education), Nanjing University, Nanjing 210023, China
\end{affiliations}


\begin{abstract}
Most galaxies host supermassive black holes (SMBHs) that remain weakly accreting or dormant for much of their lifetimes. At the lowest accretion rates, these systems may represent the transition between active nuclei and dormant black holes, but whether they can still launch collimated jets remains unclear. The nuclei in our Galaxy (\sgra) and M31 are key examples of this regime, although no clear jet structure has yet been detected in either source. Here we report multi-frequency very long baseline interferometric observations of \Msixty\ (NGC~4649), a nearby elliptical galaxy hosting a nearly dormant SMBH with an Eddington ratio of $\sim10^{-8}$. We detect a compact two-sided jet with an unusually steep synchrotron spectrum, demonstrating that collimated outflows can persist even under nearly dormant accretion conditions. The apparent radio core exhibits an unprecedentedly steep frequency-dependent position shift toward the SMBH, locating the central engine only $\sim57\,\mu$as, corresponding to a projected distance of $\sim10$ Schwarzschild radii, upstream of the 8.37-GHz core. The observed jet morphology and steep core-shift behaviour are reproduced by general relativistic magnetohydrodynamic and radiative-transfer simulations, indicating a magnetically dominated, non-equipartition jet-launching region that departs from the standard conical equipartition picture. These results provide direct observational evidence that jet production can survive near the dormant SMBHs and establish \Msixty\ as a unique laboratory for probing jet formation on event-horizon scales in the lowest-accretion SMBH regime.

\end{abstract}

Supermassive black holes (SMBHs) are believed to reside at the centers of nearly all normal galaxies. For the majority of their lifetimes, these SMBHs accrete at rates far below the Eddington limit, leaving most in a dormant state with little to no detectable nuclear activity\cite{ho2008}. Only a small fraction of SMBHs exhibit active galactic nucleus (AGN), powered by the accretion of surrounding gas\cite{yu2002}. The accretion flows are typically classified into cold-mode and hot-mode based on the mass accretion rate. In the cold mode, as seen in bright quasars and Seyfert galaxies at relatively high mass accretion rates, accretion occurs through a geometrically thin, optically thick disk that efficiently radiates energy. In contrast, at low accretion rates ($\lambda_{\rm Edd} \lesssim 10^{-2}$), the standard thin disk is expected to truncate at larger radii, giving way to a radiatively inefficient accretion flow (RIAF)—a hot, optically thin, geometrically thick structure\cite{yuan2014}. Most AGN occupy the faint end of the luminosity function and are classified as low-luminosity AGN (LLAGN), with very low accretion rates\cite{ho2008}. At the lowest end, SMBHs in an ultra-low accretion state (\(\lambda_{\rm Edd} \lesssim 10^{-7}\)) are thought to represent a transitional phase between AGN activity and completely dormant SMBHs\cite{inayoshi2020}. Investigating such systems is crucial for testing accretion and ejection theories and for understanding the long-term evolution of SMBHs in relatively inactive galaxies, such as the nearby extreme cases \sgra and M31. The detection and characterization of jets in such ultra-low accretion systems have only recently become feasible thanks to advances in high-resolution radio interferometry and sensitive multiwavelength diagnostics\cite{boccardi2017}.

Relativistic jets are a prominent feature of many LLAGN, including a few nearby radio galaxies such as M87, 3C 84, and NGC 315\cite{gu2007,hada2011,plambeck2014}. In LLAGN, the jets are thought to be launched from within tens of Schwarzschild radii (R$_{\rm S}$ = 2GM/c$^{\rm 2}$, where $G$ is the gravitational constant, $M$ the black hole mass and $c$ the speed of light) of the black hole, where energy is extracted from the spin of the black hole\cite{blandford1977}, likely in a magnetically arrested disc (MAD) configuration\cite{eht2021viii,yuan2022}. However, at the lowest accretion rates (\(\lambda_{\rm Edd} \lesssim 10^{-7}\)), as seen in nearby quiescent nuclei like \sgra\cite{yuan2003} and M31\cite{inayoshi2018}, there is no firm detection of jets, despite their accretion flows being well-described by RIAF models. Recent observations of \sgra indicate that the accretion flow may still be in a MAD state, implying that the conditions for jet launching could still be present\cite{eht2022v}. This raises a few open questions. Are jets intrinsically suppressed below a critical accretion rate, or do they remain but become too weak or diffuse to detect with current instrumental capabilities? These discrepancies challenge current models of jet production in RIAFs and highlight the need for a deeper investigation of transitional systems on the verge of nuclear quiescence.

\begin{figure}
    \centering
    \includegraphics[width=0.8\linewidth]{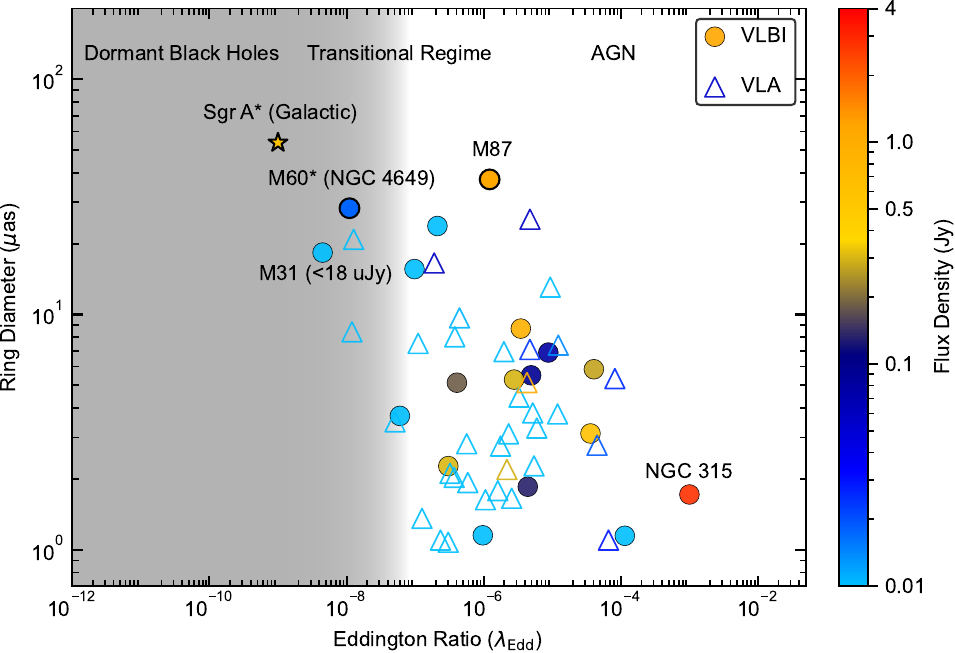}
    \caption{Photon ring diameter (5.2 $R_{\rm S}$) as a function of Eddington ratio. Filled circles and open triangles represent VLBI and VLA total flux density at 5 GHz, respectively. The color represents the total flux density. The grey region on the left denotes the dormant black holes, whereas the white region on the right denotes the active AGN. We assume the transition from \(\lambda_{\rm Edd} = 10^{-7}\). The Extended Data Table 1 lists the information for the sample of 53 sources within 200 Mpc.}
    \label{fig:sample}
\end{figure}

To identify a target source for probing jet formation on spatial scales of tens of Schwarzschild radii in extremely low accretion AGN, we systematically surveyed nearby galaxies ($\leq$ 200 Mpc) with hard X-ray luminosity and direct estimates of SMBH masses from resolved stellar or gas kinematics. We identified 53 sources (see Extended Data Table 1) with projected photon ring diameters ($\sim$ 5.2R$_{\rm S}$) larger than 1 $\rm \mu$as, corresponding to the angular scale where the jet-launching regions may begin to be resolved. Figure \ref{fig:sample} presents the relation between accretion rate and ring diameter for this sample, with radio flux densities from Very Large Array (VLA) or Very Long Baseline Interferometry (VLBI) observations encoded in the symbol color and shape. In Figure \ref{fig:sample}, \Msixty (also known as NGC~4649) emerges as the most promising extra-galactic source with a ring diameter of 28.3 $\mu$as\cite{ben2024} and an accretion rate of 10$^{\rm -8}$. \Msixty is a luminous elliptical galaxy located in a group at the eastern edge of the Virgo cluster. It harbors an SMBH with a dynamically measured mass of 4.5 $\times$ 10$^{9}$ M$_{\odot}$ (see Methods), located at a distance of D $\sim$ 16.3 Mpc\cite{lee2017}. The source is classified as an LLAGN with a nuclear hard X-ray luminosity of $\rm \sim 10^{39}$ erg $s^{-1}$ and a bolometric Eddington ratio of $\rm L_{\rm bol}/{L_{\rm Edd}} \sim 10^{-8}$ (see Methods). This nearly dormant SMBH is the best-known extragalactic analogue of \sgra.

We performed multi-frequency observations of \Msixty with the Very Long Baseline Array (VLBA) at 1.46, 1.64, 2.32, 4.87, and 8.37 GHz on 7 March and 2 April 2024, quasi-simultaneously. \Msixty lies at a high Galactic latitude (b = 74$^{\circ}$.3) and its radio emission is unaffected by scattering broadening\cite{koryukova2022}. The phase-referencing technique was used to determine the frequency-dependent positions of the radio core, with respect to the nearby calibrator J1246$+$1153 (separated by 0.78$^{\circ}$) (see Methods).
At the given observation frequency, the core is located in the region where the optical depth due to synchrotron self-absorption equals unity\cite{blandford1979}.
As such an apparent core's position is expected to move toward the central SMBH with increasing frequency — a phenomenon known as the core shift effect\cite{bartel1986,lobanov1998,OSullivan2009}.
The amount of core shifts, in particular at high frequencies, can serve as a diagnostic of whether the emission in the innermost region is dominated by an accretion flow or a jet\cite{fraga2026}, even when the source remains unresolved.
Furthermore, it provides a precise position of the central SMBH at the upstream end of a jet.
To search for a faint extended jet, we also conducted a 6-hour deep observation of \Msixty at 1.6 GHz with the European VLBI Network (EVN) on 23 May 2024.

\begin{figure}[htb]
    \centering
    \includegraphics[width=0.6\linewidth]{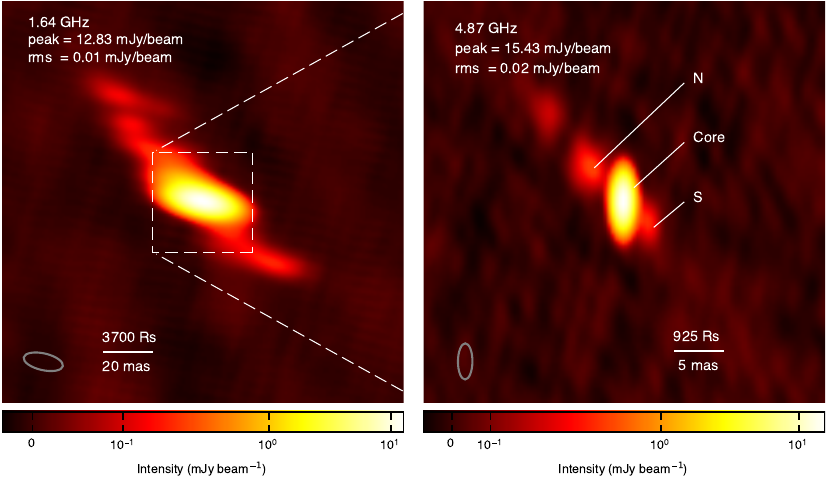}
    \includegraphics[width=0.3\linewidth]{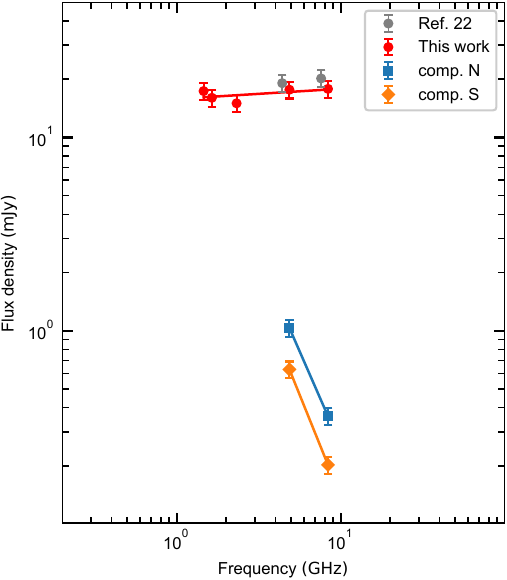}
    \caption{VLBI images of \Msixty at 1.64 and 4.87 GHz, and the quasi-simultaneous spectra. Left: 1.64 GHz naturally weighted EVN + VLBA combined CLEAN image showing the large scale of the radio jet; Middle: 4.87 GHz VLBA naturally weighted CLEAN image showing the close-up view of the sub-parsec scale jet; Right: Best-fitting power-law spectra ($S_\nu \propto \nu^{\alpha}$) for the total flux density (red) and for components N (blue) and S (orange). The two gray points for the total flux are from VLBI geodetic observations in 2022, with values taken from \cite{li2024}.}
    \label{fig:img}
\end{figure}

Our VLBI observations reveal a two-sided core-jet structure in \Msixty, with additional confirmation from the high-sensitivity EVN image. 
We identify the unresolved compact component as the radio core, which shows a flat radio spectrum with a spectral index $\alpha$ of about 0.11 ($S \propto \nu^{-\alpha}$ ) between 1.46 and 8.37 GHz. The radio core was also detected in the previous short (3-minute) VLBI geodetic observation at C/X bands in 2022\cite{li2024}.
In Figure \ref{fig:img} left, we show the deep VLBI image of \Msixty at L band, which was made by combining the EVN and VLBA visibility data.
The jet is characterized by a compact core with a two-sided jet extending $\sim$100 mas ($\sim$18000 R$_{\rm S}$) along the Northeast-Southwest direction.
The higher-resolution 4.87\,GHz image in the middle of Figure \ref{fig:img} reveals the same overall morphology.

The total flux densities we measured at 4.87 and 8.37 GHz are consistent with previous observations\cite{li2024}, suggesting that the source is in a steady or quiescent state. The spectral indices of the inner jet components between 4.87 and 8.37 GHz are extremely steep (Figure \ref{fig:img} right), $-$1.94 for component N and $-$2.10 for component S (see Methods), indicative of optically thin synchrotron emission. The jet viewing angle and intrinsic jet speed can be widely constrained using the approaching and receding jet components\cite{2012rjag.book.....B,cheng2023} (see Methods).
The estimated intrinsic speed of the inner pair of jet components is $\geq 0.07\, c$.
If we use a jet inclination angle of 30$^{\circ}$\cite{li2024}, the jet speed will be $0.08\, c$, indicating a mildly relativistic motion at a de-projected distance of r $\simeq$ 10$^{\rm 3}$ R$_{\rm S}$.
We note that the jet speed seems to be quite slower than most of the nearby LLAGN\cite{park2019,park2021}, which has a relativistic ($\geq 0.4\, c$) jet at a distance of $\simeq$ 10$^{\rm 3}$ R$_{\rm S}$.
The relatively slow, faint, and steep-spectrum jet observed in \Msixty is likely a consequence of its extremely low accretion rate, a regime where both insufficient mass loading and inefficient conversion of magnetic energy to kinetic power inherently limit the jet power. 
If the jet inclination is at a very large angle ($\rm \theta \geq 80^{\circ}$), the intrinsic jet speed could be comparable to that of nearby sources ($\approx 0.4\, c$).

\begin{figure}[ht!]
    \centering
    \includegraphics[width=0.45\linewidth]{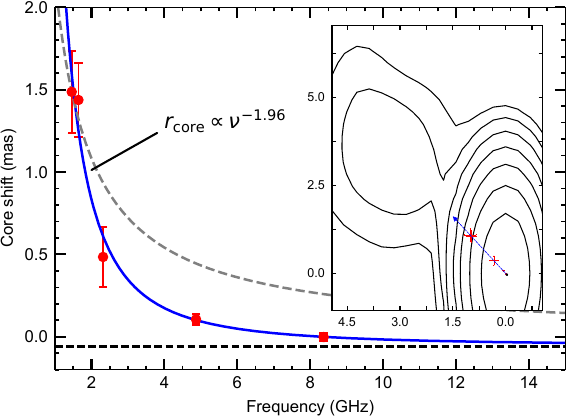}
    \includegraphics[width=0.45\linewidth]{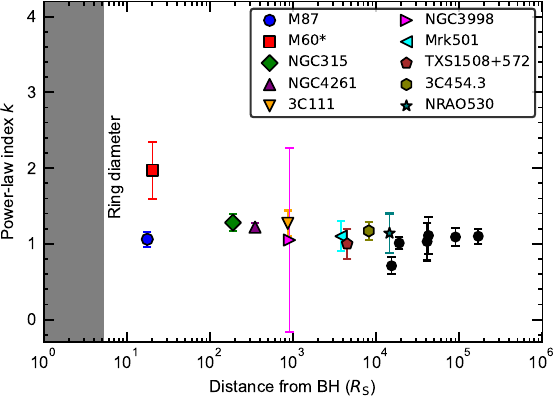}
    \caption{Left: Separation of the radio core from the reference 8.37 GHz core as a function of frequency. The blue solid curve shows the best-fit core-shift relation, while the gray dashed curve represents the conical equipartition relation. The black horizontal dashed line indicates the asymptotic position of the solid curve, located at 57\,$\mu$as upstream from the 8.37 GHz core. The inset illustrates the horizon-scale localization of the central engine through the core-shift analysis, showing the measured core positions at different frequencies (red crosses) and the inferred black-hole position (black point); Right: Power-law index k as a function of the deprojected distance from the black hole (in Schwarzschild radii) for a sample of nearby AGNs. The distances correspond to the highest-frequency core positions in the core shift studies. The data points include 1 $\sigma$ error bar, and the ten sources with the closest core–black hole separations are highlighted using distinct colors and marker styles. For NGC~3998, no viewing angle has been measured or estimated; thus, the distance represents the projected separation between the core and the black hole.}
    \label{fig:kr}
\end{figure}

The left panel of Figure \ref{fig:kr} displays the frequency-dependent core shifts measured along the jet axis.
Since the reference source J1246$+$1153 also exhibits core shifts, the observed position offsets include the contributions from both sources.
Because the jet direction of J1246$+$1153 is nearly orthogonal (~108$^{\circ}$) to that of \Msixty, the core shifts can be firmly decoupled (see Methods).
This geometry allows us to robustly measure the core shifts in \Msixty along its jet direction\cite{hada2011}.
The measured core shifts between the L and X bands are very large  and significant at levels exceeding 3$\sigma$.
The measured core positions converge smoothly with increasing frequency, tracing a trajectory that asymptotically approaches the upstream end of the jet, presumed to mark the central engine.
Using 8.37 GHz as the reference frequency, we fitted the core positions as a function of frequency using a power-law model\cite{Konigl1981}, yielding r $\rm \propto \nu^{-k}$ (k $>$ 0). 
An MCMC analysis yields a core-shift index of $k = 1.96^{+0.48}_{-0.38}$ between 1.46 and 8.37\,GHz (Fig.~\ref{fig:kr}, left). The posterior probability that $k$ exceeds the equipartition value of unity is 99.6\%.
The central SMBH location lies 57 $\pm$ 50 $\mu$as upstream of the 8.37 GHz physical core (dashed line in Fig. \ref{fig:kr} left), corresponding to a projected distance of 10 $\pm$ 9 R$_{\rm S}$.

The power-law index \(k\) characterizing the core shift reflects both the geometry of the jet and the underlying physical conditions, including the electron energy distribution and the radial profiles of magnetic field strength and particle density along the jet \cite{blandford1979}. In the case of a freely expanding conical jet under energy equipartition between magnetic fields and particles, \(k\) is expected to be unity \cite{lobanov1998}. 
Previous core-shift measurements in nearby AGNs have typically yielded power-law indices \(k\) close to unity\cite{OSullivan2009,Sokolovsky2011,hada2011}, consistent with this expectation\cite{lobanov1998}. Notable exceptions include NGC~315, which exhibits a modest deviation from equipartition with \(k = 1.28 \pm 0.11\)\cite{Ricci2025}, and 3C~454.3, where temporal variations in \(k\) from \(\sim0.6\) to \(\sim2.2\) have been reported \cite{Chamani2023}. However, in the case of 3C~454.3, these fluctuations have been attributed to opacity changes during flaring events, rather than intrinsic changes in jet structure. Our core-shift measurements for \Msixty reveal a much steeper frequency dependence, with \(k \approx 2\), representing a significant deviation from equipartition, as shown by the left panel of Figure \ref{fig:kr}. Figure~\ref{fig:kr} right shows the distribution of power-law indices as a function of the true core–black hole distance at the highest observing frequency in nearby galaxies (see Extended Data Table 2), corrected for projection effects. As shown in this Figure, \Msixty stands out as the only source with such a steep power-law index, and its core is located at a distance of only \(\sim20\,R_{\rm S}\) from the black hole—second only to M87 among known sources. The flux densities measured at 4.87 and 8.37 GHz are consistent with the previous observations, indicating a steady, non-flaring state. The VLBI images (see Methods) further confirm that the inner jet structure does not significantly contaminate the core position, suggesting that the core shift reflects the intrinsic jet properties. 

The large observed value of \(k \approx 2\) thus suggests that the jet in this ultra-low accretion rate AGN may be in a non-equipartition state, and/or the power-law index of relativistic electrons in the jet has a special value. 
It is likely that the steep core shift is the results of the combined effects of steep power-law distribution of nonthermal electrons, the magnetic dominance, and parabolic jet geometry. To test this scenario, we have performed GRMHD numerical simulation of jet formation and then using the simulation data to calculate the radiation of the jet coming from the synchrotron emission of nonthermal electrons accelerated by magnetic reconnection. The details of the simulation and the calculation of the jet radiation are presented in Methods. Our results confirm the above speculation. We find that when the accretion flow is described by an MAD around a rapidly spinning black hole, and when the power-law index of accelerated electrons is assumed to be $s\approx 6$, the synthetic images yield a steep relation, \(r_{\rm core}\propto\nu^{-k_{\rm sim}}\) with \(k_{\rm sim}\simeq1.87\), close to the observed value. In this model, the inner jet remains strongly magnetized and parabolically collimated over the scales probed by the radio core shift. The energy-density maps further show that the emitting jet region has \(u_B/u_e>1\), indicating a clear departure from equipartition. By contrast, our tests with SANE models produce a much shallower frequency dependence of the core position and therefore do not reproduce the observed steep core shift.  
We note that, the steep non-thermal electron energy distribution adopted in the GRRT calculation, \(p=6\), is also required to reproduce the extremely steep optically thin spectra of the jet, as we will describe in section 3 of Methods.

Extrapolating our fit indicates that the offset between the 8.37\,GHz core and the black hole is only $57 \pm 50\,\mu\mathrm{as}$, implying the amount of core shift between 22 and 86 GHz of less than $100\,\mu\mathrm{as}$. If we assume a similar inclination angle ($\rm \theta = 30^{\circ}$) in \sgra, this may explain the non-detection of a measurable core shift in previous observations of Sgr~A*\cite{Bower2015}. The enhanced sensitivity and angular resolution of the next-generation Event Horizon Telescope (ngEHT), particularly when using frequency phase transfer (FPT), would offer a promising path to detect such subtle core shifts in Sgr~A* and similar low-luminosity AGN.

Recent studies suggest that the magnetic field strength of M87 at a radius of a few R$\rm _{S}$ is in the range of 1–30 G\cite{eht2021,Ro2023}. Similarly, ALMA spectral modeling of \sgra indicates a magnetic field strength of 10–50 G at scales of a few R$\rm _{S}$, based on thermal synchrotron emission fits\cite{bower2019}. The new EHT result shows a consistent value of 26 G at 7 R$\rm _{S}$\cite{eht2024viii}. These results suggest that the magnetic field strengths in the innermost regions of \sgra and M87 are broadly consistent. 
In our physical scaling for M60*, the simulated comoving magnetic-field strength reaches \(B\sim0.1\)--\(1~{\rm G}\) in the innermost jet region and decreases to \(B\sim10^{-2}\)--\(10^{-1}~{\rm G}\) on scales of \(\sim10^2\)--\(10^3\,r_{\rm g}\). These values are lower than the field strengths inferred for M87 and Sgr A* at a few Schwarzschild radii, as expected for an ultra-low-accretion-rate system with \(\lambda_{\rm Edd}\sim10^{-8}\). Nevertheless, the comparison of magnetic and electron energy densities shows that the jet is still magnetically dominated over most of the emitting region. Therefore, M60* may represent a lower-power analogue of the magnetically dominated jet bases inferred in M87 and Sgr A*, extending such physics into the nearly dormant SMBH regime. Our core-shift measurements in \Msixty have reached spatial scales of a few R$\rm _{S}$ from the SMBH, making it the only known ultra-low accretion rate AGN where direct measurements of magnetic field strength at event-horizon-scale resolution are achievable. This would provide a critical observational test of accretion models developed for ultra-low accretion rate AGN, such as \sgra.

The detection of a compact but faint jet in \Msixty---despite its extremely low accretion rate---provides a unique observational perspective on the nature of outflows in nearly dormant SMBHs, such as \sgra at the center of our Milky Way. Although no extended jet has yet been directly imaged in \sgra, the steep core-shift index observed in \Msixty suggests that collimated outflows can survive even under extremely weak accretion conditions while exhibiting steep radio spectra that render them increasingly difficult to detect at mm-VLBI frequencies. This raises the possibility that a similarly weak and compact jet exists in \sgra but remains below current observational limits, whereas severe interstellar scattering toward the Galactic Center precludes a direct search for such structures with current VLBI arrays. Future facilities such as the ngVLA and ngEHT\cite{Chavez2024} may provide the sensitivity and angular resolution required to probe this regime directly. More importantly, \Msixty extends direct observational access to a previously unexplored region of parameter space at ($\rm \lambda_{\rm Edd}\lesssim10^{-8}$), where the coupling between radiatively inefficient accretion flows, magnetic-flux accumulation, and jet production remains poorly constrained. The unusually steep core-shift relation and horizon-scale localization of the central engine suggest that magnetically dominated outflows can persist even when the radiative output of the accretion flow becomes negligible. The combination of multi-frequency core-shift measurements, jet characterization, and spectral diagnostics presented here is, in principle, applicable to other nearby low-accretion LLAGN. Extending such studies to a statistically meaningful sample will provide direct observational tests of how jet production evolves across the transition from powerful radio galaxies to nearly dormant SMBHs.


\newpage

\begin{methods}
\section{Observations and data reduction}

To discriminate the dominant emission model of \Msixty, we used two independent observation methods. First, we measured the frequency-dependent core position shift, the so-called core shift, using VLBA multi-frequency phase-referencing observations.  Second, we observed the source in high-dynamic-range imaging with the EVN at the L band to detect the possible weak, short, or fully suppressed sub-pc scale jet. Here, we provide a detailed description of the VLBI observations and the applied data reduction procedures. 

We carried out phase-referenced VLBI observations at 1.46, 1.64, 2.32, 4.87, and 8.37 GHz using the VLBA in 2024 (project code: BC296). The 1.46, 1.64, and 2.32 GHz observations were conducted on 7 March 2024 using nine antennas with a recording rate of 2048 Mbps (dual polarization, 8 $\times$ 32~MHz sub-bands per polarization, 2-bit quantization). The St. Croix antenna could not participate in the experiment due to the AZ wheel failure. The 4.87 and 8.37 GHz observation was conducted on 2 April 2024 using all ten antennas with a recording rate of 4096 Mbps (dual polarization, 4 $\times$ 128~MHz sub-bands per polarization, 2-bit quantization). The data were correlated using the DiFX software correlator \cite{deller2007,deller2011} at the Array Operations Center in Socorro (New Mexico, U.S.) with 256 frequency points per channel and 1-s integration time. The target \Msixty and the phase calibrator J1246 + 1153 were observed in a 2–3–2 min nodding cycle, with the total time on source for \Msixty about 40 mins at each frequency. Because of the weakness of J1246+1153, we used the inverse phase-referencing technique. Here, \Msixty is used as the phase reference source for determining the delay and rates for the phases. The derived position offset of the faint source J1246+1153 corresponds to the inverse of the position offset of \Msixty. During the post-data reduction, a nearby blazar B1252+199 was used as the fringe finder and bandpass calibrator. 

The data were reduced following standard procedures in \aips~\cite{greisen2003}. A-prior amplitude calibration was carried out using the system temperatures and antenna gains measured at each station during the observation. Ionospheric dispersive delays were corrected from a map of total electron content provided by the Global Positioning System satellite observations and the earth orientation parameters were corrected using the measurements for the U.S. Naval Observatory database. Phase errors due to the antenna parallactic angle variations were removed. The instrumental single-band delays and phase offsets were corrected using 3 minutes of observational data from B1252+199. After inspecting the data and flagging, global fringe fitting was performed on \Msixty with a 1-minute solution interval and a point-source model by averaging over all the sub-bands. 
We further corrected the source structure phases due to the deviation of the point source model by imaging and self-calibrating the \Msixty data.
Finally, we applied the phase solutions from \Msixty to J1246+1153 by linear interpolation.

Since J1246+1153 has an extended jet structure towards the northwest (Figure \ref{fig:J1246}), the observed position shifts of J1246+1153 in these phase-referenced images are the combination of the core shifts from both sources. Also, the jet directions of the two sources are roughly perpendicular to each other. We therefore measured the relative positions between the two sources in both directions at each frequency. The brightness peak positions for each source were identified by using MCMC fitting to the visibility data: fully self-calibrated visibilities for \Msixty, whereas the phase-referenced visibilities for J1246+1153 (see the next section for details). 

The high-dynamic range imaging observation was carried out with the EVN at 1.66 GHz on 23 May 2024 (project code: EL075). We recorded data at 1024 Mbps (dual polarization, 4 $\times$ 32~MHz sub-bands per polarization, 2-bit quantization). There are 9 stations,  Jodrell Bank, Westerbork, Effelsberg, Medicina, Noto, Onsala85, Torun, Hartebeesthoek, and Irbene, participating in the 6-h observation. Because the \Msixty is relatively bright at 1.66 GHz and our high-dynamic range imaging request, the phase-referencing observation was not used. This allowed us to reach an on-source time of 286 min. The bright blazar 3C 273 was used as a fringe finder and bandpass calibrator. The correlation was done by the EVN SFXC software correlator\cite{keimpema2015} at JIVE (Joint Institute for VLBI, ERIC) using the typical correlation parameters. We calibrated the data following the EVN data reduction guide\footnote{\href{https://www.evlbi.org/evn-data-reduction-guide}{https://www.evlbi.org/evn-data-reduction-guide}} using the AIPS. The amplitude and parallactic angle calibrations were applied using the CL table provided by the EVN pipeline. Instrumental delays were corrected by running the \aips task \textsc{fring} on 3C 273. Since the L-band structure of 3C 273 could not be accurately modeled as a single-point source or Gaussian, we derived a multi-point source model from its observed visibilities and input this model into \aips for more precise delay calibration. Bandpass calibration was performed similarly using the 3C 273 data and the source model. We then performed a global fringe fitting with a 1.5-minute solution interval, applying an SNR cutoff of 4. The \textsc{calib} task was used to correct for left-right circular polarization differences. Finally, the data were split into single source mode and exported to disk for further processing.

The deconvolution, self-calibration, and imaging of the AIPS-calibrated data were carried out in Difmap. When the data were imported into Difmap, we averaged the data with 10 s to improve the SNR of the data. We first deconvolved visibility using the \textsc{modelfit} command in Difmap to roughly identify the true jet structure. We then re-performed deconvolution using the \textsc{clean} command to improve the consistency between the observed visibility and the model. We ran self-calibrations with a solution interval of 1 min. These clean models were loaded into the \aips task \textsc{fring} to improve their solutions and into the task \textsc{calib} to calibrate the errors between the RR and LL polarizations. After that, the data are reloaded to Difmap for the hybrid imaging process. The final image reaches a noise level of $6.4\,\mu\mathrm{Jy\,beam^{-1}}$.

\begin{table}[htbp]
\centering
\caption{VLBA imaging results of M60*}
\begin{tabular}{cccccc}
\toprule
Freq. & Beam & S$_{\rm peak}$ & rms & S$_{\rm tot}$ & S$_{\rm core}$  \\
(GHz) & (mas$\times$mas, deg) & (mJy beam$^{\rm -1}$) & (mJy beam$^{\rm -1}$) & (mJy) & (mJy) \\
(1)   & (2) & (3) & (4) & (5) & (6) \\
\midrule
1.46  & 19.87 $\times$ 6.00, $-$15.90 & 12.71 & 0.03 & 17.4$\pm$0.9 & 12.8$\pm$0.6 \\
1.64  & 17.61 $\times$ 5.23, $-$16.42 & 12.83 & 0.04 & 16.0$\pm$0.8 & 13.2$\pm$0.7 \\
2.32  & 14.03 $\times$ 3.81, $-$17.40 & 14.82 & 0.06 & 15.0$\pm$0.8 & 15.0$\pm$0.8 \\
4.87  & 3.64 $\times$ 1.46, $-$0.34   & 15.43 & 0.02 & 17.7$\pm$0.9 & 15.5$\pm$0.8 \\
8.37  & 2.24 $\times$ 0.86, $-$5.10   & 16.27 & 0.02 & 17.8$\pm$0.9 & 15.6$\pm$0.8 \\
\bottomrule
\end{tabular}
\vspace{10pt}
\label{tab:beamsize}
\end{table}

\section{Sources of position errors}

\subsection{Statistical position errors} 

The brightness peak positions for J1246+1153 were identified by modelfitting to the visibility data. Successful phase-referencing at all frequencies allowed us to produce J1246+1153’s phase-referenced images with sufficient signal-to-noise ratios (26, 32, 14.8, 70, and 76 at 1.46, 1.64, 2.32, 4.87, and 8.37 GHz). We estimated the statistical position errors with a Markov Chain Monte Carlo (MCMC) simulation\cite{Salafia2022}. The initial models consist of one or more circular Gaussian components obtained using the \texttt{modelfit} function in \texttt{difmap}. We enhance the weighted value of the input data by a scale factor \(1/\chi_\nu^{2}\), to provide a more accurate estimation of the statistical errors of the parameters. This results in the position errors of 81, 121, 147, 12 and 8 $\mu$as at each frequency.

\subsection{Structure of J1246+1153}  

In phase-referenced VLBI observations, the source structure in the phase calibrator can lead to systematic position shifts between different frequencies, particularly if the calibrator exhibits significant extended emission. In our observations, the phase calibrator J1246+1153 displays a highly compact core and one/two-sided jet structures (Figure \ref{fig:J1246}). To quantify the structure and its stability, we performed Gaussian model fitting to the visibility data. The derived jet axis of J1246+1153 is oriented along a position angle of $-64.7^\circ \pm 1.2^\circ$, consistent with the previous study\cite{cheng2021}. This direction is nearly orthogonal ($107.5^\circ\pm1.4^\circ$) to the jet axis of \Msixty. We found the position angles of $-65.6^\circ\pm1.2^\circ$, $-63.7^\circ\pm2.1^\circ$, at 4.9 and 8.4 GHz respectively. Based on our position measurements, we estimated the contributions of J1246+1153’s core-shift in \Msixty's jet direction to be $\sim$ 0.6 mas (between 1.46 and 8.37 GHz) to 0.05 mas (between 4.87 and 8.37 GHz).  Therefore, taking the direction error of the total core-shift of J1246+1153 from $1.2^\circ$, the potential contributions of the core-shift in the \Msixty's core-shift direction are estimated to be from 0.04 mas (between 1.46 and 8.37 GHz) to 0.004 mas (between 4.87 and 8.37 GHz).

\subsection{Ionosphere}  Propagation delays caused by the dispersive ionospheric medium are the main source of the phase errors in VLBI observations. At low frequencies (2 GHz and lower), the ionosphere can cause large unmodeled dispersive delays. Even at high frequencies (e.g., 8 GHz), the ionosphere can also be important, depending on the atmosphere during the observation. This can be of particular importance in phase referencing observations, where phases must be interpolated over weak sources. These frequency-dependent position errors contribute to the frequency-dependent position offsets that are particularly relevant in core-shift measurements. These errors are found to scale with $\nu^{-2}$\cite{hada2011}. To remove the ionospheric phase offsets, we applied a global ionospheric model derived from GPS measurements. The AIPS task TECOR corrects for the ionosphere by using such ionospheric models.  The potential remaining errors $\delta\theta$ after TECOR can be estimated through the equation
\[
\delta \theta \sim \frac{c \cdot \delta \tau_{0,\mathrm{ion}}}{B} \sec Z \tan Z \cdot \delta Z,
\]
where $\delta \tau_{0,\mathrm{ion}} = 40.3 \cdot \delta \mathrm{TEC} / \nu^2$, TEC is the total electron content, $c$ is the speed of light, Z is the zenith angle, $\delta$ is the angular separation between \Msixty and J1246+1153, and B is the baseline length. 
Owing to the nighttime observations near the 2024 solar maximum, we fitted the GPS model and found a relatively high value of total electron content (TEC) of $1.5 \times 10^{17}$\,m$^{-2}$ during our observations. Global ionospheric model based on GPS satellites have an accuracy of about 10-20\%. Adopting the uncertainty of 20\% (corresponding to $\delta$TEC $\sim$  $3 \times 10^{16}$\,m$^{-2}$), the frequency-dependent residual position uncertainties of approximately 229, 181, 91, 21, and 7\,$\mu$as at 1.46, 1.64, 2.32, 4.87, and 8.37\,GHz for $Z = 50^\circ$, $\delta Z = 0.78^\circ$, and $B = 8612$\,km.

Although the AIPS task TECOR was applied to correct for ionospheric delays using global models, the residual errors remain significant at low frequencies and dominate the total astrometric error budget.

\subsection{Troposphere} The troposphere introduces a frequency-independent delay that leads to systematic astrometric shifts in phase-referenced VLBI observations. These non-dispersive delays affect all frequencies equally, contributing to image distortion and absolute position offsets. For our observations, the zenith angle was $Z \sim 50^\circ$, with an angular separation of $\delta Z = 0.8^\circ$ between NGC4649 and J1246+1153. We adopt a typical zenith delay misestimation of $\sim$3\,cm based on the standard VLBA correlator model. Substituting this into the equation (with $c\delta\tau_{0,\mathrm{trop}} = 3 \times 10^{-2}$\,m and using the same projection geometry as in the ionospheric case) yields an absolute position uncertainty of $\sim$20\,$\mu$as. Thanks to the availability of a nearby calibrator, the separation angle is small, which helps minimize this tropospheric contribution in our measurements.

\subsection{Core identification} The core of \Msixty is compact and core-dominated, with only a faint, non-continuous jet structure (Figure \ref{fig:M60}), which minimizes the impact of jet–core blending on core identification. To estimate the position uncertainties of the core identification, we identify the core position using two independent methods. First, we performed Gaussian model fitting to the calibrated visibility data, identifying the centroid of the fitted circular Gaussian as the core position. Second, we applied an MCMC fitting to the visibility data to determine the position of the brightness peak as the core. The difference between the results from the two methods at each frequency was adopted as the uncertainty in core identification, yielding values of 36, 42, 45, 3, and 2 $\mu$as at 1.46, 1.64, 2.32, 4.87, and 8.37 GHz, respectively.

We applied the same methods to the phase-reference calibrator J1246+1153. This source also shows no significant internal jet structure, and the fitted core positions from both the model fitting and MCMC fitting methods were also consistent across all frequencies. The resulting core identification uncertainties are thus negligible in the direction of \Msixty.

\section{Jet parameters}

The VLBI morphology of \Msixty reveals a roughly symmetric two-sided jet structure. The integrated flux density of component N is 1.033 mJy at 4.87 GHz and 0.361 mJy at 8.37 GHz, corresponding to an extremely steep spectral index of $-$1.94. Similarly, component S exhibits an extremely steep spectral index of $-$2.10, with a flux density of 0.630 mJy at 4.87 GHz and 0.202 mJy at 8.37 GHz. The spectral similarity between N ($-$1.94) and S ($-$2.10) suggests negligible absorption toward the receding jet, as significant optical depth would result in a flatter or inverted spectrum. These features make \Msixty an excellent candidate for applying the approaching jet to receding jet flux ratio methods to constrain the intrinsic jet velocity and inclination angle, under the assumption of intrinsic symmetry and no absorption.

Based on the assumption of symmetry and no absorption, the observed flux density ratio $R_{\rm flux}$ between the approaching and receding jets is 
\begin{equation}
R_{\rm flux}=\frac{S_{\rm a}}{S_{\rm r}} = \left(  \frac{1 + \beta\cos\theta_{\rm v}}{1-\beta\cos\theta_{\rm v}} \right) ^{2-\alpha},
\label{eq4}
\end{equation}
where $S_{\rm a}$ and $S_{\rm r}$ are the flux densities of the approaching and receding components respectively,  $\beta$ is the intrinsic jet speed in the unit of the light speed c, $\theta_{\rm v}$ is the viewing angle between the jet axis and the line of sight, $\alpha$ is the spectral index of the jet. If we use the flux density of the inner approaching jet 0.361 mJy at 8.37 GHz, the flux density of the receding jet 0.202 mJy at 8.37 GHz, and the spectral index $-$2, we can derive the ratio of 1.79 and a constraint of $\beta\cos\theta_{\rm v}$ = $0.07\, c$. This gives $\beta >$ $0.07\, c$, $\theta_{\rm v} <$ 86$^{\circ}$.

\section{Black hole mass and accretion rate estimates}
The mass of the supermassive black hole (SMBH) at the center of the \Msixty--the optical counterpart of \Msixty--has been investigated in several studies\cite{shen2010,woo2013,Paggi2014,paggi2017}. Based on the optical stellar population synthesis model, the reported value is $\rm 4.5\pm1.0 \times 10^{9}~M_{\odot}$\cite{shen2010}. X-ray observations from \textit{Chandra/XMM-Newton}, assuming the temperature profile of the hot interstellar medium under the assumption of hydrostatic equilibrium, yield a black hole mass ranging $\rm 3.5 - 5.5 \times 10^{9}~M_{\odot}$\cite{Paggi2014,paggi2017}. The recent black hole mass (M$_{\rm BH}$)–stellar velocity dispersion ($\sigma_*$) relation study gives a consistent value of $\rm 4.7\pm1.1 \times 10^{9}~M_{\odot}$\cite{woo2013}. All three methods yield M$_{\rm BH}$ values consistent with each other and can therefore be considered accurate. As the dynamics of stellar can be directly observed, here we used the value of $M_{\rm BH} = (4.5\pm1.0)\times10^{9}~M_{\odot}$.

To compute the nucleus bolometric luminosity, the best way is measured directly from their broadband spectral energy distribution (SED). As \Msixty hosts an LLAGN, which has a faint nucleus, making its emission difficult to isolate from the host galaxy. For this study, we circumvent this difficulty by using the measure of the nuclear 2-10 keV luminosity to estimate the AGN bolometric luminosity. This is one of the most widely used bolometric correction relations as a function of hard X-ray luminosity for AGN. In practice, the bolometric correction $\rm L_{bol}/L_{2-10 keV}$ is generally using 15 for LLANGs in previous studies\cite{duras2020}. The nuclear luminosity of \Msixty is $\rm 4.2 \times 10^{38} erg s^{-1}$ in the 2 - 10 keV energy band\cite{Paggi2014}. This gives $\rm L_{bol} = 6.3 \times 10^{39} erg s^{-1}$. Given the mass of the SMBH in above, the Eddington luminosity $\rm L_{Edd}$ is  $\rm 6.3 \times 10^{47} erg s^{-1}$ and then the normalized mass accretion rate $\dot{m}$ is $\rm L_{bol}/L_{Edd} = 10^{-8}$. 

\begin{figure}
    \centering
    \includegraphics[width=0.4\linewidth]{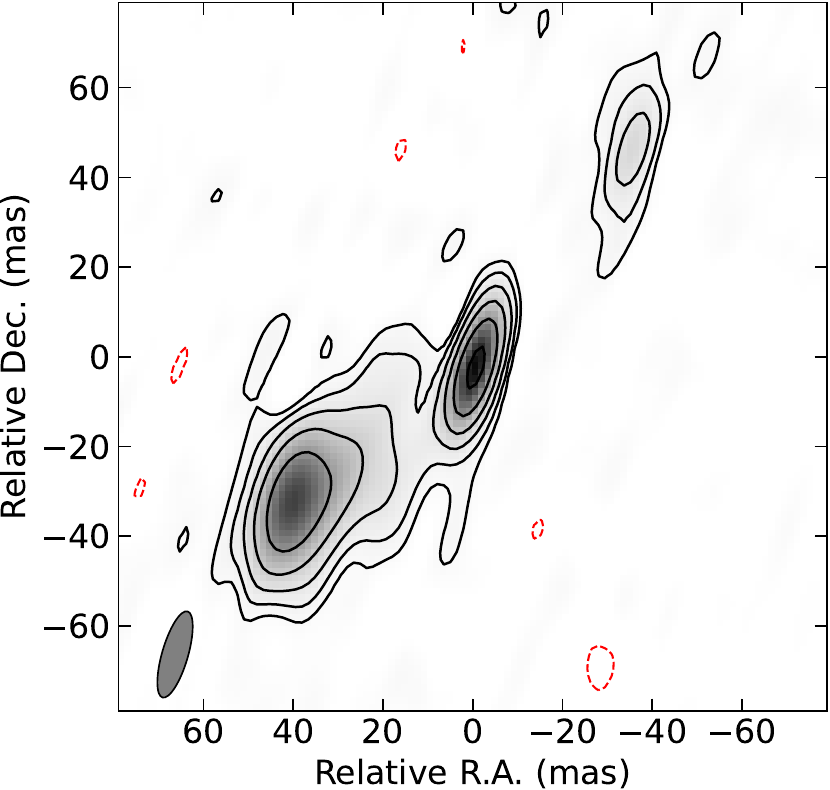} 
    \includegraphics[width=0.4\linewidth]{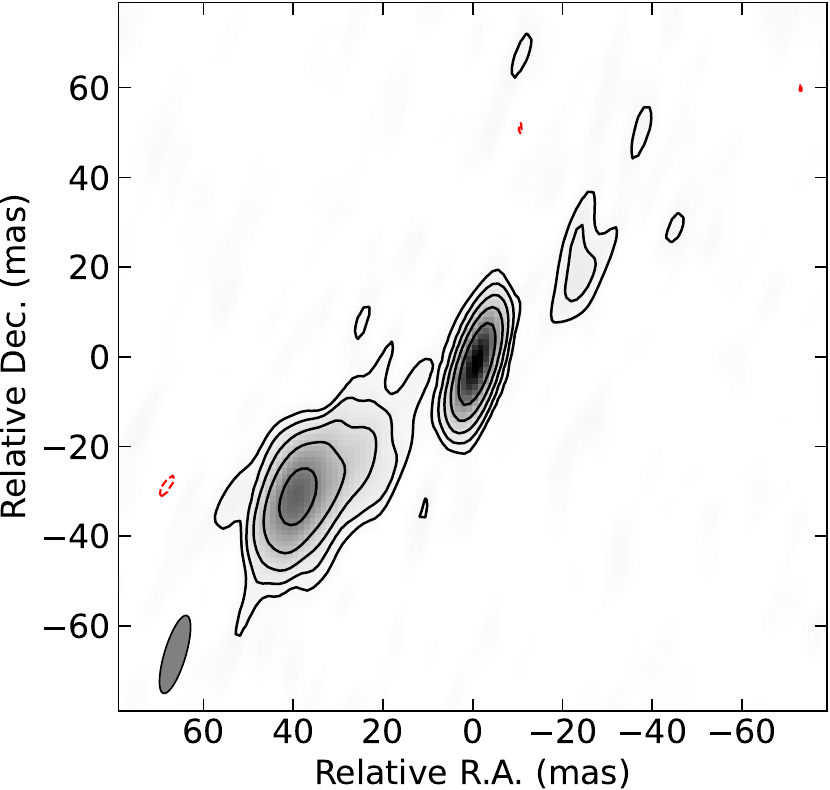} 
    \includegraphics[width=0.4\linewidth]{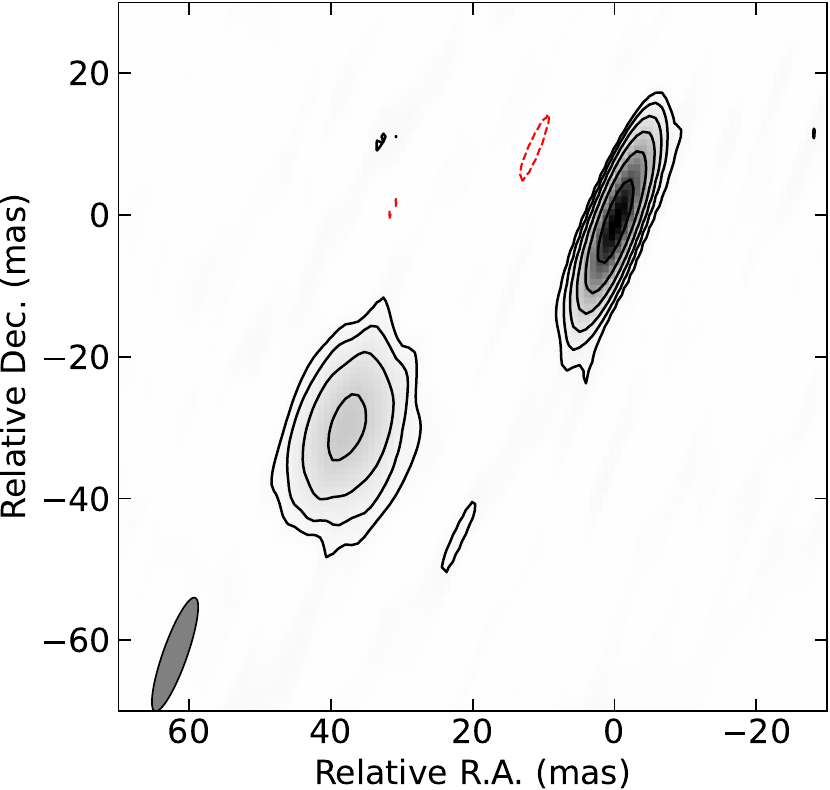} 
    \includegraphics[width=0.4\linewidth]{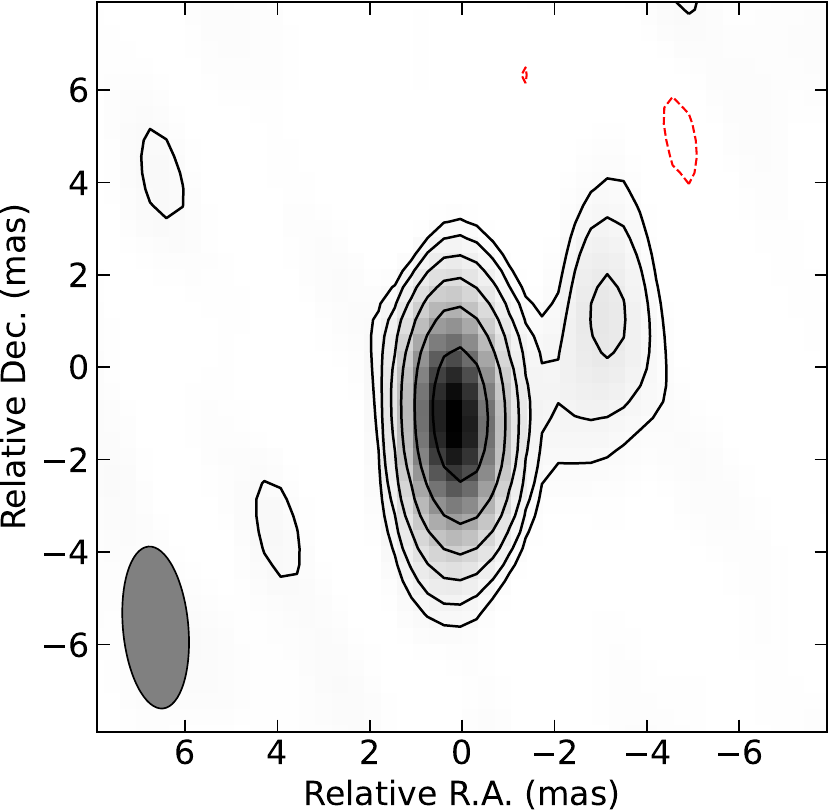} 
    \includegraphics[width=0.4\linewidth]{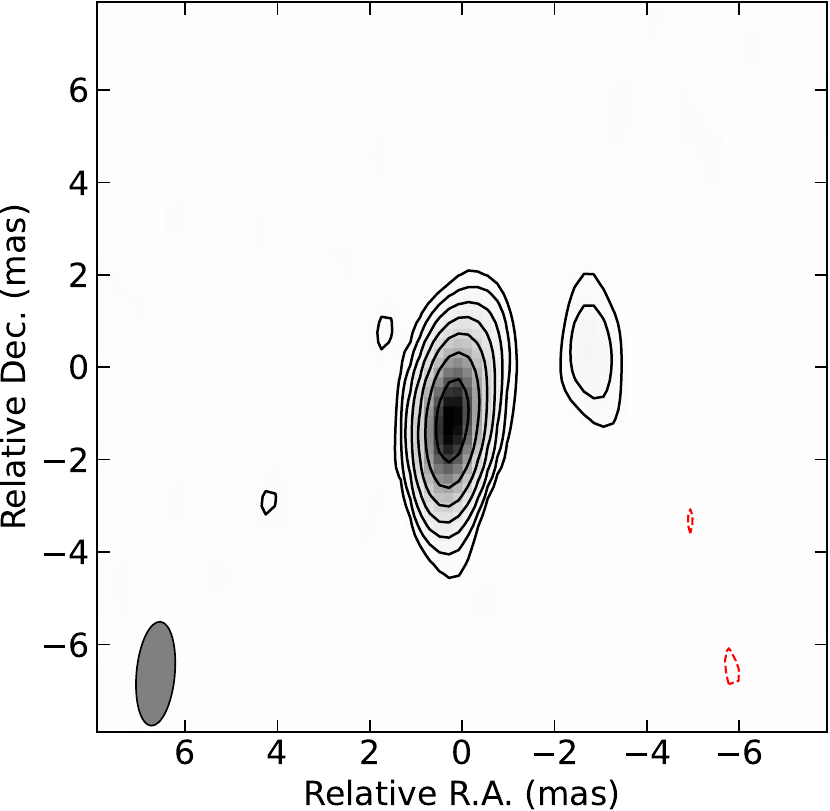}
    \caption{Self-calibrated images of J1246+1153. The top-left and top-right panels correspond to 1.46\,GHz and 1.64\,GHz, respectively; the middle-left and middle-right panels show 2.3\,GHz and 4.87\,GHz; and the bottom panel displays the 8.37\,GHz image. All images were made using natural weighting. The lowest contours are at $\pm$3 times rms noise and further positive contours are drawn at increasing steps of 2. The gray ellipse in the lower left corner of each image is the restoring beam.}
    \label{fig:J1246}
\end{figure}

\begin{figure}
    \centering
    \includegraphics[width=0.4\linewidth]{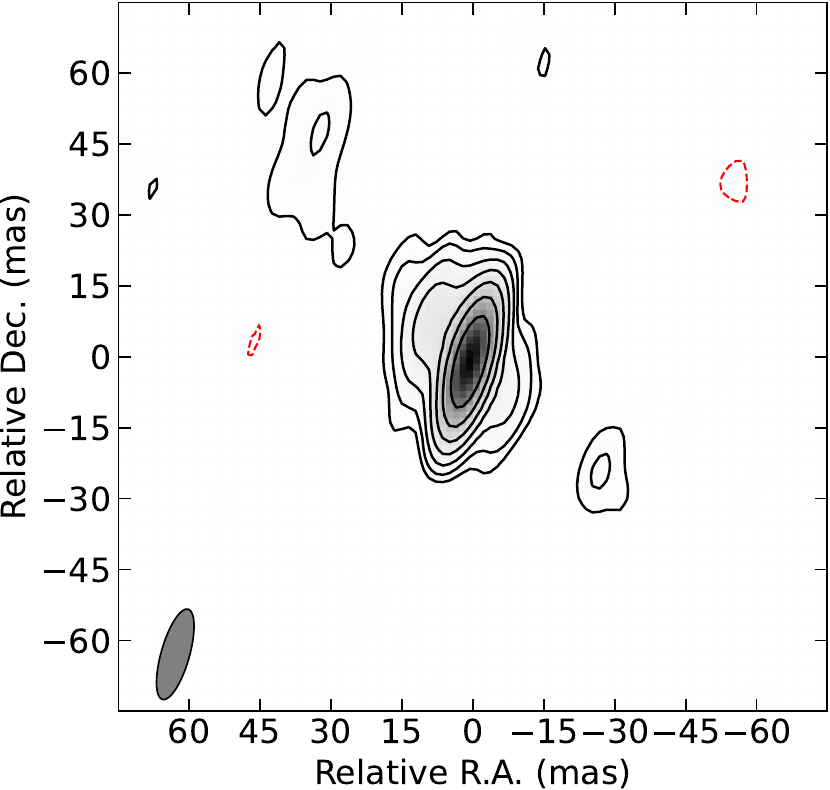} 
    \includegraphics[width=0.4\linewidth]{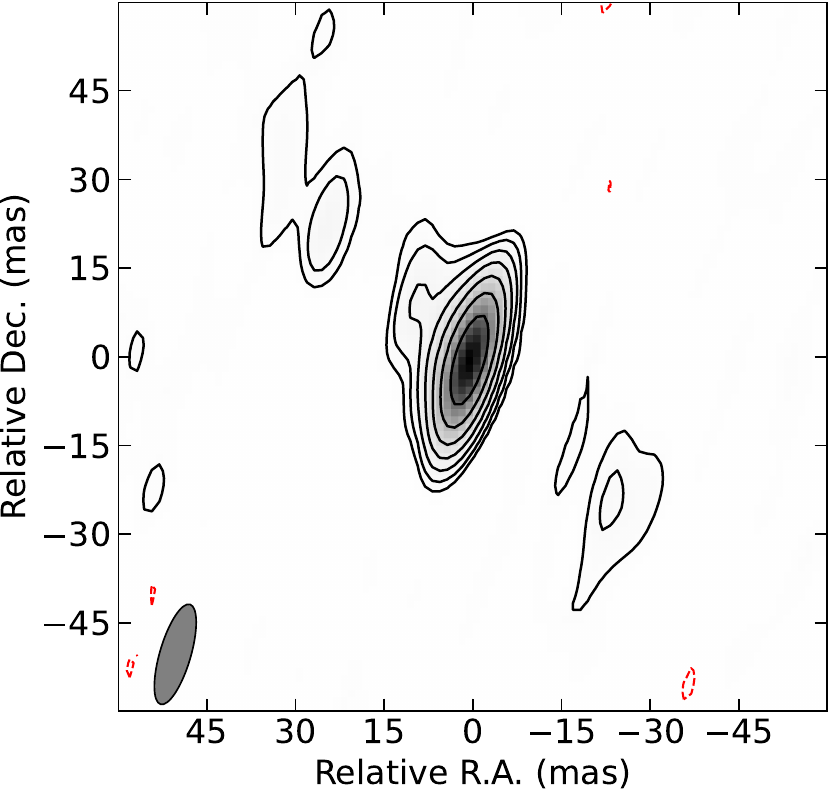} 
    \includegraphics[width=0.4\linewidth]{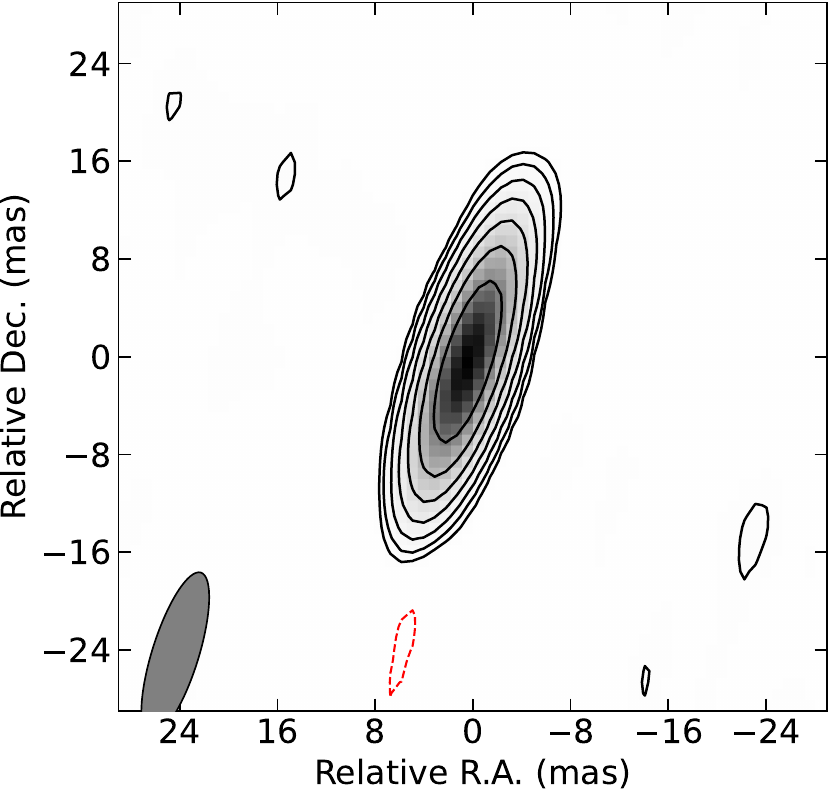} 
    \includegraphics[width=0.4\linewidth]{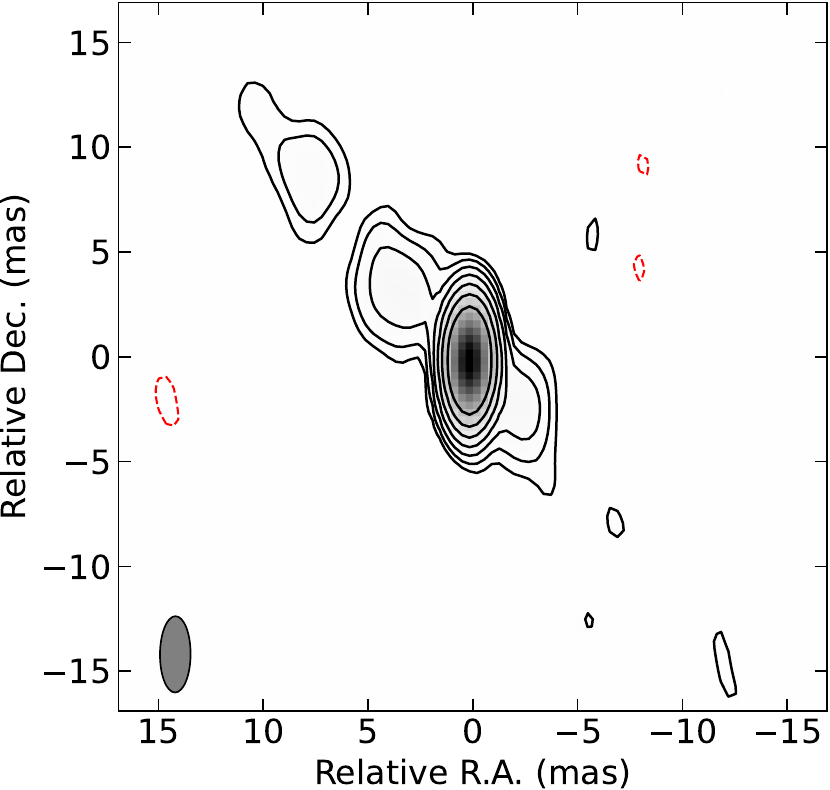} 
    \includegraphics[width=0.4\linewidth]{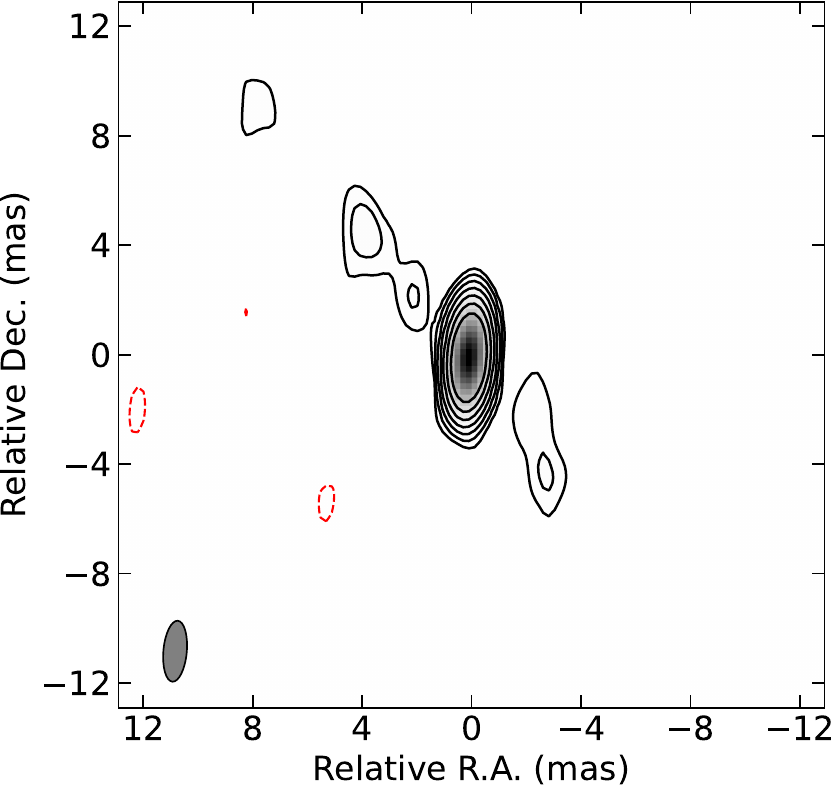}
    \caption{Self-calibrated images of \Msixty. The top-left and top-right panels correspond to 1.46\,GHz and 1.64\,GHz, respectively; the middle-left and middle-right panels show 2.3\,GHz and 4.87\,GHz; and the bottom panel displays the 8.37\,GHz image. All images were made using natural weighting. The lowest contours are at $\pm$3 times rms noise and further positive contours are drawn at increasing steps of 2. The gray ellipse in the lower left corner of each image is the restoring beam.}
    \label{fig:M60}
\end{figure}

\begin{table}[htbp]
\centering
\caption{Positional uncertainties (in $\mu$as)}
\begin{tabular}{lccccc}
\toprule
\textbf{Frequency} & 1.46 GHz & 1.64 GHz & 2.32 GHz & 4.87 GHz & 8.37 GHz \\
\midrule
Signal-to-noise                       & 81  & 121 & 147 & 12  & 8   \\
Internal structure of J1246+1153      & 45  & 34  & 29  & 6   & 4   \\
Ionospheric residuals                 & 229 & 181 & 91  & 21  & 7   \\
Tropospheric residuals                & 18  & 18  & 18  & 18  & 18  \\
Earth orientation                     & 1   & 1   & 1   & 1   & 1   \\
Antenna position                      & 2   & 2   & 2   & 2   & 2   \\
Priori source                         & 2   & 2   & 2   & 2   & 2   \\
Core identification                   & 36  & 42  & 45  & 3   & 2   \\
\midrule
\textbf{Root-sum-squared}             & \textbf{250} & \textbf{225} & \textbf{182} & \textbf{31} & \textbf{22} \\
\bottomrule
\end{tabular}
\vspace{10pt}
\end{table}
\section{GRMHD simulation of jet formation and calculation of jet radiation}



To understand the unusually steep frequency-dependent core shift in \Msixty, we first perform a three-dimensional GRMHD simulation of jet formation from black hole accretion flow and then calculate radiative transfer within the jet in the framework of general relativity. 

\subsection{GRMHD simulation of jet formation}
The GRMHD simulation was performed with Athena++ \cite{White2016,Stone2020}, which solves the equations of ideal GRMHD in a fixed Kerr spacetime. We use horizon-penetrating Kerr--Schild coordinates, \((t,r,\theta,\phi)\), and define the gravitational radius as \(r_{\rm g}=GM_{\rm BH}/c^2\). The event-horizon radius is
\begin{equation}
    r_{\rm H}=\left(1+\sqrt{1-a^2}\right)r_{\rm g},
\end{equation}
where \(a\) is the dimensionless black-hole spin. The gas is described by the rest-mass density \(\rho\), four-velocity \(u^\mu\), gas pressure \(P_{\rm gas}\), and internal energy density \(u_{\rm gas}=P_{\rm gas}/(\Gamma-1)\). We adopt an adiabatic index \(\Gamma=4/3\).

The initial condition is a Fishbone--Moncrief torus in hydrostatic equilibrium \cite{Fishbone1976}. The inner edge of the torus is located at \(40.5\,r_{\rm g}\), and the pressure maximum is at \(80\,r_{\rm g}\). An initially poloidal magnetic field is seeded in the torus following the prescription of \cite{Penna2013}. The magnetic flux is sufficiently large for the accretion flow to enter a magnetically arrested disk (MAD) state \cite{Narayan2003,Tchekhovskoy2011}. The black-hole spin is set to \(a=0.98\). The computational domain extends from \(1.1\,r_{\rm g}\) to \(1200\,r_{\rm g}\) in radius, covers the full polar range \([0,\pi]\), and spans the full azimuthal range \([0,2\pi]\). Static mesh refinement gives an effective resolution of \(1408\times512\times256\) in the \(r\), \(\theta\), and \(\phi\) directions, respectively. The simulation is evolved to \(t=40000\,r_{\rm g}/c\), corresponding to approximately 8.9 orbital periods at the initial pressure maximum. By the final time, the inflow-equilibrium radius reaches \(\sim80\,r_{\rm g}\). Further numerical details are given in Ref. \cite{yang2024}.

The preference for the MAD model is motivated by the core-shift measurement. In our GRRT tests, SANE models produced a much shallower core-shift relation than observed, whereas the MAD model yielded a core-shift index closer to the measured value. We therefore present the MAD model as the representative simulation result. This supports the interpretation that the steep core shift in M60* is associated with a magnetically dominated, non-equipartition jet base rather than with a weakly magnetized SANE-type outflow.

\subsection{Determination of thermal and non-thermal electrons}

The GRMHD simulation evolves the total gas internal energy, but synchrotron emission depends on the electron distribution. For the thermal electron component, we adopt the commonly used \(R\)-\(\beta\) prescription \cite{Moscibrodzka2016}:
\begin{equation}
\frac{T_{\rm p}}{T_{\rm e}}
=
R_{\rm low}\frac{1}{1+\beta^2}
+
R_{\rm high}\frac{\beta^2}{1+\beta^2},
\label{eq:Te}
\end{equation}
where \(T_{\rm p}\) and \(T_{\rm e}\) are the proton and electron temperatures, respectively, and \(\beta=P_{\rm gas}/P_{\rm mag}\) is the ratio of gas pressure to magnetic pressure. We set \(R_{\rm low}=1\) and \(R_{\rm high}=80\), so that electrons remain relatively hot in the magnetically dominated funnel while being cooler in the gas-pressure-dominated disk.

In addition to the thermal electrons, we also consider non-thermal population accelerated by magnetic reconnection in the jet \cite{Petersen2020,yang2024}. The non-thermal electrons are assumed to follow a power-law distribution,
 
\begin{equation} 
\frac{d n_{\rm e}}{d\gamma}=N_{\rm pl}(p-1)\gamma^{-p}, \gamma_{\rm max}> \gamma > \gamma_{\rm min} 
\label{eq:pl}
\end{equation}
where \(N_{\rm pl}\) is the non-thermal electron number density. The value of power-law index $p$ is still difficult to precisely determine from the first principle of particle acceleration by magnetic reconnection. In the present work, we therefore treat it as a free parameter. We find that the best fit parameter value is  \(p=6\). We note that this value is different from the value of $p\approx 2$ adopted in Ref. \cite{Blandford-Konigl1979}, but is more consistent with the steep optically thin radio spectra of the jet in this source. It remains interesting to investigate why this source has such a steep nonthermal electrons energy distribution. Other parameters are \(\gamma_{\max}=10^6\) and
\begin{equation}
    \gamma_{\min}=1+10\frac{k_{\rm B}T_{\rm e}}{m_{\rm e}c^2}.
\end{equation}

How to determine the number density of nonthermal electrons in the jet? In a previous work \cite{yang2024}, this question has been carefully investigated using various observational constrains such as the jet image and the jet width as a function of distance to the black hole. It is found that magnetic reconnection is the  mechanism of electrons acceleration in the jet, and the number density is determined by the current density in the reconnection current sheet. In this work, we follow this scenario to determine the spatial distribution of nonthermal electrons. The steady-state value of \(n_{\rm nth}\) is calculated by balancing current-dependent injection rate against synchrotron cooling rate:
\begin{equation} 
\eta\frac{v_{\rm A}}{r_{z}}(N_{\rm tot}-N_{\rm pl})\frac{J^{2}}{J_{0}^{2}}=\frac{N_{\rm pl}}{\tau_{cool}} 
\label{eq:nth}
\end{equation}
Here, $N_{\rm tot}$ is the total electron density including hot electrons and non-thermal electrons, which is obtained directly through simulation.  $\eta$ is a dimensionless parameter that controls the efficiency of the current in accelerating electrons into a power-law distribution of electrons. Thus the non-thermal fraction increases in regions with strong current density and long cooling time, but remains bounded by the total electron density. In our work, we set $\eta=2.0\times10^{-6}$. $v_{\rm A}$ is the  Alfvén velocity. $r_{z}$ is the typical length scale of the reconnection region. Inspired by the relationship between jet width and jet distance, we set $r_{z}=z^{1/3}$ in the jet area and $r_{z}=r$ outside the jet area. The boundaries of these two relationships about $r_{z}$ are outside the BZ jet boundary and are $R=2+2\times z^{0.7}$. $\tau_{\rm cool}$ is the radiative cooling time scale of non-thermal electrons. $J$ is the local three-current density, and the calculation formula is as follows \cite{Ball2018},
\begin{equation} 
J^{i}=\partial F^{ij} +\Gamma^{i}_{j\lambda}F^{ij} 
\label{eq:J}
\end{equation}
Here $F^{ij}$ is the electromagnetic field tensor, $\Gamma^{\lambda}_{\alpha \beta }$ is the Christoffel symbol, $i$ is equal to $t$, $r$, $\theta$, and $\varphi$. $J_{0}=c^2P_{\rm gas}/r^{2}$ is the characteristic three current density.

\subsection{Radiative transfer calculation and synthetic observables}

We compute synchrotron emission with IPOLE \cite{Moscibrodzka2018}, a polarized GRRT code designed for radiative transfer in the Kerr metric. The black-hole mass and distance are set to be
\begin{equation}
M_{\rm BH}=4.5\times10^9\,M_\odot,
\qquad
D=16.3~{\rm Mpc}.
\end{equation}
Synthetic images are produced over a field of view of \([-2,2]\) mas with \(500\times500\) pixels. The mass accretion rate is fixed by matching the observed 8.37 GHz flux density of \(0.018\) Jy. This gives a physical scaling of \(\dot M_{\rm BH}=2.2\times10^{-9}\dot M_{\rm Edd}\), where the numerical value should be obtained from the final GRRT normalization.

To compare the model directly with the VLBI measurements, the synthetic images are convolved with Gaussian beams matched to the observing frequencies. The model core position at each frequency is defined as the location of the peak intensity in the beam-convolved image. We then measure the displacement relative to the 8.37 GHz core and fit the frequency dependence with
\begin{equation}
r_{\rm core}(\nu) \propto \nu^{-k}.
\label{eq:core_shift_fit}
\end{equation}
The same procedure is applied to the observed and synthetic images to enable a direct comparison of the core-shift index \(k\). For the core-shift fitting, we use only the 1.64--8.37 GHz synthetic images, adopting the 8.37 GHz core as the reference position. The synthetic images are convolved with the corresponding VLBI restoring beams listed in Table~\ref{tab:beamsize}. For the additional frequencies, we adopt beam sizes of \(50.7\times22\) mas at 0.3 GHz, \(27\times11.7\) mas at 0.6 GHz, \(18\times7.8\) mas at 1 GHz, and \(0.72\times0.31\) mas at 15 GHz.(The position angles here are all $0^{\circ}$.) The points below 1.64 GHz are treated as lower limits and are not included in the fit.

The MAD model produces a compact, beam-smoothed radio core whose apparent position moves upstream with increasing frequency. The fitted synthetic core shift gives \(k_{\rm sim}\simeq1.87\), close to the observed value \(k_{\rm obs}=1.96\pm0.37\). This result suggests that a steep core shift can arise naturally when the optical-depth surface propagates through a magnetically dominated, parabolically collimated inner jet. The large value of \(p\) helps reproduce the steep optically thin radio spectrum, but the steep core-shift index primarily reflects the combined effects of jet geometry, magnetic dominance, and the radial stratification of synchrotron opacity.

To enable a consistent comparison with the observed jet spectrum, we post-processed the GRRT images in the same manner as the VLBI images. The synthetic images over 1--15 GHz were first convolved with a circular Gaussian beam of FWHM \(0.5\) mas. We then excluded the core-dominated region by masking all pixels with relative declination \({\rm Dec}>-1\) mas. This masking reduces contamination from the unresolved nuclear emission and allows the spectrum to be measured primarily from the extended jet. The resulting spectrum is presented in Figure~\ref{fig:simulated sectral index}. The simulated jet spectrum follows an approximately single power law, 
\(F_\nu\propto \nu^{-2.44}\). This is consistent with optically thin synchrotron 
emission from a non-thermal electron distribution \(N(\gamma)\propto\gamma^{-p}\), 
for which \(F_\nu\propto\nu^{-(p-1)/2}\). For the adopted value \(p=6\), the 
expected spectral slope is \(-(p-1)/2=-2.5\), in good agreement with the 
synthetic spectrum. 

\subsection{Magnetic and electron energy densities}

A useful diagnostic of the physical origin of the steep core shift is the ratio between magnetic and electron energy densities. In the standard Blandford--K\"onigl conical jet model, a core-shift index close to unity is expected when $p\approx 2$ and the jet is freely expanding and close to equipartition between magnetic fields and radiating particles \cite{Blandford-Konigl1979}. Therefore, a significantly steeper index provides evidence for departures from equipartition and/or from conical geometry.
 
The magnetic energy density is computed in the fluid frame as
\begin{equation}
u_B=\frac{B^2}{8\pi},
\end{equation}
where \(B\) is the comoving magnetic-field strength in cgs units. Equivalently, in the code units with \(4\pi=1\), this corresponds to \(u_B=b^2/2\), where \(b^2=b^\mu b_\mu\).

The electron energy density is written as
\begin{equation}
u_e=u_{\rm th}+u_{\rm pl}.
\end{equation}
For a relativistic Maxwell--J\"uttner distribution, the thermal electron internal energy density is\cite{Yuan-2003}
\begin{equation}
u_{\rm th}
=
a(\theta_{e})N_{\rm th}m_ec^2\theta_{e}
 ,
\qquad
\theta_e=\frac{k_{\rm B}T_e}{m_ec^2},
\end{equation}
where 
\begin{equation}
a(\theta_{e})=  \frac{1}{\theta_{e}} \left[
\frac{3K_3(1/\theta_e)+K_1(1/\theta_e)}{4K_2(1/\theta_e)}
-1
\right],
\end{equation}
where \(K_n\) are modified Bessel functions of the $n$th order.  This expression excludes the electron rest-mass energy.

For the non-thermal electrons, we use  
\begin{equation}
u_{\rm pl}
\simeq
N_{\rm pl}m_ec^2
\frac{p-1}{p-2}\gamma^{2-p}_{\min}.
\end{equation}
The resulting \(u_B/u_e\) map (Figure~\ref{fig:placeholder_ubue}) shows that the simulated jet is magnetically dominated over most of the emitting region, with \(u_B/u_e>1\). This supports the interpretation that the steep observed core shift in \Msixty is associated with a non-equipartition, magnetically dominated jet base rather than with a standard equipartition conical jet.

\begin{figure}
    \centering
    \includegraphics[width=0.3\linewidth]{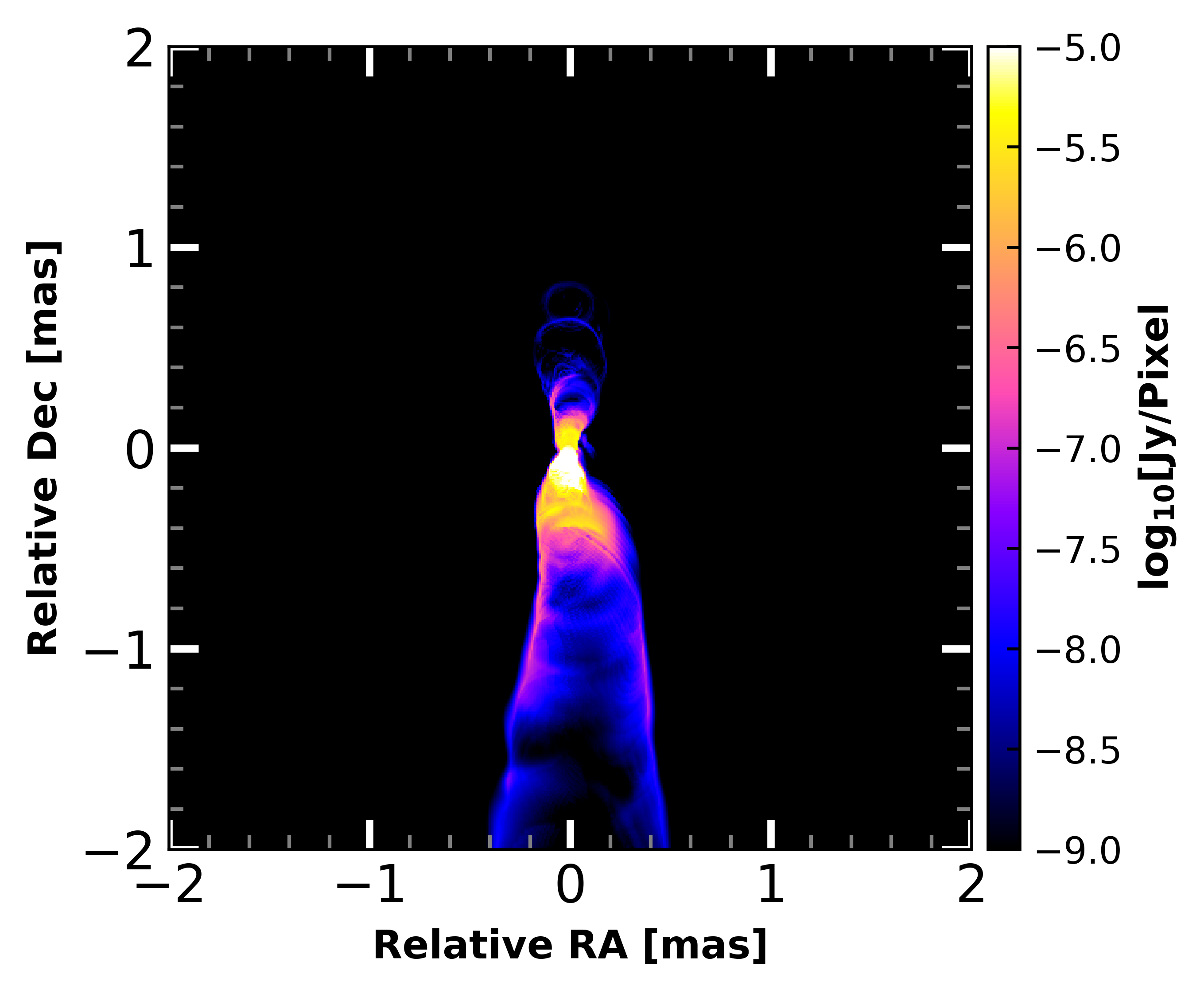}
    \includegraphics[width=0.3\linewidth]{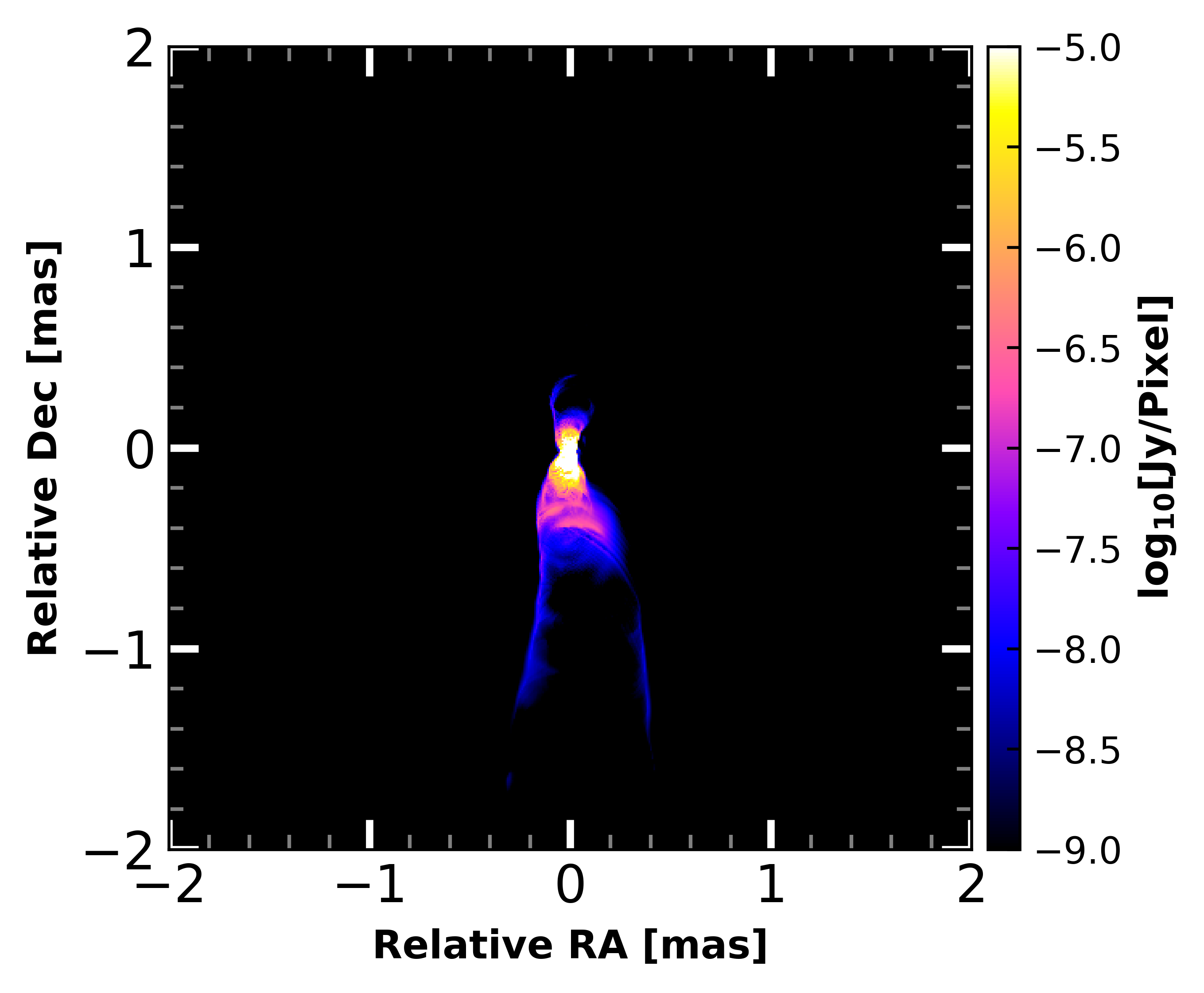}
    \includegraphics[width=0.3\linewidth]{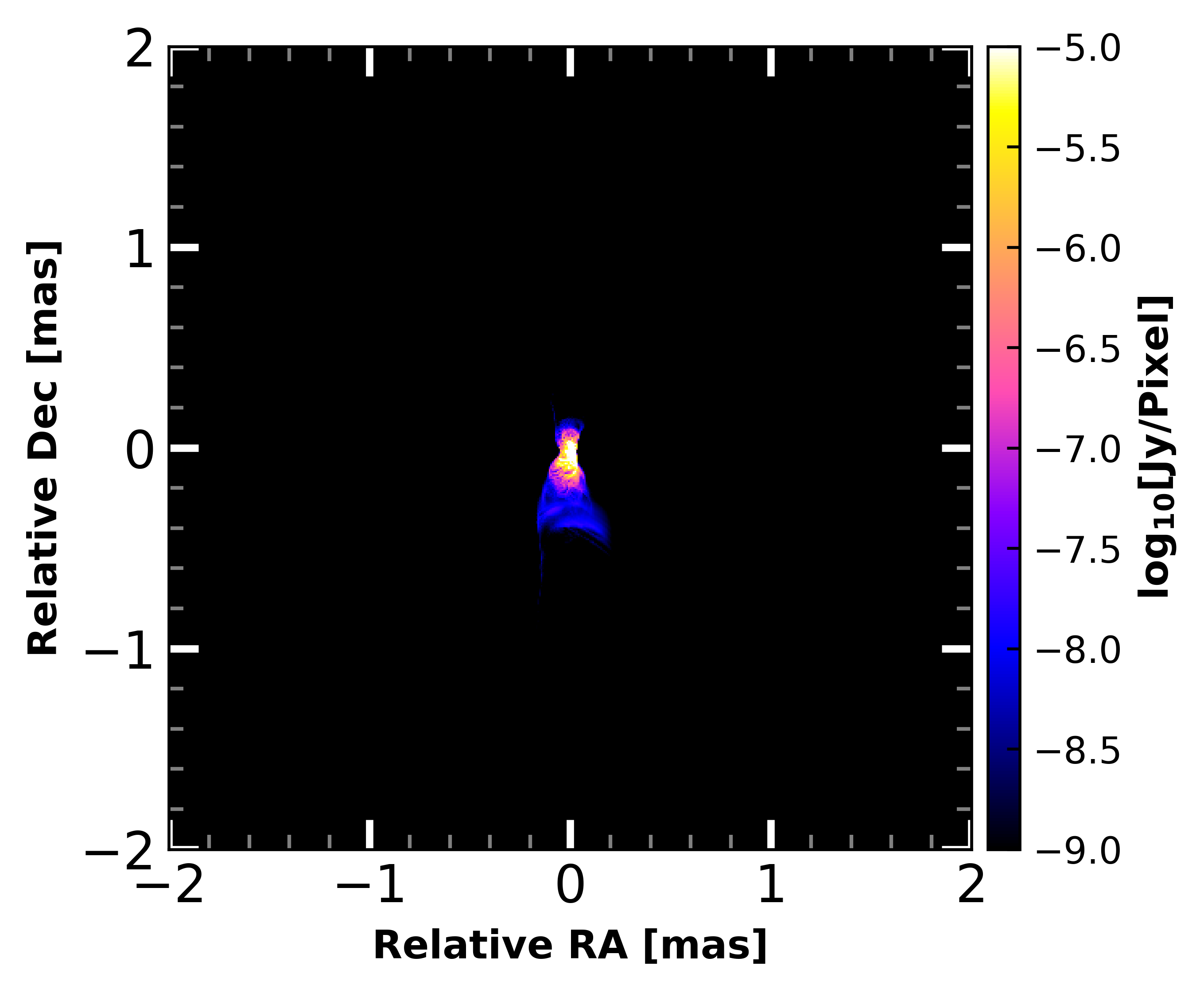}\\
    \includegraphics[width=0.3\linewidth]{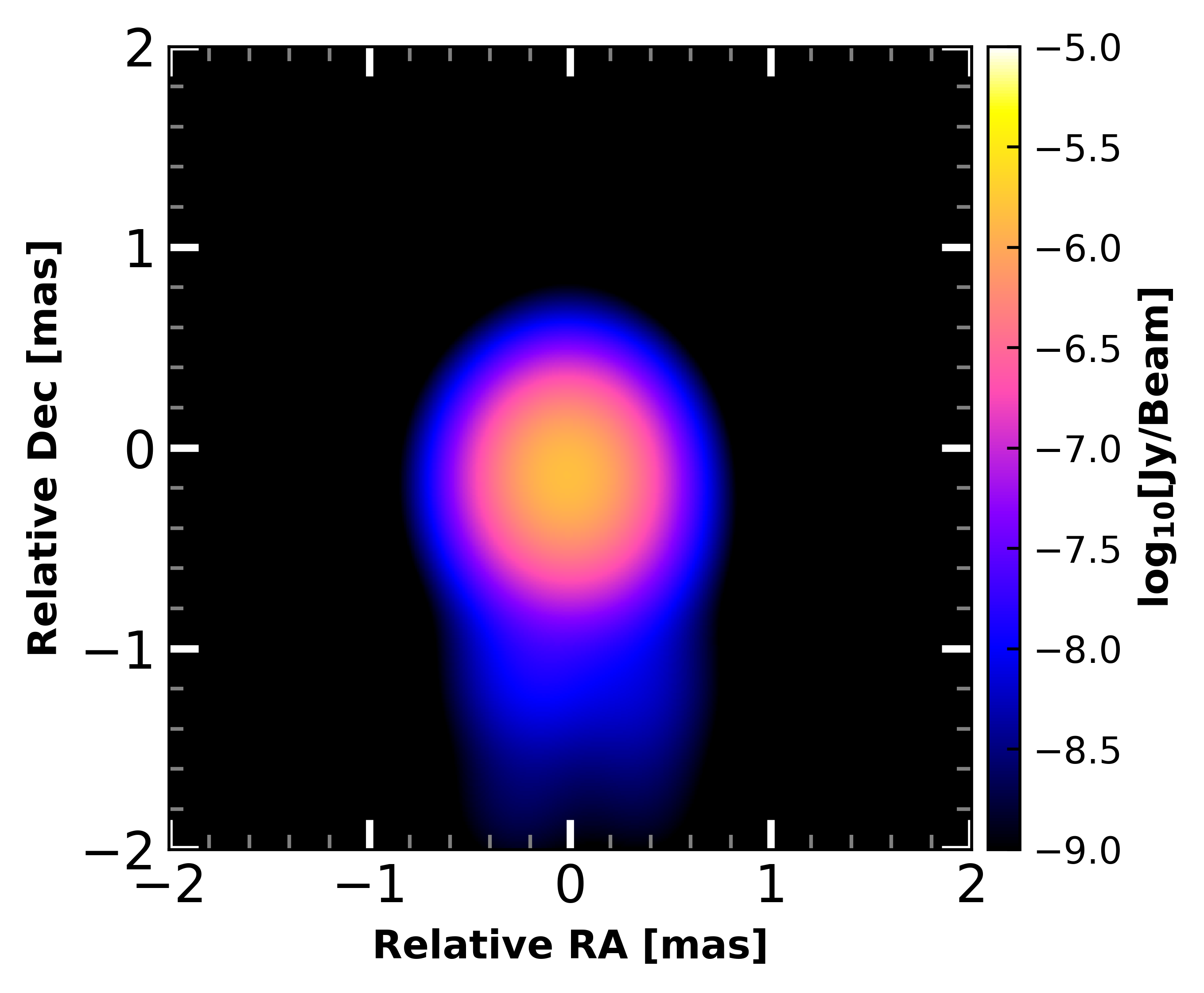}
    \includegraphics[width=0.3\linewidth]{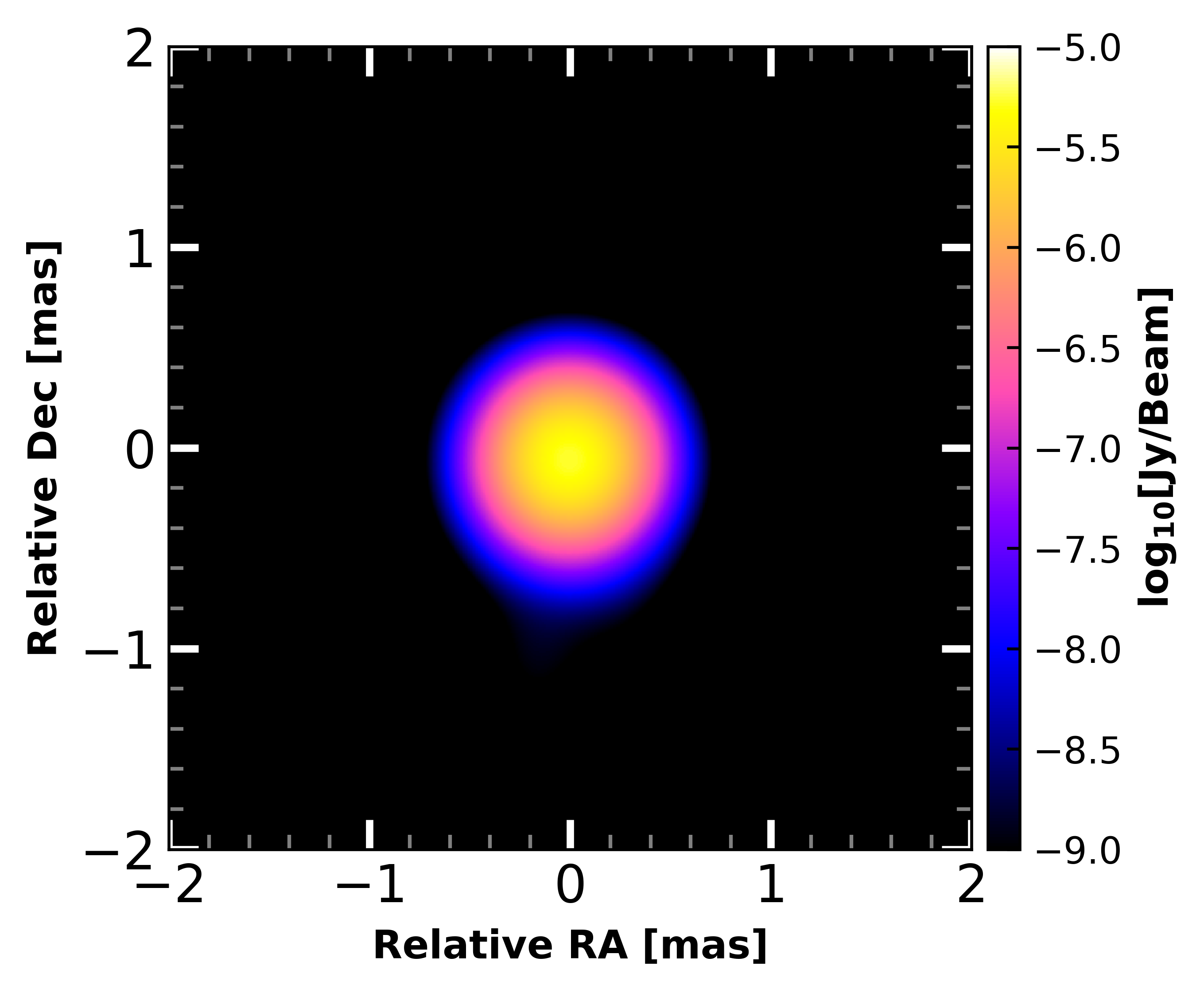}
    \includegraphics[width=0.3\linewidth]{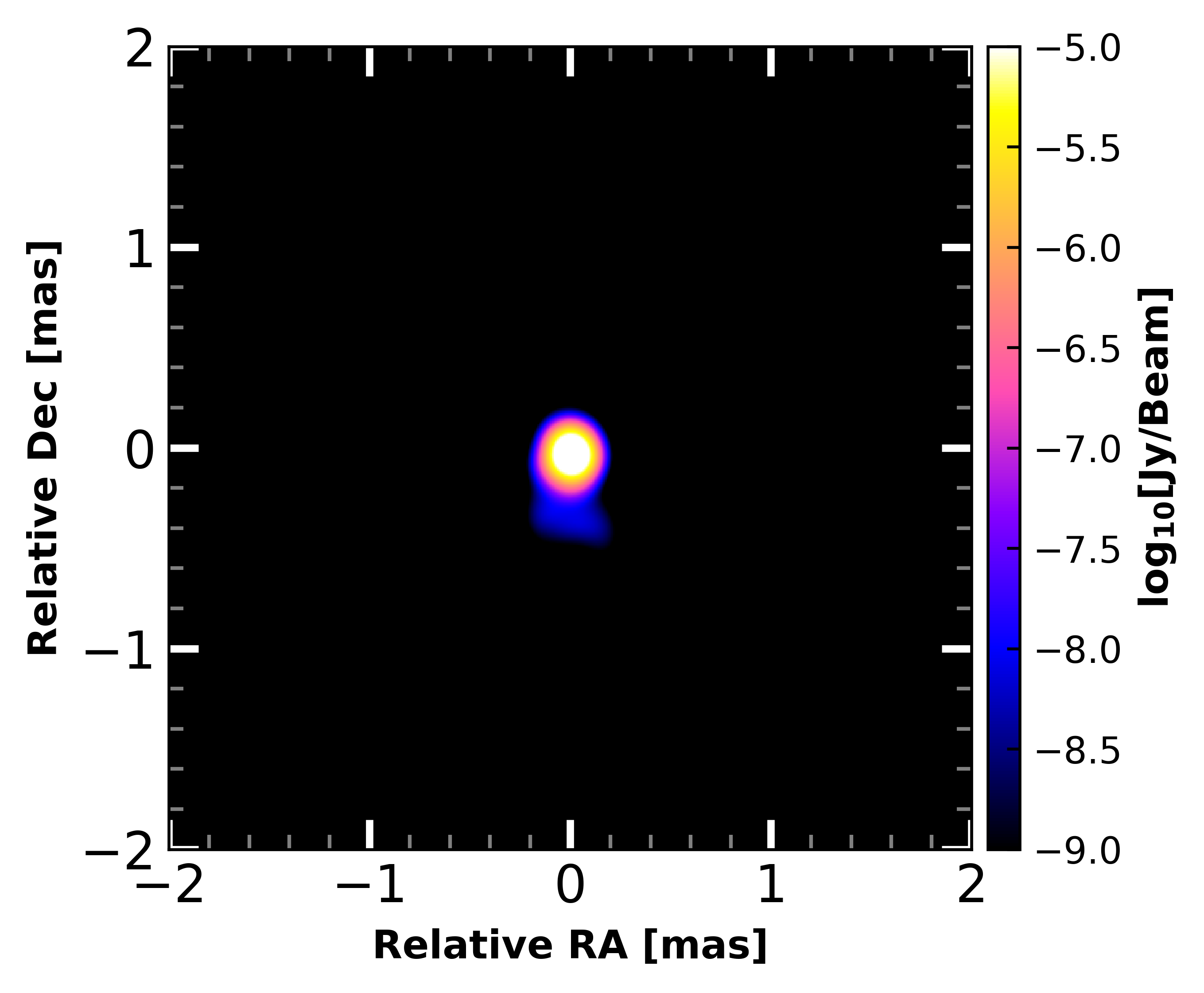}\\
    \caption{The top row presents the intrinsic simulated jet emission at 1.64 GHz (left), 4.87 GHz (middle), and 15 GHz (right). The bottom row shows the corresponding beam-convolved images at 1.6 GHz (left), 8.37 GHz (middle), and 22 GHz (right). Convolution was performed using circular Gaussian beams with full widths at half maximum (FWHMs) of 0.5 mas, 0.4 mas, and 0.1 mas, associated with 1.64 GHz, 4.87 GHz, and 15 GHz, respectively. The beam smoothing substantially suppresses the intrinsic substructure, resulting in increasingly compact emission morphologies at higher frequencies.}
    \label{fig:simulated iamges}
\end{figure}

\begin{figure}
    \centering
    \includegraphics[width=0.5\linewidth]{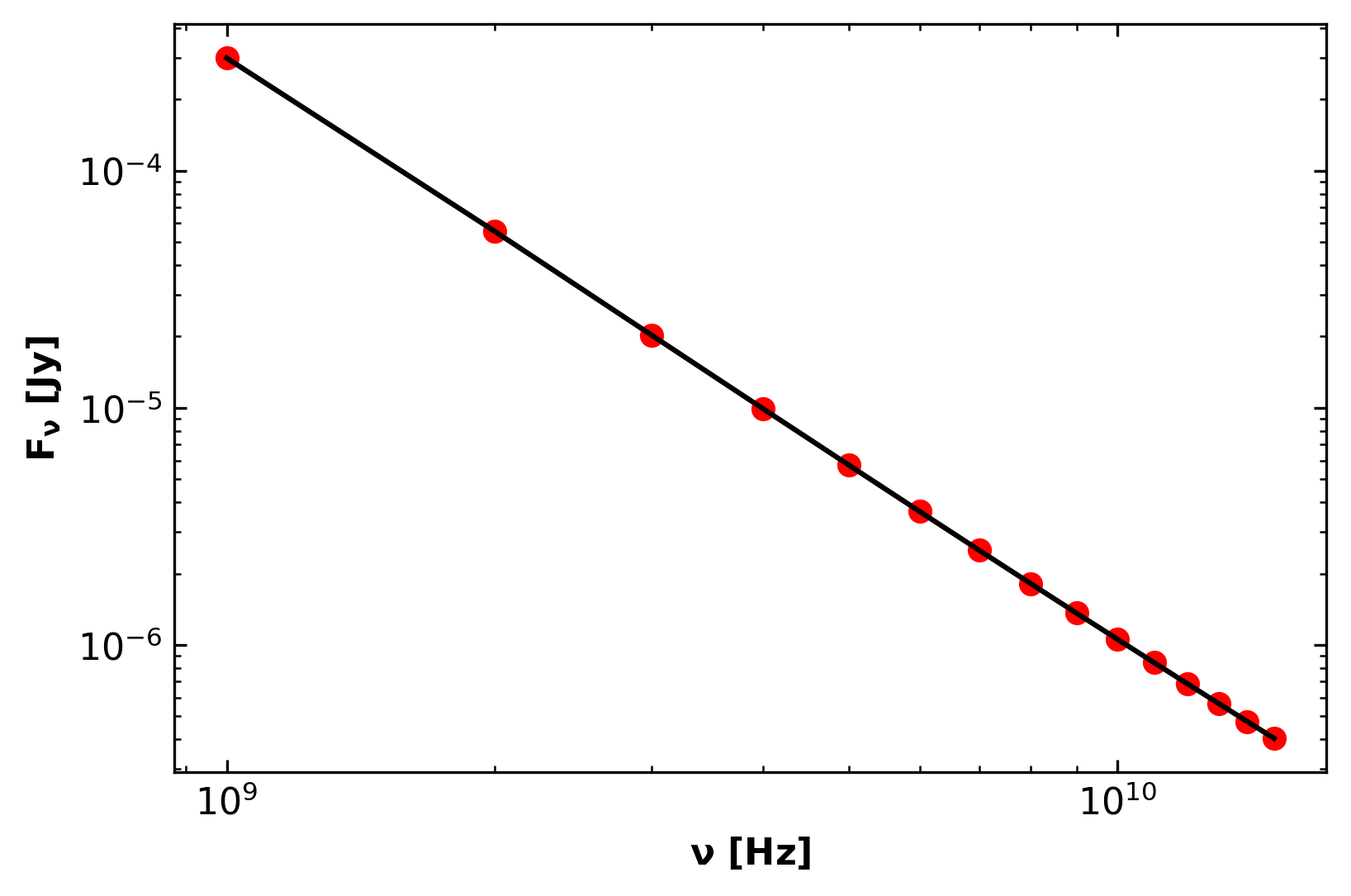}
    \caption{Simulated radio spectrum of the jet. The flux density decreases with frequency, showing a steep power-law behavior characteristic of optically thin synchrotron emission. The simulated jet spectrum follows an approximately single power law, \(F_\nu\propto \nu^{-2.44}\).}
    \label{fig:simulated sectral index}
\end{figure}

\begin{figure}
    \centering
    \includegraphics[width=0.5\linewidth]{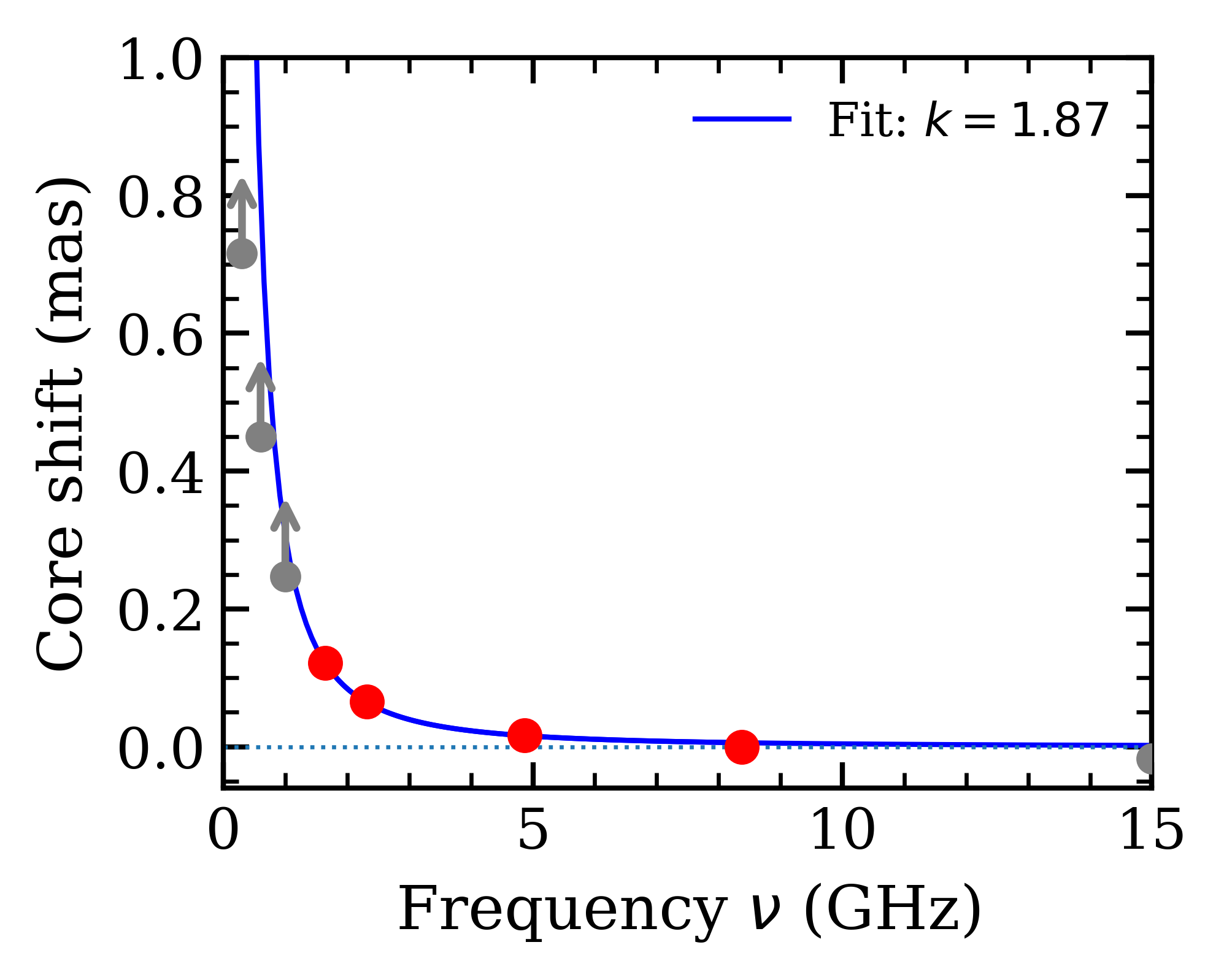}
    \caption{Simulated core shift as a function of observing frequency. The dots show the simulated measurements, and the blue line represents the best-fitting power-law, $r_{\rm core} \propto \nu^{-1.87}$, indicating a steep frequency dependence. Here we only fitted the points (red) from 1.64 to 8.37 GHz. The arrow indicates the lower limit value.}
    \label{fig:placeholder}
\end{figure}

\begin{figure}
    \centering
    \includegraphics[width=0.3\linewidth]{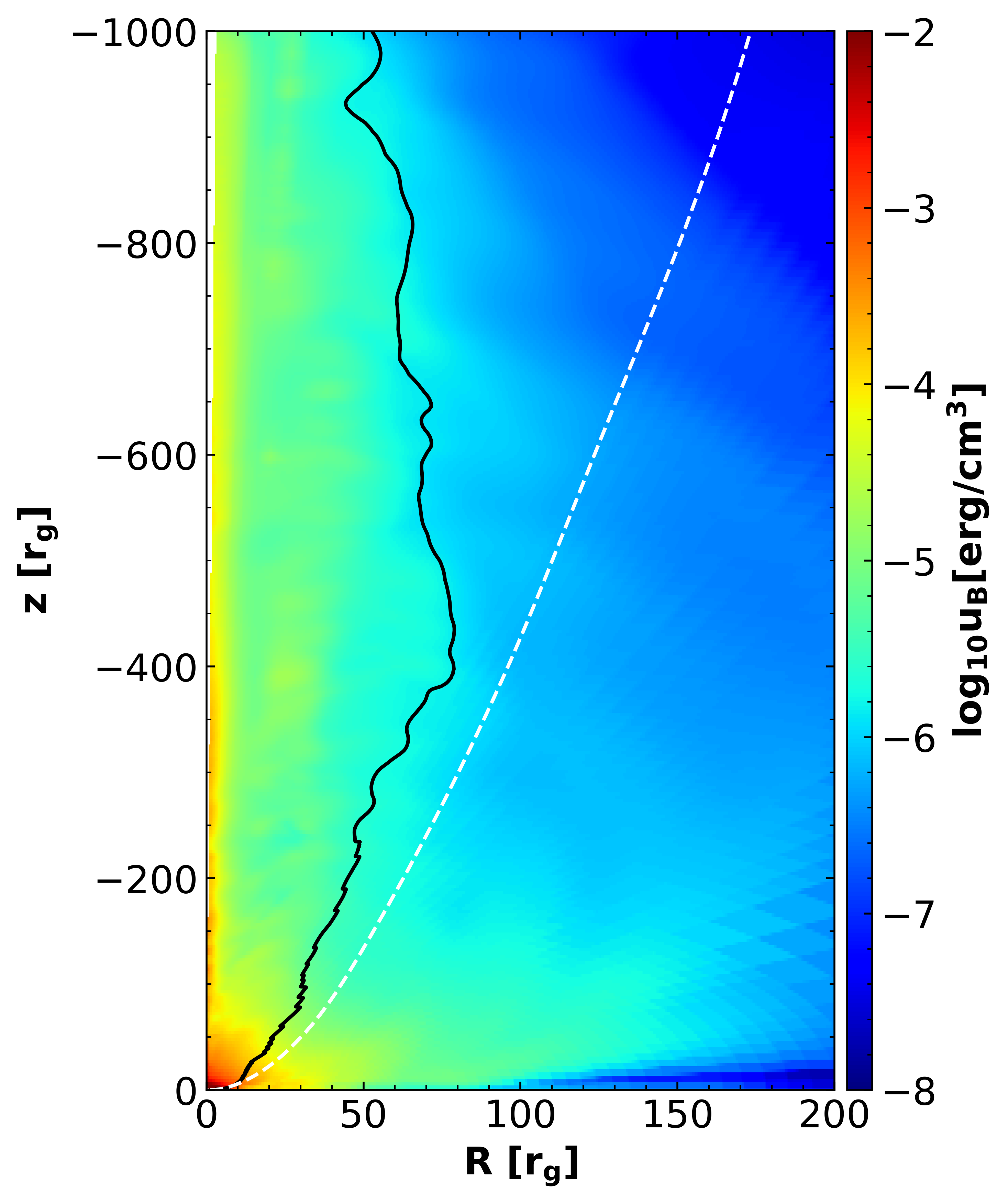}
    \includegraphics[width=0.3\linewidth]{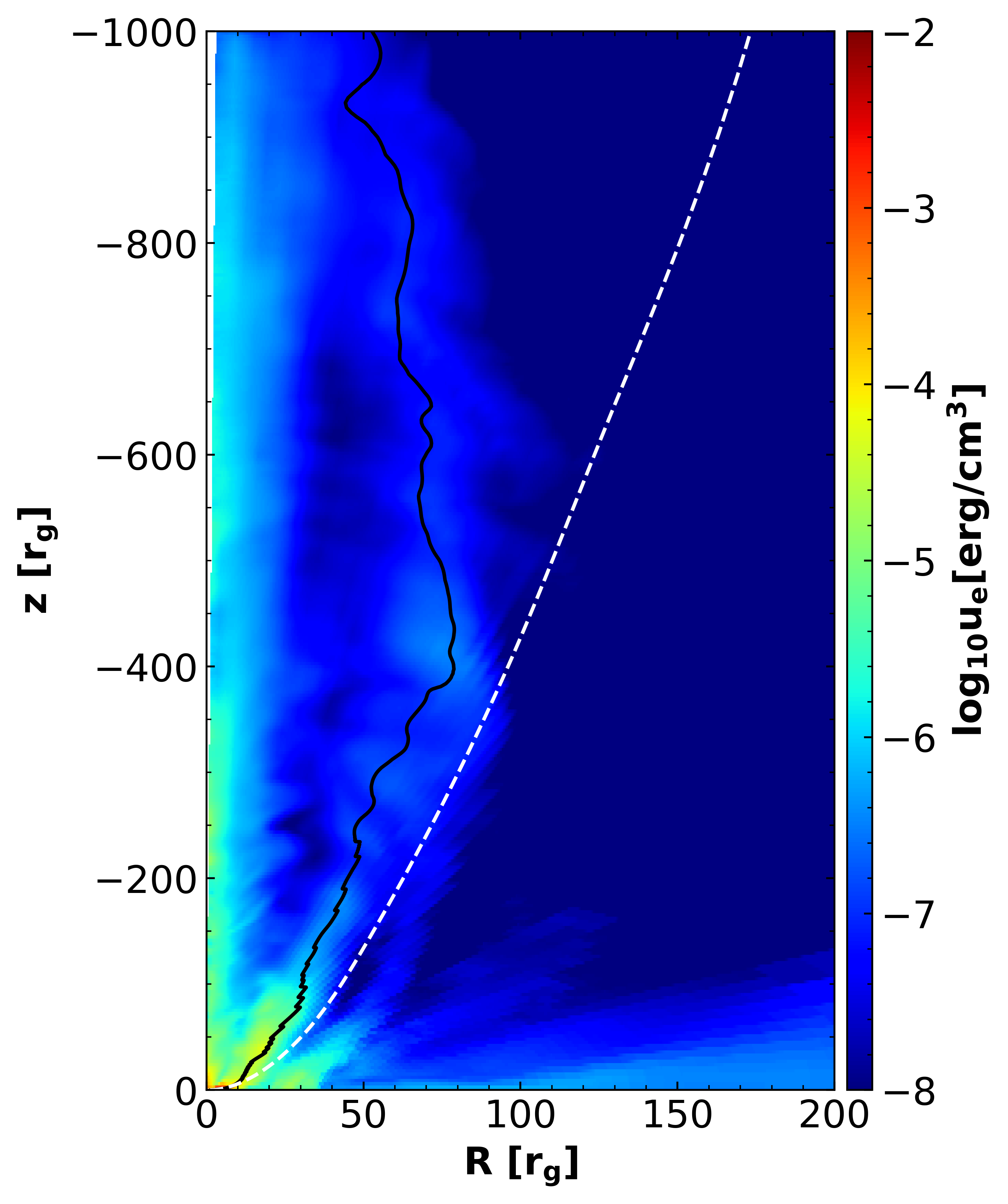}
    \includegraphics[width=0.3\linewidth]{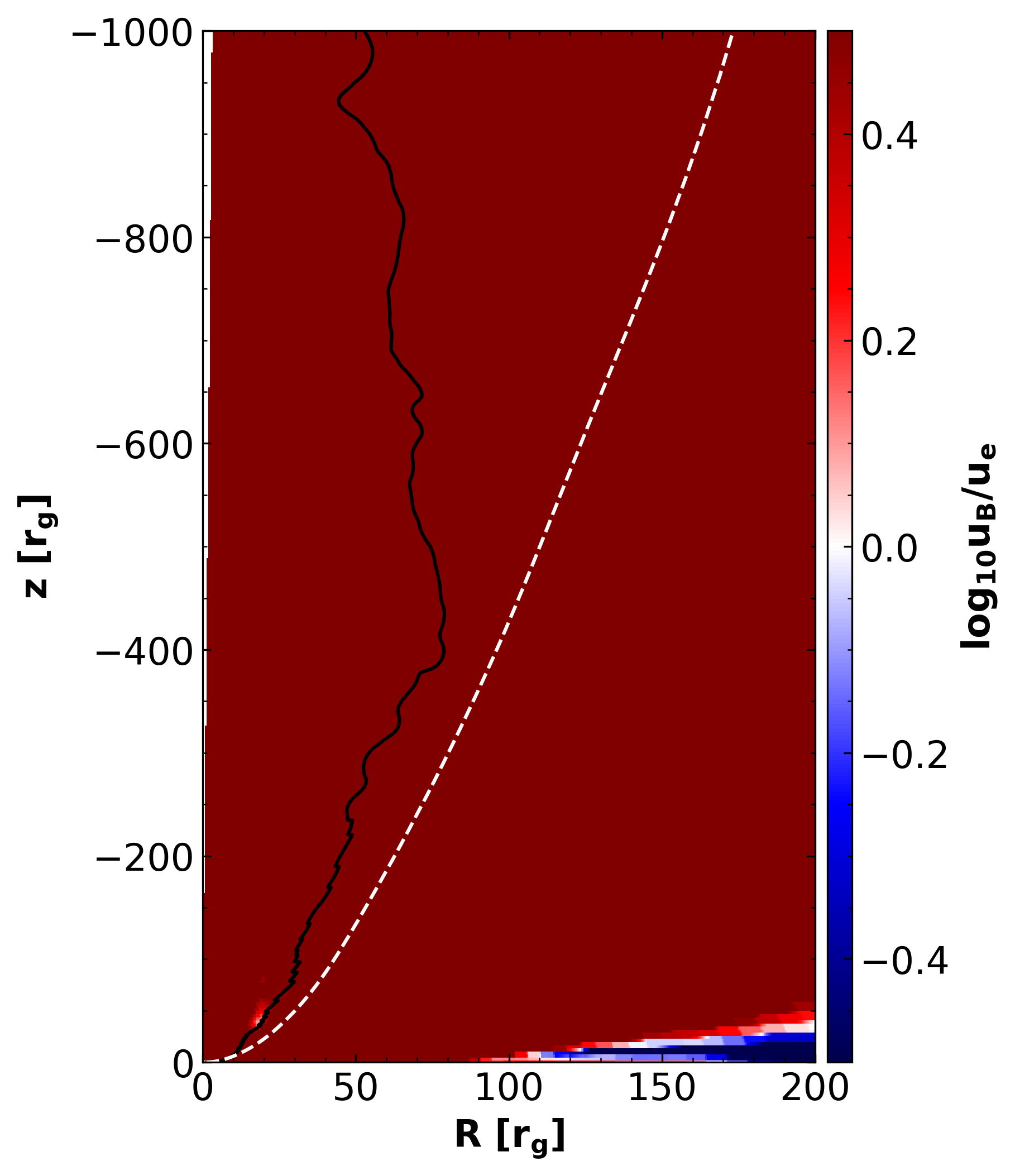}\\
    \caption{Two-dimensional distribution of the magnetization parameter, shown as $\log_{10}(U_B/U_e)$ in the $R$--$z$ plane. The black curve marks $\sigma=5$, the white dashed line shows the jet boundary. The jet is predominantly magnetically dominated ($U_B/U_e>1$) over most regions, with localized variations along the jet axis.}
    \label{fig:placeholder_ubue}
\end{figure}

\end{methods}


\vspace{0.8cm}
\bibliographystyle{arxiv}
\bibliography{M60/arxiv}

\begin{thebibliography}{100}
\expandafter\ifx\csname url\endcsname\relax
  \def\url#1{\texttt{#1}}\fi
\expandafter\ifx\csname urlprefix\endcsname\relax\def\urlprefix{URL }\fi
\providecommand{\bibinfo}[2]{#2}
\providecommand{\eprint}[2][]{\url{#2}}

\bibitem{ho2008}
\bibinfo{author}{{Ho}, L.~C.}
\newblock \bibinfo{title}{{Nuclear activity in nearby galaxies.}}
\newblock \emph{\bibinfo{journal}{\araa}} \textbf{\bibinfo{volume}{46}}, \bibinfo{pages}{475--539} (\bibinfo{year}{2008}).
\newblock \eprint{0803.2268}.

\bibitem{yu2002}
\bibinfo{author}{{Yu}, Q.} \& \bibinfo{author}{{Tremaine}, S.}
\newblock \bibinfo{title}{{Observational constraints on growth of massive black holes}}.
\newblock \emph{\bibinfo{journal}{\mnras}} \textbf{\bibinfo{volume}{335}}, \bibinfo{pages}{965--976} (\bibinfo{year}{2002}).
\newblock \eprint{astro-ph/0203082}.

\bibitem{yuan2014}
\bibinfo{author}{{Yuan}, F.} \& \bibinfo{author}{{Narayan}, R.}
\newblock \bibinfo{title}{{Hot Accretion Flows Around Black Holes}}.
\newblock \emph{\bibinfo{journal}{\araa}} \textbf{\bibinfo{volume}{52}}, \bibinfo{pages}{529--588} (\bibinfo{year}{2014}).
\newblock \eprint{1401.0586}.

\bibitem{inayoshi2020}
\bibinfo{author}{{Inayoshi}, K.}, \bibinfo{author}{{Ichikawa}, K.} \& \bibinfo{author}{{Ho}, L.~C.}
\newblock \bibinfo{title}{{Universal Transition Diagram from Dormant to Actively Accreting Supermassive Black Holes}}.
\newblock \emph{\bibinfo{journal}{\apj}} \textbf{\bibinfo{volume}{894}}, \bibinfo{pages}{141} (\bibinfo{year}{2020}).
\newblock \eprint{2001.11032}.

\bibitem{boccardi2017}
\bibinfo{author}{{Boccardi}, B.}, \bibinfo{author}{{Krichbaum}, T.~P.}, \bibinfo{author}{{Ros}, E.} \& \bibinfo{author}{{Zensus}, J.~A.}
\newblock \bibinfo{title}{{Radio observations of active galactic nuclei with mm-VLBI}}.
\newblock \emph{\bibinfo{journal}{\aapr}} \textbf{\bibinfo{volume}{25}}, \bibinfo{pages}{4} (\bibinfo{year}{2017}).
\newblock \eprint{1711.07548}.

\bibitem{gu2007}
\bibinfo{author}{{Gu}, Q.~S.}, \bibinfo{author}{{Huang}, J.~S.}, \bibinfo{author}{{Wilson}, G.} \& \bibinfo{author}{{Fazio}, G.~G.}
\newblock \bibinfo{title}{{Direct Evidence from Spitzer for a Low-Luminosity AGN at the Center of the Elliptical Galaxy NGC 315}}.
\newblock \emph{\bibinfo{journal}{\apjl}} \textbf{\bibinfo{volume}{671}}, \bibinfo{pages}{L105--L108} (\bibinfo{year}{2007}).
\newblock \eprint{0711.0051}.

\bibitem{hada2011}
\bibinfo{author}{{Hada}, K.} \emph{et~al.}
\newblock \bibinfo{title}{{An origin of the radio jet in M87 at the location of the central black hole}}.
\newblock \emph{\bibinfo{journal}{\nat}} \textbf{\bibinfo{volume}{477}}, \bibinfo{pages}{185--187} (\bibinfo{year}{2011}).

\bibitem{plambeck2014}
\bibinfo{author}{{Plambeck}, R.~L.} \emph{et~al.}
\newblock \bibinfo{title}{{Probing the Parsec-scale Accretion Flow of 3C 84 with Millimeter Wavelength Polarimetry}}.
\newblock \emph{\bibinfo{journal}{\apj}} \textbf{\bibinfo{volume}{797}}, \bibinfo{pages}{66} (\bibinfo{year}{2014}).
\newblock \eprint{1410.5887}.

\bibitem{blandford1977}
\bibinfo{author}{{Blandford}, R.~D.} \& \bibinfo{author}{{Znajek}, R.~L.}
\newblock \bibinfo{title}{{Electromagnetic extraction of energy from Kerr black holes.}}
\newblock \emph{\bibinfo{journal}{\mnras}} \textbf{\bibinfo{volume}{179}}, \bibinfo{pages}{433--456} (\bibinfo{year}{1977}).

\bibitem{eht2021viii}
\bibinfo{author}{{Event Horizon Telescope Collaboration}} \emph{et~al.}
\newblock \bibinfo{title}{{First M87 Event Horizon Telescope Results. VIII. Magnetic Field Structure near The Event Horizon}}.
\newblock \emph{\bibinfo{journal}{\apjl}} \textbf{\bibinfo{volume}{910}}, \bibinfo{pages}{L13} (\bibinfo{year}{2021}).
\newblock \eprint{2105.01173}.

\bibitem{yuan2022}
\bibinfo{author}{{Yuan}, F.}, \bibinfo{author}{{Wang}, H.} \& \bibinfo{author}{{Yang}, H.}
\newblock \bibinfo{title}{{The Accretion Flow in M87 is Really MAD}}.
\newblock \emph{\bibinfo{journal}{\apj}} \textbf{\bibinfo{volume}{924}}, \bibinfo{pages}{124} (\bibinfo{year}{2022}).
\newblock \eprint{2201.00512}.

\bibitem{yuan2003}
\bibinfo{author}{{Yuan}, F.}, \bibinfo{author}{{Quataert}, E.} \& \bibinfo{author}{{Narayan}, R.}
\newblock \bibinfo{title}{{Nonthermal Electrons in Radiatively Inefficient Accretion Flow Models of Sagittarius A*}}.
\newblock \emph{\bibinfo{journal}{\apj}} \textbf{\bibinfo{volume}{598}}, \bibinfo{pages}{301--312} (\bibinfo{year}{2003}).
\newblock \eprint{astro-ph/0304125}.

\bibitem{inayoshi2018}
\bibinfo{author}{{Inayoshi}, K.}, \bibinfo{author}{{Ostriker}, J.~P.}, \bibinfo{author}{{Haiman}, Z.} \& \bibinfo{author}{{Kuiper}, R.}
\newblock \bibinfo{title}{{Low-density, radiatively inefficient rotating-accretion flow on to a black hole}}.
\newblock \emph{\bibinfo{journal}{\mnras}} \textbf{\bibinfo{volume}{476}}, \bibinfo{pages}{1412--1426} (\bibinfo{year}{2018}).
\newblock \eprint{1709.07452}.

\bibitem{eht2022v}
\bibinfo{author}{{Event Horizon Telescope Collaboration}} \emph{et~al.}
\newblock \bibinfo{title}{{First Sagittarius A* Event Horizon Telescope Results. V. Testing Astrophysical Models of the Galactic Center Black Hole}}.
\newblock \emph{\bibinfo{journal}{\apjl}} \textbf{\bibinfo{volume}{930}}, \bibinfo{pages}{L16} (\bibinfo{year}{2022}).

\bibitem{ben2024}
\bibinfo{author}{{Ben Zineb}, Y.}, \bibinfo{author}{{Ozel}, F.} \& \bibinfo{author}{{Psaltis}, D.}
\newblock \bibinfo{title}{{Advancing Black Hole Imaging with Space-Based Interferometry}}.
\newblock \emph{\bibinfo{journal}{arXiv e-prints}} \bibinfo{pages}{arXiv:2412.01904} (\bibinfo{year}{2024}).
\newblock \eprint{2412.01904}.

\bibitem{lee2017}
\bibinfo{author}{{Lee}, M.~G.} \& \bibinfo{author}{{Jang}, I.~S.}
\newblock \bibinfo{title}{{Resolving the Discrepancy of Distance to M60, a Giant Elliptical Galaxy in Virgo}}.
\newblock \emph{\bibinfo{journal}{\apj}} \textbf{\bibinfo{volume}{841}}, \bibinfo{pages}{23} (\bibinfo{year}{2017}).
\newblock \eprint{1705.02389}.

\bibitem{koryukova2022}
\bibinfo{author}{{Koryukova}, T.~A.}, \bibinfo{author}{{Pushkarev}, A.~B.}, \bibinfo{author}{{Plavin}, A.~V.} \& \bibinfo{author}{{Kovalev}, Y.~Y.}
\newblock \bibinfo{title}{{Tracing Milky Way scattering by compact extragalactic radio sources}}.
\newblock \emph{\bibinfo{journal}{\mnras}} \textbf{\bibinfo{volume}{515}}, \bibinfo{pages}{1736--1750} (\bibinfo{year}{2022}).
\newblock \eprint{2201.04359}.

\bibitem{blandford1979}
\bibinfo{author}{{Blandford}, R.~D.} \& \bibinfo{author}{{K{\"o}nigl}, A.}
\newblock \bibinfo{title}{{Relativistic jets as compact radio sources.}}
\newblock \emph{\bibinfo{journal}{\apj}} \textbf{\bibinfo{volume}{232}}, \bibinfo{pages}{34--48} (\bibinfo{year}{1979}).

\bibitem{bartel1986}
\bibinfo{author}{{Bartel}, N.}, \bibinfo{author}{{Herring}, T.~A.}, \bibinfo{author}{{Ratner}, M.~I.}, \bibinfo{author}{{Shapiro}, I.~I.} \& \bibinfo{author}{{Corey}, B.~E.}
\newblock \bibinfo{title}{{VLBI limits on the proper motion of the `core' of the superluminal quasar 3C345}}.
\newblock \emph{\bibinfo{journal}{\nat}} \textbf{\bibinfo{volume}{319}}, \bibinfo{pages}{733--738} (\bibinfo{year}{1986}).

\bibitem{lobanov1998}
\bibinfo{author}{{Lobanov}, A.~P.}
\newblock \bibinfo{title}{{Ultracompact jets in active galactic nuclei}}.
\newblock \emph{\bibinfo{journal}{\aap}} \textbf{\bibinfo{volume}{330}}, \bibinfo{pages}{79--89} (\bibinfo{year}{1998}).
\newblock \eprint{astro-ph/9712132}.

\bibitem{OSullivan2009}
\bibinfo{author}{{O'Sullivan}, S.~P.} \& \bibinfo{author}{{Gabuzda}, D.~C.}
\newblock \bibinfo{title}{{Magnetic field strength and spectral distribution of six parsec-scale active galactic nuclei jets}}.
\newblock \emph{\bibinfo{journal}{\mnras}} \textbf{\bibinfo{volume}{400}}, \bibinfo{pages}{26--42} (\bibinfo{year}{2009}).
\newblock \eprint{0907.5211}.

\bibitem{fraga2026}
\bibinfo{author}{{Fraga-Encinas}, R.}, \bibinfo{author}{{Mo{\'s}cibrodzka}, M.} \& \bibinfo{author}{{Falcke}, H.}
\newblock \bibinfo{title}{{The core-shift of Sagittarius A* as a discriminant between disk and jet emission models with millimeter very long baseline interferometry}}.
\newblock \emph{\bibinfo{journal}{\aap}} \textbf{\bibinfo{volume}{706}}, \bibinfo{pages}{A151} (\bibinfo{year}{2026}).
\newblock \eprint{2312.12951}.

\bibitem{li2024}
\bibinfo{author}{{Li}, X.} \emph{et~al.}
\newblock \bibinfo{title}{{A Centiparsec-scale Compact Radio Core in the Nearby Galaxy M60}}.
\newblock \emph{\bibinfo{journal}{\apj}} \textbf{\bibinfo{volume}{960}}, \bibinfo{pages}{1} (\bibinfo{year}{2024}).
\newblock \eprint{2311.06126}.

\bibitem{2012rjag.book.....B}
\bibinfo{author}{{Boettcher}, M.}, \bibinfo{author}{{Harris}, D.~E.} \& \bibinfo{author}{{Krawczynski}, H.}
\newblock \emph{\bibinfo{title}{{Relativistic Jets from Active Galactic Nuclei}}} (\bibinfo{publisher}{John Wiley \& Sons, Ltd}, \bibinfo{year}{2012}).

\bibitem{cheng2023}
\bibinfo{author}{{Cheng}, X.} \emph{et~al.}
\newblock \bibinfo{title}{{Toward Microarcsecond Astrometry for the Innermost Wobbling Jet of the BL Lacertae Object OJ 287}}.
\newblock \emph{\bibinfo{journal}{\apjl}} \textbf{\bibinfo{volume}{955}}, \bibinfo{pages}{L30} (\bibinfo{year}{2023}).
\newblock \eprint{2309.03635}.

\bibitem{park2019}
\bibinfo{author}{{Park}, J.} \emph{et~al.}
\newblock \bibinfo{title}{{Kinematics of the M87 Jet in the Collimation Zone: Gradual Acceleration and Velocity Stratification}}.
\newblock \emph{\bibinfo{journal}{\apj}} \textbf{\bibinfo{volume}{887}}, \bibinfo{pages}{147} (\bibinfo{year}{2019}).
\newblock \eprint{1911.02279}.

\bibitem{park2021}
\bibinfo{author}{{Park}, J.} \emph{et~al.}
\newblock \bibinfo{title}{{Jet Collimation and Acceleration in the Giant Radio Galaxy NGC 315}}.
\newblock \emph{\bibinfo{journal}{\apj}} \textbf{\bibinfo{volume}{909}}, \bibinfo{pages}{76} (\bibinfo{year}{2021}).
\newblock \eprint{2012.14154}.

\bibitem{Konigl1981}
\bibinfo{author}{{Konigl}, A.}
\newblock \bibinfo{title}{{Relativistic jets as X-ray and gamma-ray sources.}}
\newblock \emph{\bibinfo{journal}{\apj}} \textbf{\bibinfo{volume}{243}}, \bibinfo{pages}{700--709} (\bibinfo{year}{1981}).

\bibitem{Sokolovsky2011}
\bibinfo{author}{{Sokolovsky}, K.~V.}, \bibinfo{author}{{Kovalev}, Y.~Y.}, \bibinfo{author}{{Pushkarev}, A.~B.} \& \bibinfo{author}{{Lobanov}, A.~P.}
\newblock \bibinfo{title}{{A VLBA survey of the core shift effect in AGN jets. I. Evidence of dominating synchrotron opacity}}.
\newblock \emph{\bibinfo{journal}{\aap}} \textbf{\bibinfo{volume}{532}}, \bibinfo{pages}{A38} (\bibinfo{year}{2011}).
\newblock \eprint{1103.6032}.

\bibitem{Ricci2025}
\bibinfo{author}{{Ricci}, L.} \emph{et~al.}
\newblock \bibinfo{title}{{Spectral and magnetic properties of the jet base in NGC 315}}.
\newblock \emph{\bibinfo{journal}{\aap}} \textbf{\bibinfo{volume}{693}}, \bibinfo{pages}{A172} (\bibinfo{year}{2025}).
\newblock \eprint{2411.19126}.

\bibitem{Chamani2023}
\bibinfo{author}{{Chamani}, W.} \emph{et~al.}
\newblock \bibinfo{title}{{Time variability of the core-shift effect in the blazar 3C 454.3}}.
\newblock \emph{\bibinfo{journal}{\aap}} \textbf{\bibinfo{volume}{672}}, \bibinfo{pages}{A130} (\bibinfo{year}{2023}).
\newblock \eprint{2209.13301}.

\bibitem{Bower2015}
\bibinfo{author}{{Bower}, G.~C.} \emph{et~al.}
\newblock \bibinfo{title}{{The Proper Motion of the Galactic Center Pulsar Relative to Sagittarius A*}}.
\newblock \emph{\bibinfo{journal}{\apj}} \textbf{\bibinfo{volume}{798}}, \bibinfo{pages}{120} (\bibinfo{year}{2015}).
\newblock \eprint{1411.0399}.

\bibitem{eht2021}
\bibinfo{author}{{Event Horizon Telescope Collaboration}} \emph{et~al.}
\newblock \bibinfo{title}{{First M87 Event Horizon Telescope Results. VIII. Magnetic Field Structure near The Event Horizon}}.
\newblock \emph{\bibinfo{journal}{\apjl}} \textbf{\bibinfo{volume}{910}}, \bibinfo{pages}{L13} (\bibinfo{year}{2021}).
\newblock \eprint{2105.01173}.

\bibitem{Ro2023}
\bibinfo{author}{{Ro}, H.} \emph{et~al.}
\newblock \bibinfo{title}{{Spectral analysis of a parsec-scale jet in M 87: Observational constraint on the magnetic field strengths in the jet}}.
\newblock \emph{\bibinfo{journal}{\aap}} \textbf{\bibinfo{volume}{673}}, \bibinfo{pages}{A159} (\bibinfo{year}{2023}).
\newblock \eprint{2303.01014}.

\bibitem{bower2019}
\bibinfo{author}{{Bower}, G.~C.} \emph{et~al.}
\newblock \bibinfo{title}{{ALMA Observations of the Terahertz Spectrum of Sagittarius A*}}.
\newblock \emph{\bibinfo{journal}{\apjl}} \textbf{\bibinfo{volume}{881}}, \bibinfo{pages}{L2} (\bibinfo{year}{2019}).
\newblock \eprint{1907.08319}.

\bibitem{eht2024viii}
\bibinfo{author}{{Event Horizon Telescope Collaboration}} \emph{et~al.}
\newblock \bibinfo{title}{{First Sagittarius A* Event Horizon Telescope Results. VIII. Physical Interpretation of the Polarized Ring}}.
\newblock \emph{\bibinfo{journal}{\apjl}} \textbf{\bibinfo{volume}{964}}, \bibinfo{pages}{L26} (\bibinfo{year}{2024}).

\bibitem{Chavez2024}
\bibinfo{author}{{Chavez}, E.} \emph{et~al.}
\newblock \bibinfo{title}{{Prospects of Detecting a Jet in Sagittarius A* with Very-long-baseline Interferometry}}.
\newblock \emph{\bibinfo{journal}{\apj}} \textbf{\bibinfo{volume}{974}}, \bibinfo{pages}{116} (\bibinfo{year}{2024}).
\newblock \eprint{2405.06029}.

\bibitem{deller2007}
\bibinfo{author}{{Deller}, A.~T.}, \bibinfo{author}{{Tingay}, S.~J.}, \bibinfo{author}{{Bailes}, M.} \& \bibinfo{author}{{West}, C.}
\newblock \bibinfo{title}{{DiFX: A Software Correlator for Very Long Baseline Interferometry Using Multiprocessor Computing Environments}}.
\newblock \emph{\bibinfo{journal}{\pasp}} \textbf{\bibinfo{volume}{119}}, \bibinfo{pages}{318--336} (\bibinfo{year}{2007}).
\newblock \eprint{astro-ph/0702141}.

\bibitem{deller2011}
\bibinfo{author}{{Deller}, A.~T.} \emph{et~al.}
\newblock \bibinfo{title}{{DiFX-2: A More Flexible, Efficient, Robust, and Powerful Software Correlator}}.
\newblock \emph{\bibinfo{journal}{\pasp}} \textbf{\bibinfo{volume}{123}}, \bibinfo{pages}{275} (\bibinfo{year}{2011}).
\newblock \eprint{1101.0885}.

\bibitem{greisen2003}
\bibinfo{author}{{Greisen}, E.~W.}
\newblock \bibinfo{title}{{AIPS, the VLA, and the VLBA}}.
\newblock In \bibinfo{editor}{{Heck}, A.} (ed.) \emph{\bibinfo{booktitle}{Information Handling in Astronomy - Historical Vistas}}, vol. \bibinfo{volume}{285} of \emph{\bibinfo{series}{Astrophysics and Space Science Library}}, \bibinfo{pages}{109} (\bibinfo{year}{2003}).

\bibitem{keimpema2015}
\bibinfo{author}{{Keimpema}, A.} \emph{et~al.}
\newblock \bibinfo{title}{{The SFXC software correlator for very long baseline interferometry: algorithms and implementation}}.
\newblock \emph{\bibinfo{journal}{Experimental Astronomy}} \textbf{\bibinfo{volume}{39}}, \bibinfo{pages}{259--279} (\bibinfo{year}{2015}).
\newblock \eprint{1502.00467}.

\bibitem{Salafia2022}
\bibinfo{author}{{Salafia}, O.~S.} \emph{et~al.}
\newblock \bibinfo{title}{{Multiwavelength View of the Close-by GRB 190829A Sheds Light on Gamma-Ray Burst Physics}}.
\newblock \emph{\bibinfo{journal}{\apjl}} \textbf{\bibinfo{volume}{931}}, \bibinfo{pages}{L19} (\bibinfo{year}{2022}).
\newblock \eprint{2106.07169}.

\bibitem{cheng2021}
\bibinfo{author}{{Cheng}, X.}, \bibinfo{author}{{An}, T.}, \bibinfo{author}{{Sohn}, B.~W.}, \bibinfo{author}{{Hong}, X.} \& \bibinfo{author}{{Wang}, A.}
\newblock \bibinfo{title}{{Parsec-scale properties of eight Fanaroff-Riley type 0 radio galaxies}}.
\newblock \emph{\bibinfo{journal}{\mnras}} \textbf{\bibinfo{volume}{506}}, \bibinfo{pages}{1609--1622} (\bibinfo{year}{2021}).
\newblock \eprint{2105.05396}.

\bibitem{shen2010}
\bibinfo{author}{{Shen}, J.} \& \bibinfo{author}{{Gebhardt}, K.}
\newblock \bibinfo{title}{{The Supermassive Black Hole and Dark Matter Halo of NGC 4649 (M60)}}.
\newblock \emph{\bibinfo{journal}{\apj}} \textbf{\bibinfo{volume}{711}}, \bibinfo{pages}{484--494} (\bibinfo{year}{2010}).
\newblock \eprint{0910.4168}.

\bibitem{woo2013}
\bibinfo{author}{{Woo}, J.-H.} \emph{et~al.}
\newblock \bibinfo{title}{{Do Quiescent and Active Galaxies Have Different M $_{BH}$-{\ensuremath{\sigma}}$_{*}$ Relations?}}
\newblock \emph{\bibinfo{journal}{\apj}} \textbf{\bibinfo{volume}{772}}, \bibinfo{pages}{49} (\bibinfo{year}{2013}).
\newblock \eprint{1305.2946}.

\bibitem{Paggi2014}
\bibinfo{author}{{Paggi}, A.} \emph{et~al.}
\newblock \bibinfo{title}{{Active Galactic Nucleus Feedback in the Hot Halo of NGC 4649}}.
\newblock \emph{\bibinfo{journal}{\apj}} \textbf{\bibinfo{volume}{787}}, \bibinfo{pages}{134} (\bibinfo{year}{2014}).

\bibitem{paggi2017}
\bibinfo{author}{{Paggi}, A.} \emph{et~al.}
\newblock \bibinfo{title}{{Constraining the Physical State of the Hot Gas Halos in NGC 4649 and NGC 5846}}.
\newblock \emph{\bibinfo{journal}{\apj}} \textbf{\bibinfo{volume}{844}}, \bibinfo{pages}{5} (\bibinfo{year}{2017}).
\newblock \eprint{1706.02303}.

\bibitem{duras2020}
\bibinfo{author}{{Duras}, F.} \emph{et~al.}
\newblock \bibinfo{title}{{Universal bolometric corrections for active galactic nuclei over seven luminosity decades}}.
\newblock \emph{\bibinfo{journal}{\aap}} \textbf{\bibinfo{volume}{636}}, \bibinfo{pages}{A73} (\bibinfo{year}{2020}).
\newblock \eprint{2001.09984}.

\bibitem{White2016}
\bibinfo{author}{{White}, C.~J.}, \bibinfo{author}{{Stone}, J.~M.} \& \bibinfo{author}{{Gammie}, C.~F.}
\newblock \bibinfo{title}{{An Extension of the Athena++ Code Framework for GRMHD Based on Advanced Riemann Solvers and Staggered-mesh Constrained Transport}}.
\newblock \emph{\bibinfo{journal}{ApJS}} \textbf{\bibinfo{volume}{225}}, \bibinfo{pages}{22} (\bibinfo{year}{2016}).
\newblock \eprint{1511.00943}.

\bibitem{Stone2020}
\bibinfo{author}{{Stone}, J.~M.}, \bibinfo{author}{{Tomida}, K.}, \bibinfo{author}{{White}, C.~J.} \& \bibinfo{author}{{Felker}, K.~G.}
\newblock \bibinfo{title}{{The Athena++ Adaptive Mesh Refinement Framework: Design and Magnetohydrodynamic Solvers}}.
\newblock \emph{\bibinfo{journal}{ApJS}} \textbf{\bibinfo{volume}{249}}, \bibinfo{pages}{4} (\bibinfo{year}{2020}).
\newblock \eprint{2005.06651}.

\bibitem{Fishbone1976}
\bibinfo{author}{{Fishbone}, L.~G.} \& \bibinfo{author}{{Moncrief}, V.}
\newblock \bibinfo{title}{{Relativistic fluid disks in orbit around Kerr black holes.}}
\newblock \emph{\bibinfo{journal}{ApJ}} \textbf{\bibinfo{volume}{207}}, \bibinfo{pages}{962--976} (\bibinfo{year}{1976}).

\bibitem{Penna2013}
\bibinfo{author}{{Penna}, R.~F.}, \bibinfo{author}{{Kulkarni}, A.} \& \bibinfo{author}{{Narayan}, R.}
\newblock \bibinfo{title}{{A new equilibrium torus solution and GRMHD initial conditions}}.
\newblock \emph{\bibinfo{journal}{AAP}} \textbf{\bibinfo{volume}{559}}, \bibinfo{pages}{A116} (\bibinfo{year}{2013}).
\newblock \eprint{1309.3680}.

\bibitem{Narayan2003}
\bibinfo{author}{{Narayan}, R.}, \bibinfo{author}{{Igumenshchev}, I.~V.} \& \bibinfo{author}{{Abramowicz}, M.~A.}
\newblock \bibinfo{title}{{Magnetically Arrested Disk: an Energetically Efficient Accretion Flow}}.
\newblock \emph{\bibinfo{journal}{PASJ}} \textbf{\bibinfo{volume}{55}}, \bibinfo{pages}{L69--L72} (\bibinfo{year}{2003}).
\newblock \eprint{astro-ph/0305029}.

\bibitem{Tchekhovskoy2011}
\bibinfo{author}{{Tchekhovskoy}, A.}, \bibinfo{author}{{Narayan}, R.} \& \bibinfo{author}{{McKinney}, J.~C.}
\newblock \bibinfo{title}{{Efficient generation of jets from magnetically arrested accretion on a rapidly spinning black hole}}.
\newblock \emph{\bibinfo{journal}{\mnras}} \textbf{\bibinfo{volume}{418}}, \bibinfo{pages}{L79--L83} (\bibinfo{year}{2011}).
\newblock \eprint{1108.0412}.

\bibitem{yang2024}
\bibinfo{author}{{Yang}, H.} \emph{et~al.}
\newblock \bibinfo{title}{{Modeling the inner part of the jet in M87: Confronting jet morphology with theory}}.
\newblock \emph{\bibinfo{journal}{Science Advances}} \textbf{\bibinfo{volume}{10}}, \bibinfo{pages}{eadn3544} (\bibinfo{year}{2024}).
\newblock \eprint{2403.15950}.

\bibitem{Moscibrodzka2016}
\bibinfo{author}{{Mo{\'s}cibrodzka}, M.}, \bibinfo{author}{{Falcke}, H.} \& \bibinfo{author}{{Shiokawa}, H.}
\newblock \bibinfo{title}{{General relativistic magnetohydrodynamical simulations of the jet in M 87}}.
\newblock \emph{\bibinfo{journal}{\aap}} \textbf{\bibinfo{volume}{586}}, \bibinfo{pages}{A38} (\bibinfo{year}{2016}).
\newblock \eprint{1510.07243}.

\bibitem{Petersen2020}
\bibinfo{author}{{Petersen}, E.} \& \bibinfo{author}{{Gammie}, C.}
\newblock \bibinfo{title}{{Non-thermal models for infrared flares from Sgr A*}}.
\newblock \emph{\bibinfo{journal}{MNRAS}} \textbf{\bibinfo{volume}{494}}, \bibinfo{pages}{5923--5935} (\bibinfo{year}{2020}).

\bibitem{Blandford-Konigl1979}
\bibinfo{author}{{Blandford}, R.~D.} \& \bibinfo{author}{{K{\"o}nigl}, A.}
\newblock \bibinfo{title}{{Relativistic jets as compact radio sources.}}
\newblock \emph{\bibinfo{journal}{\apj}} \textbf{\bibinfo{volume}{232}}, \bibinfo{pages}{34--48} (\bibinfo{year}{1979}).

\bibitem{Ball2018}
\bibinfo{author}{{Ball}, D.}, \bibinfo{author}{{{\"O}zel}, F.}, \bibinfo{author}{{Psaltis}, D.}, \bibinfo{author}{{Chan}, C.-K.} \& \bibinfo{author}{{Sironi}, L.}
\newblock \bibinfo{title}{{The Properties of Reconnection Current Sheets in GRMHD Simulations of Radiatively Inefficient Accretion Flows}}.
\newblock \emph{\bibinfo{journal}{\apj}} \textbf{\bibinfo{volume}{853}}, \bibinfo{pages}{184} (\bibinfo{year}{2018}).
\newblock \eprint{1705.06293}.

\bibitem{Moscibrodzka2018}
\bibinfo{author}{{Mo{\'s}cibrodzka}, M.} \& \bibinfo{author}{{Gammie}, C.~F.}
\newblock \bibinfo{title}{{IPOLE - semi-analytic scheme for relativistic polarized radiative transport}}.
\newblock \emph{\bibinfo{journal}{\mnras}} \textbf{\bibinfo{volume}{475}}, \bibinfo{pages}{43--54} (\bibinfo{year}{2018}).
\newblock \eprint{1712.03057}.

\bibitem{Yuan-2003}
\bibinfo{author}{{Yuan}, F.}, \bibinfo{author}{{Quataert}, E.} \& \bibinfo{author}{{Narayan}, R.}
\newblock \bibinfo{title}{{Nonthermal Electrons in Radiatively Inefficient Accretion Flow Models of Sagittarius A*}}.
\newblock \emph{\bibinfo{journal}{\apj}} \textbf{\bibinfo{volume}{598}}, \bibinfo{pages}{301--312} (\bibinfo{year}{2003}).
\newblock \eprint{astro-ph/0304125}.

\bibitem{GRAVITY2019}
\bibinfo{author}{{GRAVITY Collaboration}} \emph{et~al.}
\newblock \bibinfo{title}{{A geometric distance measurement to the Galactic center black hole with 0.3\% uncertainty}}.
\newblock \emph{\bibinfo{journal}{\aap}} \textbf{\bibinfo{volume}{625}}, \bibinfo{pages}{L10} (\bibinfo{year}{2019}).
\newblock \eprint{1904.05721}.

\bibitem{Ghez2008}
\bibinfo{author}{{Ghez}, A.~M.} \emph{et~al.}
\newblock \bibinfo{title}{{Measuring Distance and Properties of the Milky Way's Central Supermassive Black Hole with Stellar Orbits}}.
\newblock \emph{\bibinfo{journal}{\apj}} \textbf{\bibinfo{volume}{689}}, \bibinfo{pages}{1044--1062} (\bibinfo{year}{2008}).
\newblock \eprint{0808.2870}.

\bibitem{eht2022vi}
\bibinfo{author}{{Event Horizon Telescope Collaboration}} \emph{et~al.}
\newblock \bibinfo{title}{{First Sagittarius A* Event Horizon Telescope Results. VI. Testing the Black Hole Metric}}.
\newblock \emph{\bibinfo{journal}{\apjl}} \textbf{\bibinfo{volume}{930}}, \bibinfo{pages}{L17} (\bibinfo{year}{2022}).

\bibitem{Lo1998}
\bibinfo{author}{{Lo}, K.~Y.}, \bibinfo{author}{{Shen}, Z.-Q.}, \bibinfo{author}{{Zhao}, J.-H.} \& \bibinfo{author}{{Ho}, P. T.~P.}
\newblock \bibinfo{title}{{Intrinsic Size of Sagittarius A*: 72 Schwarzschild Radii}}.
\newblock \emph{\bibinfo{journal}{\apjl}} \textbf{\bibinfo{volume}{508}}, \bibinfo{pages}{L61--L64} (\bibinfo{year}{1998}).
\newblock \eprint{astro-ph/9809222}.

\bibitem{eht2019i}
\bibinfo{author}{{Event Horizon Telescope Collaboration}} \emph{et~al.}
\newblock \bibinfo{title}{{First M87 Event Horizon Telescope Results. I. The Shadow of the Supermassive Black Hole}}.
\newblock \emph{\bibinfo{journal}{\apjl}} \textbf{\bibinfo{volume}{875}}, \bibinfo{pages}{L1} (\bibinfo{year}{2019}).
\newblock \eprint{1906.11238}.

\bibitem{gonzalez2009}
\bibinfo{author}{{Gonz{\'a}lez-Mart{\'\i}n}, O.}, \bibinfo{author}{{Masegosa}, J.}, \bibinfo{author}{{M{\'a}rquez}, I.}, \bibinfo{author}{{Guainazzi}, M.} \& \bibinfo{author}{{Jim{\'e}nez-Bail{\'o}n}, E.}
\newblock \bibinfo{title}{{An X-ray view of 82 LINERs with Chandra and XMM-Newton data}}.
\newblock \emph{\bibinfo{journal}{\aap}} \textbf{\bibinfo{volume}{506}}, \bibinfo{pages}{1107--1121} (\bibinfo{year}{2009}).
\newblock \eprint{0905.2973}.

\bibitem{fomalont2000}
\bibinfo{author}{{Fomalont}, E.~B.} \emph{et~al.}
\newblock \bibinfo{title}{{The VSOP 5 GHz Continuum Survey: The Prelaunch VLBA Observations}}.
\newblock \emph{\bibinfo{journal}{\apjs}} \textbf{\bibinfo{volume}{131}}, \bibinfo{pages}{95--183} (\bibinfo{year}{2000}).

\bibitem{veale2017}
\bibinfo{author}{{Veale}, M.} \emph{et~al.}
\newblock \bibinfo{title}{{The MASSIVE Survey - V. Spatially resolved stellar angular momentum, velocity dispersion, and higher moments of the 41 most massive local early-type galaxies}}.
\newblock \emph{\bibinfo{journal}{\mnras}} \textbf{\bibinfo{volume}{464}}, \bibinfo{pages}{356--384} (\bibinfo{year}{2017}).
\newblock \eprint{1609.00391}.

\bibitem{thomas2016}
\bibinfo{author}{{Thomas}, J.} \emph{et~al.}
\newblock \bibinfo{title}{{A 17-billion-solar-mass black hole in a group galaxy with a diffuse core}}.
\newblock \emph{\bibinfo{journal}{\nat}} \textbf{\bibinfo{volume}{532}}, \bibinfo{pages}{340--342} (\bibinfo{year}{2016}).
\newblock \eprint{1604.01400}.

\bibitem{Beuing1999}
\bibinfo{author}{{Beuing}, J.}, \bibinfo{author}{{Dobereiner}, S.}, \bibinfo{author}{{Bohringer}, H.} \& \bibinfo{author}{{Bender}, R.}
\newblock \bibinfo{title}{{X-ray luminosities for a magnitude-limited sample of early-type galaxies from the ROSAT All-Sky Survey}}.
\newblock \emph{\bibinfo{journal}{\mnras}} \textbf{\bibinfo{volume}{302}}, \bibinfo{pages}{209--221} (\bibinfo{year}{1999}).

\bibitem{brown2011}
\bibinfo{author}{{Brown}, M. J.~I.}, \bibinfo{author}{{Jannuzi}, B.~T.}, \bibinfo{author}{{Floyd}, D. J.~E.} \& \bibinfo{author}{{Mould}, J.~R.}
\newblock \bibinfo{title}{{The Ubiquitous Radio Continuum Emission from the Most Massive Early-type Galaxies}}.
\newblock \emph{\bibinfo{journal}{\apjl}} \textbf{\bibinfo{volume}{731}}, \bibinfo{pages}{L41} (\bibinfo{year}{2011}).
\newblock \eprint{1103.2828}.

\bibitem{beasley2018}
\bibinfo{author}{{Beasley}, M.~A.}, \bibinfo{author}{{Trujillo}, I.}, \bibinfo{author}{{Leaman}, R.} \& \bibinfo{author}{{Montes}, M.}
\newblock \bibinfo{title}{{A single population of red globular clusters around the massive compact galaxy NGC 1277}}.
\newblock \emph{\bibinfo{journal}{\nat}} \textbf{\bibinfo{volume}{555}}, \bibinfo{pages}{483--486} (\bibinfo{year}{2018}).
\newblock \eprint{1803.04893}.

\bibitem{van2012}
\bibinfo{author}{{van den Bosch}, R. C.~E.} \emph{et~al.}
\newblock \bibinfo{title}{{An over-massive black hole in the compact lenticular galaxy NGC 1277}}.
\newblock \emph{\bibinfo{journal}{\nat}} \textbf{\bibinfo{volume}{491}}, \bibinfo{pages}{729--731} (\bibinfo{year}{2012}).
\newblock \eprint{1211.6429}.

\bibitem{Scharw2016}
\bibinfo{author}{{Scharw{\"a}chter}, J.}, \bibinfo{author}{{Combes}, F.}, \bibinfo{author}{{Salom{\'e}}, P.}, \bibinfo{author}{{Sun}, M.} \& \bibinfo{author}{{Krips}, M.}
\newblock \bibinfo{title}{{The overmassive black hole in NGC 1277: new constraints from molecular gas kinematics}}.
\newblock \emph{\bibinfo{journal}{\mnras}} \textbf{\bibinfo{volume}{457}}, \bibinfo{pages}{4272--4284} (\bibinfo{year}{2016}).
\newblock \eprint{1507.02292}.

\bibitem{park2017}
\bibinfo{author}{{Park}, S.}, \bibinfo{author}{{Yang}, J.}, \bibinfo{author}{{Oonk}, J.~B.~R.} \& \bibinfo{author}{{Paragi}, Z.}
\newblock \bibinfo{title}{{Discovery of five low-luminosity active galactic nuclei at the centre of the Perseus cluster}}.
\newblock \emph{\bibinfo{journal}{\mnras}} \textbf{\bibinfo{volume}{465}}, \bibinfo{pages}{3943--3948} (\bibinfo{year}{2017}).
\newblock \eprint{1611.05986}.

\bibitem{mcconnell2012}
\bibinfo{author}{{McConnell}, N.~J.} \emph{et~al.}
\newblock \bibinfo{title}{{Dynamical Measurements of Black Hole Masses in Four Brightest Cluster Galaxies at 100 Mpc}}.
\newblock \emph{\bibinfo{journal}{\apj}} \textbf{\bibinfo{volume}{756}}, \bibinfo{pages}{179} (\bibinfo{year}{2012}).
\newblock \eprint{1203.1620}.

\bibitem{Sanders2014}
\bibinfo{author}{{Sanders}, J.~S.} \emph{et~al.}
\newblock \bibinfo{title}{{The X-ray coronae of the two brightest galaxies in the Coma cluster}}.
\newblock \emph{\bibinfo{journal}{\mnras}} \textbf{\bibinfo{volume}{439}}, \bibinfo{pages}{1182--1192} (\bibinfo{year}{2014}).
\newblock \eprint{1401.3131}.

\bibitem{Breuval2023}
\bibinfo{author}{{Breuval}, L.} \emph{et~al.}
\newblock \bibinfo{title}{{A 1.3\% Distance to M33 from Hubble Space Telescope Cepheid Photometry}}.
\newblock \emph{\bibinfo{journal}{\apj}} \textbf{\bibinfo{volume}{951}}, \bibinfo{pages}{118} (\bibinfo{year}{2023}).
\newblock \eprint{2304.00037}.

\bibitem{bender2005}
\bibinfo{author}{{Bender}, R.} \emph{et~al.}
\newblock \bibinfo{title}{{HST STIS Spectroscopy of the Triple Nucleus of M31: Two Nested Disks in Keplerian Rotation around a Supermassive Black Hole}}.
\newblock \emph{\bibinfo{journal}{\apj}} \textbf{\bibinfo{volume}{631}}, \bibinfo{pages}{280--300} (\bibinfo{year}{2005}).
\newblock \eprint{astro-ph/0509839}.

\bibitem{Garcia2010}
\bibinfo{author}{{Garcia}, M.~R.} \emph{et~al.}
\newblock \bibinfo{title}{{X-ray and Radio Variability of M31*, The Andromeda Galaxy Nuclear Supermassive Black Hole}}.
\newblock \emph{\bibinfo{journal}{\apj}} \textbf{\bibinfo{volume}{710}}, \bibinfo{pages}{755--763} (\bibinfo{year}{2010}).
\newblock \eprint{0907.4977}.

\bibitem{peng2023}
\bibinfo{author}{{Peng}, S.} \emph{et~al.}
\newblock \bibinfo{title}{{Searching for Radio Outflows from M31* with VLBI Observations}}.
\newblock \emph{\bibinfo{journal}{\apj}} \textbf{\bibinfo{volume}{953}}, \bibinfo{pages}{12} (\bibinfo{year}{2023}).
\newblock \eprint{2306.07189}.

\bibitem{tonry2001}
\bibinfo{author}{{Tonry}, J.~L.} \emph{et~al.}
\newblock \bibinfo{title}{{The SBF Survey of Galaxy Distances. IV. SBF Magnitudes, Colors, and Distances}}.
\newblock \emph{\bibinfo{journal}{\apj}} \textbf{\bibinfo{volume}{546}}, \bibinfo{pages}{681--693} (\bibinfo{year}{2001}).
\newblock \eprint{astro-ph/0011223}.

\bibitem{rusli2013}
\bibinfo{author}{{Rusli}, S.~P.} \emph{et~al.}
\newblock \bibinfo{title}{{The Influence of Dark Matter Halos on Dynamical Estimates of Black Hole Mass: 10 New Measurements for High-{\ensuremath{\sigma}} Early-type Galaxies}}.
\newblock \emph{\bibinfo{journal}{\aj}} \textbf{\bibinfo{volume}{146}}, \bibinfo{pages}{45} (\bibinfo{year}{2013}).
\newblock \eprint{1306.1124}.

\bibitem{Pellegrini2005}
\bibinfo{author}{{Pellegrini}, S.}
\newblock \bibinfo{title}{{Nuclear Accretion in Galaxies of the Local Universe: Clues from Chandra Observations}}.
\newblock \emph{\bibinfo{journal}{\apj}} \textbf{\bibinfo{volume}{624}}, \bibinfo{pages}{155--161} (\bibinfo{year}{2005}).
\newblock \eprint{astro-ph/0502035}.

\bibitem{giacintucci2012}
\bibinfo{author}{{Giacintucci}, S.} \emph{et~al.}
\newblock \bibinfo{title}{{Recurrent Radio Outbursts at the Center of the NGC 1407 Galaxy Group}}.
\newblock \emph{\bibinfo{journal}{\apj}} \textbf{\bibinfo{volume}{755}}, \bibinfo{pages}{172} (\bibinfo{year}{2012}).
\newblock \eprint{1206.5751}.

\bibitem{blakeslee2009}
\bibinfo{author}{{Blakeslee}, J.~P.} \emph{et~al.}
\newblock \bibinfo{title}{{The ACS Fornax Cluster Survey. V. Measurement and Recalibration of Surface Brightness Fluctuations and a Precise Value of the Fornax-Virgo Relative Distance}}.
\newblock \emph{\bibinfo{journal}{\apj}} \textbf{\bibinfo{volume}{694}}, \bibinfo{pages}{556--572} (\bibinfo{year}{2009}).
\newblock \eprint{0901.1138}.

\bibitem{Panessa2006}
\bibinfo{author}{{Panessa}, F.} \emph{et~al.}
\newblock \bibinfo{title}{{On the X-ray, optical emission line and black hole mass properties of local Seyfert galaxies}}.
\newblock \emph{\bibinfo{journal}{\aap}} \textbf{\bibinfo{volume}{455}}, \bibinfo{pages}{173--185} (\bibinfo{year}{2006}).
\newblock \eprint{astro-ph/0605236}.

\bibitem{nagar2005}
\bibinfo{author}{{Nagar}, N.~M.}, \bibinfo{author}{{Falcke}, H.} \& \bibinfo{author}{{Wilson}, A.~S.}
\newblock \bibinfo{title}{{Radio sources in low-luminosity active galactic nuclei. IV. Radio luminosity function, importance of jet power, and radio properties of the complete Palomar sample}}.
\newblock \emph{\bibinfo{journal}{\aap}} \textbf{\bibinfo{volume}{435}}, \bibinfo{pages}{521--543} (\bibinfo{year}{2005}).
\newblock \eprint{astro-ph/0502551}.

\bibitem{Gultekin2019}
\bibinfo{author}{{G{\"u}ltekin}, K.} \emph{et~al.}
\newblock \bibinfo{title}{{The Fundamental Plane of Black Hole Accretion and Its Use as a Black Hole-Mass Estimator}}.
\newblock \emph{\bibinfo{journal}{\apj}} \textbf{\bibinfo{volume}{871}}, \bibinfo{pages}{80} (\bibinfo{year}{2019}).
\newblock \eprint{1901.02530}.

\bibitem{sadler1989}
\bibinfo{author}{{Sadler}, E.~M.}, \bibinfo{author}{{Jenkins}, C.~R.} \& \bibinfo{author}{{Kotanyi}, C.~G.}
\newblock \bibinfo{title}{{Low-luminosity radio sources in early-type galaxies.}}
\newblock \emph{\bibinfo{journal}{\mnras}} \textbf{\bibinfo{volume}{240}}, \bibinfo{pages}{591--635} (\bibinfo{year}{1989}).

\bibitem{sorce2013}
\bibinfo{author}{{Sorce}, J.~G.} \emph{et~al.}
\newblock \bibinfo{title}{{Calibration of the Mid-infrared Tully-Fisher Relation}}.
\newblock \emph{\bibinfo{journal}{\apj}} \textbf{\bibinfo{volume}{765}}, \bibinfo{pages}{94} (\bibinfo{year}{2013}).
\newblock \eprint{1301.4833}.

\bibitem{Caglar2017}
\bibinfo{author}{{Caglar}, T.} \& \bibinfo{author}{{Hudaverdi}, M.}
\newblock \bibinfo{title}{{XMM-Newton view of X-ray overdensities from nearby galaxy clusters: the environmental dependencies}}.
\newblock \emph{\bibinfo{journal}{\mnras}} \textbf{\bibinfo{volume}{471}}, \bibinfo{pages}{4990--5007} (\bibinfo{year}{2017}).
\newblock \eprint{1709.00117}.

\bibitem{feretti1994}
\bibinfo{author}{{Feretti}, L.} \& \bibinfo{author}{{Giovannini}, G.}
\newblock \bibinfo{title}{{Structures of small-size radio galaxies in clusters.}}
\newblock \emph{\bibinfo{journal}{\aap}} \textbf{\bibinfo{volume}{281}}, \bibinfo{pages}{375--387} (\bibinfo{year}{1994}).

\bibitem{ferrarese2000}
\bibinfo{author}{{Ferrarese}, L.} \emph{et~al.}
\newblock \bibinfo{title}{{The Hubble Space Telescope Key Project on the Extragalactic Distance Scale. XXVI. The Calibration of Population II Secondary Distance Indicators and the Value of the Hubble Constant}}.
\newblock \emph{\bibinfo{journal}{\apj}} \textbf{\bibinfo{volume}{529}}, \bibinfo{pages}{745--767} (\bibinfo{year}{2000}).
\newblock \eprint{astro-ph/9908192}.

\bibitem{emsellem1999}
\bibinfo{author}{{Emsellem}, E.}, \bibinfo{author}{{Dejonghe}, H.} \& \bibinfo{author}{{Bacon}, R.}
\newblock \bibinfo{title}{{Dynamical models of NGC 3115}}.
\newblock \emph{\bibinfo{journal}{\mnras}} \textbf{\bibinfo{volume}{303}}, \bibinfo{pages}{495--514} (\bibinfo{year}{1999}).
\newblock \eprint{astro-ph/9810306}.

\bibitem{Pellegrini2010}
\bibinfo{author}{{Pellegrini}, S.}
\newblock \bibinfo{title}{{The Nuclear X-ray Emission of Nearby Early-type Galaxies}}.
\newblock \emph{\bibinfo{journal}{\apj}} \textbf{\bibinfo{volume}{717}}, \bibinfo{pages}{640--652} (\bibinfo{year}{2010}).
\newblock \eprint{1005.2344}.

\bibitem{wrobel2012}
\bibinfo{author}{{Wrobel}, J.~M.} \& \bibinfo{author}{{Nyland}, K.}
\newblock \bibinfo{title}{{Discovery of a Flat-spectrum Radio Nucleus in NGC 3115}}.
\newblock \emph{\bibinfo{journal}{\aj}} \textbf{\bibinfo{volume}{144}}, \bibinfo{pages}{160} (\bibinfo{year}{2012}).
\newblock \eprint{1209.6602}.

\bibitem{cappellari2002}
\bibinfo{author}{{Cappellari}, M.} \emph{et~al.}
\newblock \bibinfo{title}{{The Counterrotating Core and the Black Hole Mass of IC 1459}}.
\newblock \emph{\bibinfo{journal}{\apj}} \textbf{\bibinfo{volume}{578}}, \bibinfo{pages}{787--805} (\bibinfo{year}{2002}).
\newblock \eprint{astro-ph/0202155}.

\bibitem{tingay2015}
\bibinfo{author}{{Tingay}, S.~J.} \& \bibinfo{author}{{Edwards}, P.~G.}
\newblock \bibinfo{title}{{The multifrequency parsec-scale structure of PKS 2254-367 (IC 1459): a luminosity-dependent break in morphology for the precursors of radio galaxies?}}
\newblock \emph{\bibinfo{journal}{\mnras}} \textbf{\bibinfo{volume}{448}}, \bibinfo{pages}{252--257} (\bibinfo{year}{2015}).
\newblock \eprint{1501.04393}.

\bibitem{theureau2007}
\bibinfo{author}{{Theureau}, G.}, \bibinfo{author}{{Hanski}, M.~O.}, \bibinfo{author}{{Coudreau}, N.}, \bibinfo{author}{{Hallet}, N.} \& \bibinfo{author}{{Martin}, J.~M.}
\newblock \bibinfo{title}{{Kinematics of the Local Universe. XIII. 21-cm line measurements of 452 galaxies with the Nan{\c{c}}ay radiotelescope, JHK Tully-Fisher relation, and preliminary maps of the peculiar velocity field}}.
\newblock \emph{\bibinfo{journal}{\aap}} \textbf{\bibinfo{volume}{465}}, \bibinfo{pages}{71--85} (\bibinfo{year}{2007}).
\newblock \eprint{astro-ph/0611626}.

\bibitem{Trinchieri2012}
\bibinfo{author}{{Trinchieri}, G.}, \bibinfo{author}{{Marino}, A.}, \bibinfo{author}{{Mazzei}, P.}, \bibinfo{author}{{Rampazzo}, R.} \& \bibinfo{author}{{Wolter}, A.}
\newblock \bibinfo{title}{{Hot gas in groups: NGC 5328 and the intriguing case of NGC 4756 with XMM-Newton}}.
\newblock \emph{\bibinfo{journal}{\aap}} \textbf{\bibinfo{volume}{545}}, \bibinfo{pages}{A140} (\bibinfo{year}{2012}).
\newblock \eprint{1208.1408}.

\bibitem{dunn2010}
\bibinfo{author}{{Dunn}, R.~J.~H.} \emph{et~al.}
\newblock \bibinfo{title}{{The radio properties of a complete, X-ray selected sample of nearby, massive elliptical galaxies}}.
\newblock \emph{\bibinfo{journal}{\mnras}} \textbf{\bibinfo{volume}{404}}, \bibinfo{pages}{180--197} (\bibinfo{year}{2010}).
\newblock \eprint{1001.1630}.

\bibitem{tully2013}
\bibinfo{author}{{Tully}, R.~B.} \emph{et~al.}
\newblock \bibinfo{title}{{Cosmicflows-2: The Data}}.
\newblock \emph{\bibinfo{journal}{\aj}} \textbf{\bibinfo{volume}{146}}, \bibinfo{pages}{86} (\bibinfo{year}{2013}).
\newblock \eprint{1307.7213}.

\bibitem{Lakhchaura2018}
\bibinfo{author}{{Lakhchaura}, K.} \emph{et~al.}
\newblock \bibinfo{title}{{Thermodynamic properties, multiphase gas, and AGN feedback in a large sample of giant ellipticals}}.
\newblock \emph{\bibinfo{journal}{\mnras}} \textbf{\bibinfo{volume}{481}}, \bibinfo{pages}{4472--4504} (\bibinfo{year}{2018}).
\newblock \eprint{1806.00455}.

\bibitem{slee1994}
\bibinfo{author}{{Slee}, O.~B.}, \bibinfo{author}{{Sadler}, E.~M.}, \bibinfo{author}{{Reynolds}, J.~E.} \& \bibinfo{author}{{Ekers}, R.~D.}
\newblock \bibinfo{title}{{Parsec-scale radio cores in early-type galaxies.}}
\newblock \emph{\bibinfo{journal}{\mnras}} \textbf{\bibinfo{volume}{269}}, \bibinfo{pages}{928--946} (\bibinfo{year}{1994}).

\bibitem{waters2024}
\bibinfo{author}{{Waters}, T.~K.}, \bibinfo{author}{{G{\"u}ltekin}, K.}, \bibinfo{author}{{Gebhardt}, K.}, \bibinfo{author}{{Nagar}, N.} \& \bibinfo{author}{{{\'A}vila}, V.}
\newblock \bibinfo{title}{{A Stellar Dynamical Mass Measurement of the Supermassive Black Hole in NGC 3258}}.
\newblock \emph{\bibinfo{journal}{\apj}} \textbf{\bibinfo{volume}{971}}, \bibinfo{pages}{149} (\bibinfo{year}{2024}).
\newblock \eprint{2406.14623}.

\bibitem{osullivan2001}
\bibinfo{author}{{O'Sullivan}, E.}, \bibinfo{author}{{Forbes}, D.~A.} \& \bibinfo{author}{{Ponman}, T.~J.}
\newblock \bibinfo{title}{{A catalogue and analysis of X-ray luminosities of early-type galaxies}}.
\newblock \emph{\bibinfo{journal}{\mnras}} \textbf{\bibinfo{volume}{328}}, \bibinfo{pages}{461--484} (\bibinfo{year}{2001}).
\newblock \eprint{astro-ph/0108181}.

\bibitem{romana2022}
\bibinfo{author}{{Grossov{\'a}}, R.} \emph{et~al.}
\newblock \bibinfo{title}{{Very Large Array Radio Study of a Sample of Nearby X-Ray and Optically Bright Early-type Galaxies}}.
\newblock \emph{\bibinfo{journal}{\apjs}} \textbf{\bibinfo{volume}{258}}, \bibinfo{pages}{30} (\bibinfo{year}{2022}).
\newblock \eprint{2111.02430}.

\bibitem{sabbi2018}
\bibinfo{author}{{Sabbi}, E.} \emph{et~al.}
\newblock \bibinfo{title}{{The Resolved Stellar Populations in the LEGUS Galaxies1}}.
\newblock \emph{\bibinfo{journal}{\apjs}} \textbf{\bibinfo{volume}{235}}, \bibinfo{pages}{23} (\bibinfo{year}{2018}).
\newblock \eprint{1801.05467}.

\bibitem{jardel2011}
\bibinfo{author}{{Jardel}, J.~R.} \emph{et~al.}
\newblock \bibinfo{title}{{Orbit-based Dynamical Models of the Sombrero Galaxy (NGC 4594)}}.
\newblock \emph{\bibinfo{journal}{\apj}} \textbf{\bibinfo{volume}{739}}, \bibinfo{pages}{21} (\bibinfo{year}{2011}).
\newblock \eprint{1107.1238}.

\bibitem{petrov2006}
\bibinfo{author}{{Petrov}, L.}, \bibinfo{author}{{Kovalev}, Y.~Y.}, \bibinfo{author}{{Fomalont}, E.~B.} \& \bibinfo{author}{{Gordon}, D.}
\newblock \bibinfo{title}{{The Fourth VLBA Calibrator Survey: VCS4}}.
\newblock \emph{\bibinfo{journal}{\aj}} \textbf{\bibinfo{volume}{131}}, \bibinfo{pages}{1872--1879} (\bibinfo{year}{2006}).
\newblock \eprint{astro-ph/0508506}.

\bibitem{walsh2012}
\bibinfo{author}{{Walsh}, J.~L.}, \bibinfo{author}{{van den Bosch}, R. C.~E.}, \bibinfo{author}{{Barth}, A.~J.} \& \bibinfo{author}{{Sarzi}, M.}
\newblock \bibinfo{title}{{A Stellar Dynamical Mass Measurement of the Black Hole in NGC 3998 from Keck Adaptive Optics Observations}}.
\newblock \emph{\bibinfo{journal}{\apj}} \textbf{\bibinfo{volume}{753}}, \bibinfo{pages}{79} (\bibinfo{year}{2012}).
\newblock \eprint{1205.0816}.

\bibitem{helmboldt2007}
\bibinfo{author}{{Helmboldt}, J.~F.} \emph{et~al.}
\newblock \bibinfo{title}{{The VLBA Imaging and Polarimetry Survey at 5 GHz}}.
\newblock \emph{\bibinfo{journal}{\apj}} \textbf{\bibinfo{volume}{658}}, \bibinfo{pages}{203--216} (\bibinfo{year}{2007}).
\newblock \eprint{astro-ph/0611459}.

\bibitem{willick1997}
\bibinfo{author}{{Willick}, J.~A.} \emph{et~al.}
\newblock \bibinfo{title}{{Homogeneous Velocity-Distance Data for Peculiar Velocity Analysis. III. The Mark III Catalog of Galaxy Peculiar Velocities}}.
\newblock \emph{\bibinfo{journal}{\apjs}} \textbf{\bibinfo{volume}{109}}, \bibinfo{pages}{333--366} (\bibinfo{year}{1997}).
\newblock \eprint{astro-ph/9610202}.

\bibitem{liepold2020}
\bibinfo{author}{{Liepold}, E.~R.} \emph{et~al.}
\newblock \bibinfo{title}{{The MASSIVE Survey. XV. A Stellar Dynamical Mass Measurement of the Supermassive Black Hole in Massive Elliptical Galaxy NGC 1453}}.
\newblock \emph{\bibinfo{journal}{\apj}} \textbf{\bibinfo{volume}{891}}, \bibinfo{pages}{4} (\bibinfo{year}{2020}).
\newblock \eprint{2001.08753}.

\bibitem{freedman2001}
\bibinfo{author}{{Freedman}, W.~L.} \emph{et~al.}
\newblock \bibinfo{title}{{Final Results from the Hubble Space Telescope Key Project to Measure the Hubble Constant}}.
\newblock \emph{\bibinfo{journal}{\apj}} \textbf{\bibinfo{volume}{553}}, \bibinfo{pages}{47--72} (\bibinfo{year}{2001}).
\newblock \eprint{astro-ph/0012376}.

\bibitem{zhang2025}
\bibinfo{author}{{Zhang}, H.} \emph{et~al.}
\newblock \bibinfo{title}{{WISDOM Project - XXII. A 5 per cent precision CO-dynamical supermassive black hole mass measurement in the galaxy NGC 383}}.
\newblock \emph{\bibinfo{journal}{\mnras}} \textbf{\bibinfo{volume}{537}}, \bibinfo{pages}{520--536} (\bibinfo{year}{2025}).
\newblock \eprint{2501.06303}.

\bibitem{wang2016}
\bibinfo{author}{{Wang}, S.} \emph{et~al.}
\newblock \bibinfo{title}{{CHANDRA ACIS Survey of X-Ray Point Sources: The Source Catalog}}.
\newblock \emph{\bibinfo{journal}{\apjs}} \textbf{\bibinfo{volume}{224}}, \bibinfo{pages}{40} (\bibinfo{year}{2016}).
\newblock \eprint{1603.08353}.

\bibitem{healey2007}
\bibinfo{author}{{Healey}, S.~E.} \emph{et~al.}
\newblock \bibinfo{title}{{CRATES: An All-Sky Survey of Flat-Spectrum Radio Sources}}.
\newblock \emph{\bibinfo{journal}{\apjs}} \textbf{\bibinfo{volume}{171}}, \bibinfo{pages}{61--71} (\bibinfo{year}{2007}).
\newblock \eprint{astro-ph/0702346}.

\bibitem{ferrarese1996}
\bibinfo{author}{{Ferrarese}, L.}, \bibinfo{author}{{Ford}, H.~C.} \& \bibinfo{author}{{Jaffe}, W.}
\newblock \bibinfo{title}{{Evidence for a Massive Black Hole in the Active Galaxy NGC 4261 from Hubble Space Telescope Images and Spectra}}.
\newblock \emph{\bibinfo{journal}{\apj}} \textbf{\bibinfo{volume}{470}}, \bibinfo{pages}{444} (\bibinfo{year}{1996}).

\bibitem{Yan2025}
\bibinfo{author}{{Yan}, X.}, \bibinfo{author}{{Cui}, L.} \& \bibinfo{author}{{Ho}, L.~C.}
\newblock \bibinfo{title}{{Multifrequency Very Long Baseline Interferometry Study of Emission and Absorption in the Two-sided Jets of NGC 3998}}.
\newblock \emph{\bibinfo{journal}{\apj}} \textbf{\bibinfo{volume}{983}}, \bibinfo{pages}{169} (\bibinfo{year}{2025}).
\newblock \eprint{2503.09084}.

\bibitem{walsh2010}
\bibinfo{author}{{Walsh}, J.~L.}, \bibinfo{author}{{Barth}, A.~J.} \& \bibinfo{author}{{Sarzi}, M.}
\newblock \bibinfo{title}{{The Supermassive Black Hole in M84 Revisited}}.
\newblock \emph{\bibinfo{journal}{\apj}} \textbf{\bibinfo{volume}{721}}, \bibinfo{pages}{762--776} (\bibinfo{year}{2010}).
\newblock \eprint{1008.0005}.

\bibitem{tonry1997}
\bibinfo{author}{{Tonry}, J.~L.}, \bibinfo{author}{{Blakeslee}, J.~P.}, \bibinfo{author}{{Ajhar}, E.~A.} \& \bibinfo{author}{{Dressler}, A.}
\newblock \bibinfo{title}{{The SBF Survey of Galaxy Distances. I. Sample Selection, Photometric Calibration, and the Hubble Constant}}.
\newblock \emph{\bibinfo{journal}{\apj}} \textbf{\bibinfo{volume}{475}}, \bibinfo{pages}{399--413} (\bibinfo{year}{1997}).
\newblock \eprint{astro-ph/9609113}.

\bibitem{Kormendy2013}
\bibinfo{author}{{Kormendy}, J.} \& \bibinfo{author}{{Ho}, L.~C.}
\newblock \bibinfo{title}{{Coevolution (Or Not) of Supermassive Black Holes and Host Galaxies}}.
\newblock \emph{\bibinfo{journal}{\araa}} \textbf{\bibinfo{volume}{51}}, \bibinfo{pages}{511--653} (\bibinfo{year}{2013}).
\newblock \eprint{1304.7762}.

\bibitem{killeen1988}
\bibinfo{author}{{Killeen}, N.~E.~B.}, \bibinfo{author}{{Bicknell}, G.~V.} \& \bibinfo{author}{{Ekers}, R.~D.}
\newblock \bibinfo{title}{{The Thermally Confined Radio Source in NGC 1399}}.
\newblock \emph{\bibinfo{journal}{\apj}} \textbf{\bibinfo{volume}{325}}, \bibinfo{pages}{180} (\bibinfo{year}{1988}).

\bibitem{salaris1998}
\bibinfo{author}{{Salaris}, M.} \& \bibinfo{author}{{Cassisi}, S.}
\newblock \bibinfo{title}{{A new analysis of the red giant branch `tip' distance scale and the value of the Hubble constant}}.
\newblock \emph{\bibinfo{journal}{\mnras}} \textbf{\bibinfo{volume}{298}}, \bibinfo{pages}{166--178} (\bibinfo{year}{1998}).
\newblock \eprint{astro-ph/9803103}.

\bibitem{vandenbosch2010}
\bibinfo{author}{{van den Bosch}, R. C.~E.} \& \bibinfo{author}{{de Zeeuw}, P.~T.}
\newblock \bibinfo{title}{{Estimating black hole masses in triaxial galaxies}}.
\newblock \emph{\bibinfo{journal}{\mnras}} \textbf{\bibinfo{volume}{401}}, \bibinfo{pages}{1770--1780} (\bibinfo{year}{2010}).
\newblock \eprint{0910.0844}.

\bibitem{nyland2016}
\bibinfo{author}{{Nyland}, K.} \emph{et~al.}
\newblock \bibinfo{title}{{The ATLAS$^{3D}$ Project - XXXI. Nuclear radio emission in nearby early-type galaxies}}.
\newblock \emph{\bibinfo{journal}{\mnras}} \textbf{\bibinfo{volume}{458}}, \bibinfo{pages}{2221--2268} (\bibinfo{year}{2016}).
\newblock \eprint{1602.05579}.

\bibitem{mei2007}
\bibinfo{author}{{Mei}, S.} \emph{et~al.}
\newblock \bibinfo{title}{{The ACS Virgo Cluster Survey. XIII. SBF Distance Catalog and the Three-dimensional Structure of the Virgo Cluster}}.
\newblock \emph{\bibinfo{journal}{\apj}} \textbf{\bibinfo{volume}{655}}, \bibinfo{pages}{144--162} (\bibinfo{year}{2007}).
\newblock \eprint{astro-ph/0702510}.

\bibitem{kormendy1997}
\bibinfo{author}{{Kormendy}, J.} \emph{et~al.}
\newblock \bibinfo{title}{{Spectroscopic Evidence for a Supermassive Black Hole in NCG 4486B}}.
\newblock \emph{\bibinfo{journal}{\apjl}} \textbf{\bibinfo{volume}{482}}, \bibinfo{pages}{L139--L142} (\bibinfo{year}{1997}).
\newblock \eprint{astro-ph/9703188}.

\bibitem{capetti2009}
\bibinfo{author}{{Capetti}, A.}, \bibinfo{author}{{Kharb}, P.}, \bibinfo{author}{{Axon}, D.~J.}, \bibinfo{author}{{Merritt}, D.} \& \bibinfo{author}{{Baldi}, R.~D.}
\newblock \bibinfo{title}{{A Very Large Array Radio Survey of Early-Type Galaxies in the Virgo Cluster}}.
\newblock \emph{\bibinfo{journal}{\aj}} \textbf{\bibinfo{volume}{138}}, \bibinfo{pages}{1990--1997} (\bibinfo{year}{2009}).
\newblock \eprint{0910.4102}.

\bibitem{schulze2011}
\bibinfo{author}{{Schulze}, A.} \& \bibinfo{author}{{Gebhardt}, K.}
\newblock \bibinfo{title}{{Effect of a Dark Matter Halo on the Determination of Black Hole Masses}}.
\newblock \emph{\bibinfo{journal}{\apj}} \textbf{\bibinfo{volume}{729}}, \bibinfo{pages}{21} (\bibinfo{year}{2011}).
\newblock \eprint{1011.5077}.

\bibitem{Williams2022}
\bibinfo{author}{{Williams}, D.~R.~A.} \emph{et~al.}
\newblock \bibinfo{title}{{LeMMINGs - IV. The X-ray properties of a statistically complete sample of the nuclei in active and inactive galaxies from the Palomar sample}}.
\newblock \emph{\bibinfo{journal}{\mnras}} \textbf{\bibinfo{volume}{510}}, \bibinfo{pages}{4909--4928} (\bibinfo{year}{2022}).
\newblock \eprint{2111.09077}.

\bibitem{krajnovic2009}
\bibinfo{author}{{Krajnovi{\'c}}, D.}, \bibinfo{author}{{McDermid}, R.~M.}, \bibinfo{author}{{Cappellari}, M.} \& \bibinfo{author}{{Davies}, R.~L.}
\newblock \bibinfo{title}{{Determination of masses of the central black holes in NGC 524 and 2549 using laser guide star adaptive optics}}.
\newblock \emph{\bibinfo{journal}{\mnras}} \textbf{\bibinfo{volume}{399}}, \bibinfo{pages}{1839--1857} (\bibinfo{year}{2009}).
\newblock \eprint{0907.3748}.

\bibitem{jensen2003}
\bibinfo{author}{{Jensen}, J.~B.} \emph{et~al.}
\newblock \bibinfo{title}{{Measuring Distances and Probing the Unresolved Stellar Populations of Galaxies Using Infrared Surface Brightness Fluctuations}}.
\newblock \emph{\bibinfo{journal}{\apj}} \textbf{\bibinfo{volume}{583}}, \bibinfo{pages}{712--726} (\bibinfo{year}{2003}).
\newblock \eprint{astro-ph/0210129}.

\bibitem{kundu2001}
\bibinfo{author}{{Kundu}, A.} \& \bibinfo{author}{{Whitmore}, B.~C.}
\newblock \bibinfo{title}{{New Insights from Hubble Space Telescope Studies of Globular Cluster Systems. II. Analysis of 29 S0 Systems}}.
\newblock \emph{\bibinfo{journal}{\aj}} \textbf{\bibinfo{volume}{122}}, \bibinfo{pages}{1251--1270} (\bibinfo{year}{2001}).
\newblock \eprint{astro-ph/0105198}.

\bibitem{barth2016}
\bibinfo{author}{{Barth}, A.~J.} \emph{et~al.}
\newblock \bibinfo{title}{{Measurement of the Black Hole Mass in NGC 1332 from ALMA Observations at 0.044 arcsecond Resolution}}.
\newblock \emph{\bibinfo{journal}{\apjl}} \textbf{\bibinfo{volume}{822}}, \bibinfo{pages}{L28} (\bibinfo{year}{2016}).
\newblock \eprint{1605.01346}.

\bibitem{ueda2001}
\bibinfo{author}{{Ueda}, Y.}, \bibinfo{author}{{Ishisaki}, Y.}, \bibinfo{author}{{Takahashi}, T.}, \bibinfo{author}{{Makishima}, K.} \& \bibinfo{author}{{Ohashi}, T.}
\newblock \bibinfo{title}{{The ASCA Medium Sensitivity Survey (the GIS Catalog Project): Source Catalog}}.
\newblock \emph{\bibinfo{journal}{\apjs}} \textbf{\bibinfo{volume}{133}}, \bibinfo{pages}{1--52} (\bibinfo{year}{2001}).
\newblock \eprint{astro-ph/9908128}.

\bibitem{boizelle2021}
\bibinfo{author}{{Boizelle}, B.~D.} \emph{et~al.}
\newblock \bibinfo{title}{{Black Hole Mass Measurements of Radio Galaxies NGC 315 and NGC 4261 Using ALMA CO Observations}}.
\newblock \emph{\bibinfo{journal}{\apj}} \textbf{\bibinfo{volume}{908}}, \bibinfo{pages}{19} (\bibinfo{year}{2021}).
\newblock \eprint{2012.04669}.

\bibitem{davis2013}
\bibinfo{author}{{Davis}, T.~A.}, \bibinfo{author}{{Bureau}, M.}, \bibinfo{author}{{Cappellari}, M.}, \bibinfo{author}{{Sarzi}, M.} \& \bibinfo{author}{{Blitz}, L.}
\newblock \bibinfo{title}{{A black-hole mass measurement from molecular gas kinematics in NGC4526}}.
\newblock \emph{\bibinfo{journal}{\nat}} \textbf{\bibinfo{volume}{494}}, \bibinfo{pages}{328--330} (\bibinfo{year}{2013}).
\newblock \eprint{1301.7184}.

\bibitem{ueda2005}
\bibinfo{author}{{Ueda}, Y.}, \bibinfo{author}{{Ishisaki}, Y.}, \bibinfo{author}{{Takahashi}, T.}, \bibinfo{author}{{Makishima}, K.} \& \bibinfo{author}{{Ohashi}, T.}
\newblock \bibinfo{title}{{The ASCA Medium Sensitivity Survey (The GIS Catalog Project): Source Catalog II.}}
\newblock \emph{\bibinfo{journal}{\apjs}} \textbf{\bibinfo{volume}{161}}, \bibinfo{pages}{185--223} (\bibinfo{year}{2005}).

\bibitem{mcconnell2011}
\bibinfo{author}{{McConnell}, N.~J.} \emph{et~al.}
\newblock \bibinfo{title}{{The Black Hole Mass in the Brightest Cluster Galaxy NGC 6086}}.
\newblock \emph{\bibinfo{journal}{\apj}} \textbf{\bibinfo{volume}{728}}, \bibinfo{pages}{100} (\bibinfo{year}{2011}).
\newblock \eprint{1009.0750}.

\bibitem{candini2023}
\bibinfo{author}{{Candini}, S.} \emph{et~al.}
\newblock \bibinfo{title}{{New filamentary remnant radio emission and duty cycle constraints in the radio galaxy NGC 6086}}.
\newblock \emph{\bibinfo{journal}{\aap}} \textbf{\bibinfo{volume}{677}}, \bibinfo{pages}{A4} (\bibinfo{year}{2023}).
\newblock \eprint{2305.18077}.

\bibitem{murgia2011}
\bibinfo{author}{{Murgia}, M.} \emph{et~al.}
\newblock \bibinfo{title}{{Dying radio galaxies in clusters}}.
\newblock \emph{\bibinfo{journal}{\aap}} \textbf{\bibinfo{volume}{526}}, \bibinfo{pages}{A148} (\bibinfo{year}{2011}).
\newblock \eprint{1011.0567}.

\bibitem{van2015}
\bibinfo{author}{{van den Bosch}, R. C.~E.}, \bibinfo{author}{{Gebhardt}, K.}, \bibinfo{author}{{G{\"u}ltekin}, K.}, \bibinfo{author}{{Y{\i}ld{\i}r{\i}m}, A.} \& \bibinfo{author}{{Walsh}, J.~L.}
\newblock \bibinfo{title}{{Hunting for Supermassive Black Holes in Nearby Galaxies With the Hobby-Eberly Telescope}}.
\newblock \emph{\bibinfo{journal}{\apjs}} \textbf{\bibinfo{volume}{218}}, \bibinfo{pages}{10} (\bibinfo{year}{2015}).
\newblock \eprint{1502.00632}.

\bibitem{cohn2021}
\bibinfo{author}{{Cohn}, J.~H.} \emph{et~al.}
\newblock \bibinfo{title}{{An ALMA Gas-dynamical Mass Measurement of the Supermassive Black Hole in the Local Compact Galaxy UGC 2698}}.
\newblock \emph{\bibinfo{journal}{\apj}} \textbf{\bibinfo{volume}{919}}, \bibinfo{pages}{77} (\bibinfo{year}{2021}).
\newblock \eprint{2104.07779}.

\bibitem{condon2002}
\bibinfo{author}{{Condon}, J.~J.}, \bibinfo{author}{{Cotton}, W.~D.} \& \bibinfo{author}{{Broderick}, J.~J.}
\newblock \bibinfo{title}{{Radio Sources and Star Formation in the Local Universe}}.
\newblock \emph{\bibinfo{journal}{\aj}} \textbf{\bibinfo{volume}{124}}, \bibinfo{pages}{675--689} (\bibinfo{year}{2002}).

\bibitem{springob2014}
\bibinfo{author}{{Springob}, C.~M.} \emph{et~al.}
\newblock \bibinfo{title}{{The 6dF Galaxy Survey: peculiar velocity field and cosmography}}.
\newblock \emph{\bibinfo{journal}{\mnras}} \textbf{\bibinfo{volume}{445}}, \bibinfo{pages}{2677--2697} (\bibinfo{year}{2014}).
\newblock \eprint{1409.6161}.

\bibitem{dominiak2024}
\bibinfo{author}{{Dominiak}, P.} \emph{et~al.}
\newblock \bibinfo{title}{{The MASSIVE survey - XIX. Molecular gas measurements of the supermassive black hole masses in the elliptical galaxies NGC 1684 and NGC 0997}}.
\newblock \emph{\bibinfo{journal}{\mnras}} \textbf{\bibinfo{volume}{529}}, \bibinfo{pages}{1597--1616} (\bibinfo{year}{2024}).
\newblock \eprint{2401.16376}.

\bibitem{griffith1995}
\bibinfo{author}{{Griffith}, M.~R.}, \bibinfo{author}{{Wright}, A.~E.}, \bibinfo{author}{{Burke}, B.~F.} \& \bibinfo{author}{{Ekers}, R.~D.}
\newblock \bibinfo{title}{{The Parkes-MIT-NRAO (PMN) Surveys. VI. Source Catalog for the Equatorial Survey (-9.5 degrees < delta < +10.0 degrees )}}.
\newblock \emph{\bibinfo{journal}{\apjs}} \textbf{\bibinfo{volume}{97}}, \bibinfo{pages}{347} (\bibinfo{year}{1995}).

\bibitem{thater2019}
\bibinfo{author}{{Thater}, S.} \emph{et~al.}
\newblock \bibinfo{title}{{Six new supermassive black hole mass determinations from adaptive-optics assisted SINFONI observations}}.
\newblock \emph{\bibinfo{journal}{\aap}} \textbf{\bibinfo{volume}{625}}, \bibinfo{pages}{A62} (\bibinfo{year}{2019}).
\newblock \eprint{1902.10175}.

\bibitem{filho2002}
\bibinfo{author}{{Filho}, M.~E.}, \bibinfo{author}{{Barthel}, P.~D.} \& \bibinfo{author}{{Ho}, L.~C.}
\newblock \bibinfo{title}{{The Radio Properties of Composite LINER/H II Galaxies}}.
\newblock \emph{\bibinfo{journal}{\apjs}} \textbf{\bibinfo{volume}{142}}, \bibinfo{pages}{223--238} (\bibinfo{year}{2002}).
\newblock \eprint{astro-ph/0205196}.

\bibitem{defrancesco2008}
\bibinfo{author}{{de Francesco}, G.}, \bibinfo{author}{{Capetti}, A.} \& \bibinfo{author}{{Marconi}, A.}
\newblock \bibinfo{title}{{Measuring supermassive black holes with gas kinematics. II. The LINERs <ASTROBJ>IC 989</ASTROBJ>, <ASTROBJ>NGC 5077</ASTROBJ>, and <ASTROBJ>NGC 6500</ASTROBJ>}}.
\newblock \emph{\bibinfo{journal}{\aap}} \textbf{\bibinfo{volume}{479}}, \bibinfo{pages}{355--363} (\bibinfo{year}{2008}).
\newblock \eprint{0801.0064}.

\bibitem{Balmaverde2015}
\bibinfo{author}{{Balmaverde}, B.} \& \bibinfo{author}{{Capetti}, A.}
\newblock \bibinfo{title}{{The naked nuclei of low ionization nuclear emission line regions}}.
\newblock \emph{\bibinfo{journal}{\aap}} \textbf{\bibinfo{volume}{581}}, \bibinfo{pages}{A76} (\bibinfo{year}{2015}).

\bibitem{cretton1999}
\bibinfo{author}{{Cretton}, N.} \& \bibinfo{author}{{van den Bosch}, F.~C.}
\newblock \bibinfo{title}{{Evidence for a Massive Black Hole in the S0 Galaxy NGC 4342}}.
\newblock \emph{\bibinfo{journal}{\apj}} \textbf{\bibinfo{volume}{514}}, \bibinfo{pages}{704--724} (\bibinfo{year}{1999}).
\newblock \eprint{astro-ph/9805324}.

\bibitem{gerke2011}
\bibinfo{author}{{Gerke}, J.~R.}, \bibinfo{author}{{Kochanek}, C.~S.}, \bibinfo{author}{{Prieto}, J.~L.}, \bibinfo{author}{{Stanek}, K.~Z.} \& \bibinfo{author}{{Macri}, L.~M.}
\newblock \bibinfo{title}{{A Study of Cepheids in M81 with the Large Binocular Telescope (Efficiently Calibrated with Hubble Space Telescope)}}.
\newblock \emph{\bibinfo{journal}{\apj}} \textbf{\bibinfo{volume}{743}}, \bibinfo{pages}{176} (\bibinfo{year}{2011}).
\newblock \eprint{1103.0549}.

\bibitem{devereux2003}
\bibinfo{author}{{Devereux}, N.}, \bibinfo{author}{{Ford}, H.}, \bibinfo{author}{{Tsvetanov}, Z.} \& \bibinfo{author}{{Jacoby}, G.}
\newblock \bibinfo{title}{{STIS Spectroscopy of the Central 10 Parsecs of M81: Evidence for a Massive Black Hole}}.
\newblock \emph{\bibinfo{journal}{\aj}} \textbf{\bibinfo{volume}{125}}, \bibinfo{pages}{1226--1235} (\bibinfo{year}{2003}).

\bibitem{Jang2014}
\bibinfo{author}{{Jang}, I.}, \bibinfo{author}{{Gliozzi}, M.}, \bibinfo{author}{{Hughes}, C.} \& \bibinfo{author}{{Titarchuk}, L.}
\newblock \bibinfo{title}{{Constraining black hole masses in low-accreting active galactic nuclei using X-ray spectra}}.
\newblock \emph{\bibinfo{journal}{\mnras}} \textbf{\bibinfo{volume}{443}}, \bibinfo{pages}{72--85} (\bibinfo{year}{2014}).
\newblock \eprint{1405.5729}.

\bibitem{cappellari2011}
\bibinfo{author}{{Cappellari}, M.} \emph{et~al.}
\newblock \bibinfo{title}{{The ATLAS$^{3D}$ project - I. A volume-limited sample of 260 nearby early-type galaxies: science goals and selection criteria}}.
\newblock \emph{\bibinfo{journal}{\mnras}} \textbf{\bibinfo{volume}{413}}, \bibinfo{pages}{813--836} (\bibinfo{year}{2011}).
\newblock \eprint{1012.1551}.

\bibitem{onishi2017}
\bibinfo{author}{{Onishi}, K.} \emph{et~al.}
\newblock \bibinfo{title}{{WISDOM project - I. Black hole mass measurement using molecular gas kinematics in NGC 3665}}.
\newblock \emph{\bibinfo{journal}{\mnras}} \textbf{\bibinfo{volume}{468}}, \bibinfo{pages}{4663--4674} (\bibinfo{year}{2017}).
\newblock \eprint{1703.05247}.

\bibitem{gultekin2009}
\bibinfo{author}{{G{\"u}ltekin}, K.} \emph{et~al.}
\newblock \bibinfo{title}{{A Quintet of Black Hole Mass Determinations}}.
\newblock \emph{\bibinfo{journal}{\apj}} \textbf{\bibinfo{volume}{695}}, \bibinfo{pages}{1577--1590} (\bibinfo{year}{2009}).
\newblock \eprint{0901.4162}.

\bibitem{ferrarese2007}
\bibinfo{author}{{Ferrarese}, L.} \emph{et~al.}
\newblock \bibinfo{title}{{The Discovery of Cepheids and a Distance to NGC 5128}}.
\newblock \emph{\bibinfo{journal}{\apj}} \textbf{\bibinfo{volume}{654}}, \bibinfo{pages}{186--218} (\bibinfo{year}{2007}).
\newblock \eprint{astro-ph/0605707}.

\bibitem{cappellari2009}
\bibinfo{author}{{Cappellari}, M.} \emph{et~al.}
\newblock \bibinfo{title}{{The mass of the black hole in Centaurus A from SINFONI AO-assisted integral-field observations of stellar kinematics}}.
\newblock \emph{\bibinfo{journal}{\mnras}} \textbf{\bibinfo{volume}{394}}, \bibinfo{pages}{660--674} (\bibinfo{year}{2009}).
\newblock \eprint{0812.1000}.

\bibitem{horiuchi2006}
\bibinfo{author}{{Horiuchi}, S.}, \bibinfo{author}{{Meier}, D.~L.}, \bibinfo{author}{{Preston}, R.~A.} \& \bibinfo{author}{{Tingay}, S.~J.}
\newblock \bibinfo{title}{{Ten Milliparsec-Scale Structure of the Nucleus Region in Centaurus A}}.
\newblock \emph{\bibinfo{journal}{\pasj}} \textbf{\bibinfo{volume}{58}}, \bibinfo{pages}{211--216} (\bibinfo{year}{2006}).
\newblock \eprint{astro-ph/0508445}.

\bibitem{liuzzo2010}
\bibinfo{author}{{Liuzzo}, E.}, \bibinfo{author}{{Giovannini}, G.}, \bibinfo{author}{{Giroletti}, M.} \& \bibinfo{author}{{Taylor}, G.~B.}
\newblock \bibinfo{title}{{Parsec-scale properties of brightest cluster galaxies}}.
\newblock \emph{\bibinfo{journal}{\aap}} \textbf{\bibinfo{volume}{516}}, \bibinfo{pages}{A1} (\bibinfo{year}{2010}).
\newblock \eprint{1002.1380}.

\bibitem{davis2017}
\bibinfo{author}{{Davis}, T.~A.} \emph{et~al.}
\newblock \bibinfo{title}{{WISDOM Project - II. Molecular gas measurement of the supermassive black hole mass in NGC 4697}}.
\newblock \emph{\bibinfo{journal}{\mnras}} \textbf{\bibinfo{volume}{468}}, \bibinfo{pages}{4675--4690} (\bibinfo{year}{2017}).
\newblock \eprint{1703.05248}.

\bibitem{wrobel2008}
\bibinfo{author}{{Wrobel}, J.~M.}, \bibinfo{author}{{Terashima}, Y.} \& \bibinfo{author}{{Ho}, L.~C.}
\newblock \bibinfo{title}{{Outflow-dominated Emission from the Quiescent Massive Black Holes in NGC 4621 and NGC 4697}}.
\newblock \emph{\bibinfo{journal}{\apj}} \textbf{\bibinfo{volume}{675}}, \bibinfo{pages}{1041--1047} (\bibinfo{year}{2008}).
\newblock \eprint{0712.1308}.

\bibitem{barth2001}
\bibinfo{author}{{Barth}, A.~J.} \emph{et~al.}
\newblock \bibinfo{title}{{Evidence for a Supermassive Black Hole in the S0 Galaxy NGC 3245}}.
\newblock \emph{\bibinfo{journal}{\apj}} \textbf{\bibinfo{volume}{555}}, \bibinfo{pages}{685--708} (\bibinfo{year}{2001}).
\newblock \eprint{astro-ph/0012213}.

\bibitem{filho2004}
\bibinfo{author}{{Filho}, M.~E.} \emph{et~al.}
\newblock \bibinfo{title}{{Further clues to the nature of composite LINER/H II galaxies}}.
\newblock \emph{\bibinfo{journal}{\aap}} \textbf{\bibinfo{volume}{418}}, \bibinfo{pages}{429--443} (\bibinfo{year}{2004}).
\newblock \eprint{astro-ph/0401593}.

\bibitem{Nguyen2020}
\bibinfo{author}{{Nguyen}, D.~D.} \emph{et~al.}
\newblock \bibinfo{title}{{The MBHBM$_{{\ensuremath{\star}}}$ Project. I. Measurement of the Central Black Hole Mass in Spiral Galaxy NGC 3504 Using Molecular Gas Kinematics}}.
\newblock \emph{\bibinfo{journal}{\apj}} \textbf{\bibinfo{volume}{892}}, \bibinfo{pages}{68} (\bibinfo{year}{2020}).
\newblock \eprint{1902.03813}.

\bibitem{Haga2015}
\bibinfo{author}{{Haga}, T.} \emph{et~al.}
\newblock \bibinfo{title}{{Determination of Central Engine Position and Accretion Disk Structure in NGC 4261 by Core Shift Measurements}}.
\newblock \emph{\bibinfo{journal}{\apj}} \textbf{\bibinfo{volume}{807}}, \bibinfo{pages}{15} (\bibinfo{year}{2015}).
\newblock \eprint{1510.00734}.

\bibitem{Fromm2015}
\bibinfo{author}{{Fromm}, C.~M.}, \bibinfo{author}{{Perucho}, M.}, \bibinfo{author}{{Ros}, E.}, \bibinfo{author}{{Savolainen}, T.} \& \bibinfo{author}{{Zensus}, J.~A.}
\newblock \bibinfo{title}{{On the location of the supermassive black hole in CTA 102}}.
\newblock \emph{\bibinfo{journal}{\aap}} \textbf{\bibinfo{volume}{576}}, \bibinfo{pages}{A43} (\bibinfo{year}{2015}).
\newblock \eprint{1412.1317}.

\bibitem{Benke2024}
\bibinfo{author}{{Benke}, P.} \emph{et~al.}
\newblock \bibinfo{title}{{Very-long-baseline interferometry study of the flaring blazar TXS 1508+572 in the early Universe}}.
\newblock \emph{\bibinfo{journal}{\aap}} \textbf{\bibinfo{volume}{689}}, \bibinfo{pages}{A43} (\bibinfo{year}{2024}).
\newblock \eprint{2406.03135}.

\bibitem{Croke2010}
\bibinfo{author}{{Croke}, S.~M.}, \bibinfo{author}{{O'Sullivan}, S.~P.} \& \bibinfo{author}{{Gabuzda}, D.~C.}
\newblock \bibinfo{title}{{The parsec-scale distributions of intensity, linear polarization and Faraday rotation in the core and jet of Mrk501 at 8.4-1.6 GHz}}.
\newblock \emph{\bibinfo{journal}{\mnras}} \textbf{\bibinfo{volume}{402}}, \bibinfo{pages}{259--270} (\bibinfo{year}{2010}).

\bibitem{Kovalev2008}
\bibinfo{author}{{Kovalev}, Y.~Y.}, \bibinfo{author}{{Lobanov}, A.~P.} \& \bibinfo{author}{{Pushkarev}, A.~B.}
\newblock \bibinfo{title}{{Physics of the central region in the quasar 0850+581}}.
\newblock \emph{\bibinfo{journal}{\memsai}} \textbf{\bibinfo{volume}{79}}, \bibinfo{pages}{1153} (\bibinfo{year}{2008}).
\newblock \eprint{0810.2240}.

\bibitem{Bartolini2025}
\bibinfo{author}{{Bartolini}, V.} \emph{et~al.}
\newblock \bibinfo{title}{{Multifrequency simultaneous VLBA view of the radio source 3C 111}}.
\newblock \emph{\bibinfo{journal}{\aap}} \textbf{\bibinfo{volume}{698}}, \bibinfo{pages}{A123} (\bibinfo{year}{2025}).
\newblock \eprint{2503.18621}.

\bibitem{Lisakov2017}
\bibinfo{author}{{Lisakov}, M.~M.}, \bibinfo{author}{{Kovalev}, Y.~Y.}, \bibinfo{author}{{Savolainen}, T.}, \bibinfo{author}{{Hovatta}, T.} \& \bibinfo{author}{{Kutkin}, A.~M.}
\newblock \bibinfo{title}{{A connection between {\ensuremath{\gamma}}-ray and parsec-scale radio flares in the blazar 3C 273}}.
\newblock \emph{\bibinfo{journal}{\mnras}} \textbf{\bibinfo{volume}{468}}, \bibinfo{pages}{4478--4493} (\bibinfo{year}{2017}).
\newblock \eprint{1703.07976}.

\bibitem{Pushkarev2019}
\bibinfo{author}{{Pushkarev}, A.~B.}, \bibinfo{author}{{Butuzova}, M.~S.}, \bibinfo{author}{{Kovalev}, Y.~Y.} \& \bibinfo{author}{{Hovatta}, T.}
\newblock \bibinfo{title}{{Multifrequency study of the gamma-ray flaring BL Lacertae object PKS 2233-148 in 2009-2012}}.
\newblock \emph{\bibinfo{journal}{\mnras}} \textbf{\bibinfo{volume}{482}}, \bibinfo{pages}{2336--2353} (\bibinfo{year}{2019}).
\newblock \eprint{1808.06138}.

\bibitem{Lisakov2025}
\bibinfo{author}{{Lisakov}, M.} \emph{et~al.}
\newblock \bibinfo{title}{{Kilogauss magnetic field and jet dynamics in the quasar NRAO 530}}.
\newblock \emph{\bibinfo{journal}{\aap}} \textbf{\bibinfo{volume}{693}}, \bibinfo{pages}{A9} (\bibinfo{year}{2025}).
\newblock \eprint{2411.03446}.

\bibitem{Roder2024}
\bibinfo{author}{{R{\"o}der}, J.}, \bibinfo{author}{{Ros}, E.}, \bibinfo{author}{{Schinzel}, F.~K.} \& \bibinfo{author}{{Lobanov}, A.~P.}
\newblock \bibinfo{title}{{Up around the bend: A multiwavelength view of the quasar 3C 345}}.
\newblock \emph{\bibinfo{journal}{\aap}} \textbf{\bibinfo{volume}{684}}, \bibinfo{pages}{A211} (\bibinfo{year}{2024}).
\newblock \eprint{2402.17006}.

\end{thebibliography}


\begin{addendum}
 \item X.C. and B.-W.S. were supported by the Brain Pool Program through the National Research Foundation of Korea (NRF) funded by the Ministry of Science and ICT (RS-2024-00407499). This research was supported by funding from the Korean government (Korea AeroSpace Administration, KASA; grant RS-2026-25587698). H.Y. acknowledges support from the Polish National Science Center grant 2023/48/Q/ST9/00138. he Very Long Baseline Array (VLBA) is a facility of the National Science Foundation operated under cooperative agreement by Associated Universities, Inc. The European VLBI Network (EVN) is a joint facility of independent European, African, Asian, and North American radio astronomy institutes. Scientific results from data presented in this publication are derived from the EVN project EL075.
 \item[Author contributions] 
 X.C. coordinated the project, led the VLBA observing proposal, carried out the data reduction and image analysis, and wrote the first draft of the manuscript. H.Y. carried out the numerical simulations, analysed the simulation results, and wrote the simulation-related sections of the manuscript. J.Y. co-led the VLBA observing proposal, contributed to the observational design and scientific interpretation, and substantially revised the manuscript. F.Y. supervised the simulation work, contributed to the physical interpretation of the results, and revised the manuscript. X.L. provided the EVN observations and contributed to the VLBI data analysis. R.L., H.R., B.-W.S., L.F., Y.Z., W.C., N.L., J.E.C., and T.J. contributed to the scientific discussion, interpretation of the results, and manuscript review. All authors discussed the results and commented on the manuscript.
\end{addendum}


\clearpage



\begin{table*}[htbp]
\centering
\scriptsize
\captionsetup{labelformat=empty}
\caption*{Extended Data Table 1: Nearby galaxies sample}
\begin{tabular}{lcccccc}
\hline
Name & Distance (Mpc) & Log$_{10}$M (M$_{\rm \odot}$) &  Ring diameter ($\mu$as) & $\rm Log (L_{\rm bol}/L_{\rm Edd})$ & VLBI (mJy) & VLA (mJy) \\
(1) & (2) & (3) & (4) & (5) & (6) & (7) \\
\hline
\sgra    & 0.0082\cite{GRAVITY2019} & 6.6\cite{Ghez2008} & 53.1 & $-$9.0\cite{eht2022vi} & 600\cite{Lo1998}  &  \\
M87      & 16.8\cite{eht2019i}   & 9.8\cite{eht2019i} & 37.8 & $-$5.9\cite{gonzalez2009} & 1116.0\cite{fomalont2000} &  \\
\Msixty  & 16.3\cite{lee2017}   & 9.7\cite{shen2010} & 28.3 & $-$8.0\cite{Paggi2014} & 17.7 &  \\
NGC 1600 & 68.6\cite{veale2017}   & 10.2\cite{thomas2016} & 27.2 & $-$5.3\cite{Beuing1999} &  & 33.1\cite{brown2011} \\
NGC 1277 & 73.3\cite{beasley2018}   & 10.2\cite{van2012} & 23.9 & $-$7.0\cite{Scharw2016} & 0.84\cite{park2017} &  \\
NGC 4889 & 103.2\cite{mcconnell2012}  & 10.3\cite{mcconnell2012} & 20.9 & $-$7.9\cite{Sanders2014} &  & 1.2\cite{brown2011}  \\
M31      & 0.8\cite{Breuval2023}    & 8.2\cite{bender2005} & 19.0 & $-$8.3\cite{Garcia2010} & 0.018\cite{peng2023} & \\
NGC 1407 & 28.8\cite{tonry2001}   & 9.7\cite{rusli2013} & 16.4 & $-$6.7\cite{Pellegrini2005} &  & 34.0\cite{giacintucci2012} \\
NGC 4472 & 16.7\cite{blakeslee2009}   & 9.4\cite{rusli2013} & 15.6 & $-$7.0\cite{Panessa2006} & 2.3\cite{nagar2005}  &  \\
NGC 4751 & 26.9\cite{rusli2013}   & 9.5\cite{rusli2013} & 13.1 & $-$5.0\cite{Gultekin2019} &  & 3.0\cite{sadler1989}  \\
NGC 3842 & 96.4\cite{sorce2013}   & 10.0\cite{mcconnell2012} & 10.1 & $-$6.4\cite{Caglar2017} &  & 10.5\cite{feretti1994} \\
NGC 3115 & 10.9\cite{ferrarese2000}   & 9.0\cite{emsellem1999} & 9.6  & $-$6.9\cite{Pellegrini2010} &  & 0.3\cite{wrobel2012}  \\
IC 1459  & 29.2\cite{tonry2001}   & 9.4\cite{cappellari2002} & 8.8  & $-$5.7\cite{gonzalez2009} & 900.1\cite{tingay2015} &\\
NGC 5328 & 60.6\cite{theureau2007}   & 9.7\cite{rusli2013} & 7.6  & $-$6.4\cite{Trinchieri2012} &  & 0.9\cite{brown2011}  \\
NGC 1550 & 53.6\cite{theureau2007}   & 9.6\cite{rusli2013} & 7.6  & $-$3.5\cite{Beuing1999} &  & 12.7\cite{dunn2010}  \\
NGC 6861 & 28.6\cite{tully2013}   & 9.3\cite{rusli2013} & 7.5  & $-$6.4\cite{Lakhchaura2018} &  & 6.0\cite{slee1994} \\
NGC 3258 & 32.5\cite{tully2013}   & 9.4\cite{waters2024} & 7.2  & $-$5.3\cite{osullivan2001} &  & 18.3\cite{sadler1989} \\
NGC 3091 & 54.9\cite{theureau2007}   & 9.6\cite{rusli2013} & 7.2  & $-$5.1\cite{Beuing1999} &  & 0.8\cite{romana2022}  \\
NGC 4594 & 9.9\cite{sabbi2018}    & 8.8\cite{jardel2011} & 6.9  & $-$5.2\cite{gonzalez2009} & 80.0\cite{petrov2006} &   \\
NGC 3998 & 14.2\cite{tully2013}   & 8.9\cite{walsh2012} & 6.1  & $-$4.4\cite{gonzalez2009} & 276.8\cite{helmboldt2007} & \\
NGC 1453 & 55.7\cite{willick1997}   & 9.5\cite{liepold2020} & 5.8  & $-$4.1\cite{osullivan2001} &  & 19\cite{slee1994} \\
NGC 383  & 66.6\cite{freedman2001}   & 9.6\cite{zhang2025} & 5.5  & $-$5.3\cite{wang2016} & 92.0\cite{healey2007} &  \\
NGC 4261 & 32.4\cite{tully2013}   & 9.2\cite{ferrarese1996} & 5.5  & $-$5.0\cite{gonzalez2009} & 300.0\cite{Yan2025} & \\
NGC 4374 & 18.5\cite{blakeslee2009}   & 9.0\cite{walsh2010} & 5.1  & $-$6.4\cite{gonzalez2009} & 186.0\cite{nagar2005} & \\
NGC 7619 & 53.0\cite{tonry2001}   & 9.4\cite{rusli2013} & 4.4  & $-$5.5\cite{osullivan2001} &  & 1.9\cite{Gultekin2019} \\
NGC 1399 & 17.6\cite{tonry1997}   & 8.9\cite{Kormendy2013} & 4.3  & $-$5.4\cite{osullivan2001} &  & 230.0\cite{killeen1988} \\
NGC 3379 & 12.2\cite{salaris1998}   & 8.6\cite{vandenbosch2010} & 4.0  & $-$7.3\cite{gonzalez2009} &  & 0.7\cite{nyland2016}  \\
NGC 4486B& 16.3\cite{mei2007}   & 8.8\cite{kormendy1997} & 3.8  & $-$4.9\cite{Pellegrini2010} &  & 0.2\cite{capetti2009}  \\
NGC 4291 & 26.3\cite{tully2013}   & 9.0\cite{schulze2011} & 3.8  & $-$6.3\cite{Williams2022} &  & 0.5\cite{brown2011}  \\
NGC 524  & 24.0\cite{tully2013}   & 8.9\cite{krajnovic2009} & 3.7  & $-$7.3\cite{Pellegrini2010} & 1.5\cite{nagar2005} &   \\
NGC 1374 & 18.4\cite{jensen2003}   & 8.8\cite{rusli2013} & 3.1  & $-$5.2\cite{osullivan2001} &  & 0.8\cite{brown2011}  \\
NGC 1332 & 21.9\cite{kundu2001}   & 8.8\cite{barth2016} & 3.1  & $-$5.6\cite{ueda2001} &  & 4.6\cite{brown2011}  \\
NGC 315  & 68.4\cite{Ricci2025}   & 9.3\cite{boizelle2021} & 2.9  & $-$4.4\cite{Williams2022} & 594.0\cite{Ricci2025} & \\
NGC 4526 & 16.4\cite{tonry2001}   & 8.7\cite{davis2013} & 2.8  & $-$6.2\cite{ueda2005} &  & 1.5\cite{nyland2016}  \\
NGC 6086 & 138.0\cite{mcconnell2011}  & 9.6\cite{mcconnell2011} & 2.8  & $-$4.3\cite{candini2023} &  & 15.5\cite{murgia2011} \\
UGC 2698 & 91.5\cite{van2015}   & 9.4\cite{cohn2021} & 2.8  & $-$5.8\cite{ueda2005} &  & 2.9\cite{condon2002}  \\
NGC 1684 & 65.5\cite{springob2014}   & 9.1\cite{dominiak2024} & 2.3  & $-$5.7\cite{ueda2005} &  & 101.0\cite{griffith1995} \\
NGC 4281 & 24.4\cite{tully2013}   & 8.7\cite{thater2019} & 2.3  & $-$5.3\cite{osullivan2001} &  & 0.5\cite{filho2002}  \\
NGC 5077 & 38.7\cite{Kormendy2013}   & 8.9\cite{defrancesco2008} & 2.3  & $-$6.2\cite{Balmaverde2015} & 341.0\cite{petrov2006} & \\
NGC 3608 & 22.6\cite{tully2013}   & 8.7\cite{schulze2011} & 2.1 & $-$6.5\cite{Panessa2006} &  & 0.26\cite{nyland2016}  \\
NGC 4342 & 22.9\cite{Kormendy2013}   & 8.7\cite{cretton1999} & 2.0 & $-$6.9\cite{Pellegrini2010} &  & 0.5\cite{nyland2016}   \\
NGC 5845 & 25.9\cite{tonry2001}   & 8.7\cite{schulze2011} & 1.9 & $-$6.5\cite{Pellegrini2010} &  & 0.1\cite{nyland2016}   \\
M81      & 3.6\cite{gerke2011}    & 7.8\cite{devereux2003} & 1.9 & $-$4.5\cite{Jang2014} & 132.0\cite{nagar2005} &  \\
NGC 3665 & 33.1\cite{cappellari2011}   & 8.8\cite{onishi2017} & 1.8 & $-$5.8\cite{Williams2022} &  & 9.0\cite{nyland2016}   \\
NGC 3377 & 11.2\cite{tonry2001}   & 8.3\cite{schulze2011} & 1.7 & $-$6.9\cite{Jang2014} &  & 0.2\cite{brown2011}   \\
NGC 3585 & 20.4\cite{tully2013}   & 8.5\cite{gultekin2009} & 1.6 & $-$6.7\cite{Jang2014} &  & 0.6\cite{brown2011}   \\
NGC 5128 & 3.4\cite{ferrarese2007}    & 7.8\cite{cappellari2009} & 1.6 & $-$3.0\cite{Pellegrini2010} & 2781.0\cite{horiuchi2006} & \\
NGC 4026 & 13.6\cite{tonry2001}   & 8.3\cite{gultekin2009} & 1.4 & $-$6.9\cite{Williams2022} &  & 1.3\cite{brown2011}   \\
NGC 7768 & 120.0\cite{tully2013}  & 9.1\cite{mcconnell2012} & 1.2 & $-$3.9\cite{Beuing1999} & 1.3\cite{liuzzo2010} &    \\
NGC 4697 & 12.4\cite{tully2013}   & 8.1\cite{davis2017} & 1.2 & $-$6.5\cite{Jang2014} &  & 0.1\cite{wrobel2008}   \\
NGC 3245 & 21.3\cite{tully2013}   & 8.4\cite{barth2001} & 1.1 & $-$6.0\cite{Williams2022} & 0.5\cite{filho2004} &    \\
NGC 7049 & 29.9\cite{tully2013}   & 8.5\cite{thater2019} & 1.1 & $-$4.2\cite{Beuing1999} &  & 21.0\cite{slee1994}  \\
NGC 5576 & 25.5\cite{tully2013}   & 8.3\cite{gultekin2009} & 1.1 & $-$6.3\cite{Jang2014} &  & 0.1\cite{brown2011}   \\
\hline
\end{tabular}
\vspace{1ex}
\parbox{0.95\textwidth}{\footnotesize
\textbf{Note.} (1) Galaxy name; (2) Luminosity distance; (3) Black hole mass in logarithmic scale, the values are from the direct estimates from the stellar or gas kinematics, the example references are \cite{Kormendy2013,onishi2017,Nguyen2020,cohn2021}; (4) Ring diameter (5.2 R$\rm _{S}$); (5) Eddington ratio estimated from bolometric luminosity, the bolometric luminosity are calculated based on the hard X-ray luminosity (bolometric correction using 15) from NED; (6) VLBI total flux density at 5 GHz from NED; (7) VLA total flux density at 5 GHz from NED.}
\end{table*}

\begin{table*}[htbp]
\centering
\captionsetup{labelformat=empty}
\caption*{Extended Data Table 2: Measured core-shift power-law indices $k$ in nearby AGNs}
\begin{tabular}{lcccc}
\hline
Name & Frequency (GHz) & Power index $k$ & Distance to BH ($r$) & Reference \\
(1) & (2) & (3) & (4) & (5) \\
\hline
\Msixty        & 8   & $1.96 \pm 0.37$   & 20       & This work \\
M87             & 43  & $1.06 \pm 0.10$   & 17       & Hada11 \\
NGC 315         & 43  & $1.28 \pm 0.11$   & 188      & Ricci25 \\
BL Lac          & 43  & $1.01 \pm 0.08$   & 18900    & O'Sullivan09 \\
NGC 4261        & 43  & $1.22 \pm 0.06$   & 347      & Haga15 \\
3C 454.3        & 43  & $1.17 \pm 0.12$   & 8135     & Chamani23 \\
CTA 102         & 86  & $1.11 \pm 0.25$   & 42500    & Fromm15 \\
NGC 3998        & 5   & $1.05 \pm 1.21$   & 900      & Yan25 \\
TXS 1508+572    & 43  & $1.00 \pm 0.20$   & 4462     & Benke24 \\
Mrk 501         & 8   & $1.10 \pm 0.20$   & 3767     & Croke10 \\
4C +58.17       & 43  & $1.10 \pm 0.10$   & 169200   & Kovalev08 \\
3C 111          & 86  & $1.27 \pm 0.17$   & 860      & Bartolini25 \\
3C 273          & 43  & $1.03 \pm 0.25$   & 40950    & Lisakov17 \\
PKS 2233$-$148  & 43  & $1.09 \pm 0.12$   & 90000    & Pushkarev19 \\
NRAO 530        & 227 & $1.14 \pm 0.26$   & 14426    & Lisakov25 \\
3C 345          & 86  & $0.71 \pm 0.11$   & 15282    & R$\rm \ddot{o}$der24 \\
\hline
\end{tabular}
\vspace{1ex}
\parbox{0.95\textwidth}{\footnotesize
\textbf{Note.} (1) Source name; (2) Highest observing frequency used in the core-shift measurement; (3) Power-law index $k$; (4) Deprojected core distance from the black hole at the highest frequency, expressed in units of Schwarzschild radii. Distances are estimated based on core-shift measurements; (5) reference for the core shift study. References: Hada11, ref.\cite{hada2011}; Ricci25, ref.\cite{Ricci2025}; O'Sullivan09, ref.\cite{OSullivan2009}; Haga15, ref.\cite{Haga2015}; Chamani23, ref.\cite{Chamani2023}; Fromm15, ref.\cite{Fromm2015}; Yan25, ref.\cite{Yan2025}; Benke24, ref.\cite{Benke2024}; Croke10, ref.\cite{Croke2010}; Kovalev08, ref.\cite{Kovalev2008}; Bartolini25, ref.\cite{Bartolini2025}; Lisakov17, ref.\cite{Lisakov2017}; Pushkarev19, ref.\cite{Pushkarev2019}; Lisakov25, ref.\cite{Lisakov2025}; R$\rm \ddot{o}$der24, ref.\cite{Roder2024}.
}
\end{table*}

\end{document}